%% file: main.tex
\documentclass[fleqn,usenatbib]{mnras} 

\usepackage{newtxtext,newtxmath}
\usepackage[T1]{fontenc}

\DeclareRobustCommand{\VAN}[3]{#2}
\let\VANthebibliography\thebibliography
\def\thebibliography{\DeclareRobustCommand{\VAN}[3]{##3}\VANthebibliography}

\usepackage{graphicx}	% Including figure files
\usepackage{subfigure}
\usepackage{color}
\usepackage{tabularx}
\usepackage{longtable}
\usepackage{bm}
\usepackage{here}
\usepackage{threeparttable}
\usepackage{float}
\usepackage{amsmath,amsfonts} 
\usepackage{natbib}
\usepackage{multicol,caption}
\usepackage{lscape}
\usepackage[normalem]{ulem}
\usepackage{verbatim}

\newcommand{\code}[1]{{\texttt{#1}}}
\newcommand{\tess}{\textit{TESS}}

\newcommand{\gaia}{\textit{Gaia}}
\newcommand{\wise}{\textit{WISE}}

\newcommand{\tar}{{TOI-1846}}
\providecommand{\msun}{\ensuremath{\rm\,M_\odot}}
\providecommand{\rsun}{\ensuremath{\rm\,R_\odot}}
\providecommand{\me}{\ensuremath{\rm\,M_\oplus}}
\providecommand{\re}{\ensuremath{\rm\,R_\oplus}}
\providecommand{\TESS}{\textit{TESS}}
\providecommand{\Gaia}{\textit{Gaia}}

\providecommand{\bjdtdb}{\ensuremath{\rm {BJD_{TDB}}}}

\providecommand{\msun}{\ensuremath{\,M_\Sun}}
\providecommand{\rsun}{\ensuremath{\,R_\Sun}}

\providecommand{\me}{\ensuremath{\,M_{\rm E}}}
\providecommand{\re}{\ensuremath{\,R_{\rm E}}}
\providecommand{\fave}{\langle F \rangle}
\providecommand{\fluxcgs}{10$^9$ erg s$^{-1}$ cm$^{-2}$}
\title[TOI-1846b]{TOI-1846\,b: A super-Earth in the radius valley orbiting a nearby M dwarf}

\author[A. Soubkiou et al.]
{
Abderahmane Soubkiou$^1$\thanks{E-mail: abdousoubkiou@gmail.com},
Khalid Barkaoui$^{2,3,4}$\thanks{E-mail: Khalid.Barkaoui@uliege.be},
Zouhair Benkhaldoun$^{1}$,
Mourad Ghachoui$^{1,2}$,  
\newauthor
Jamila Chouqar$^{1}$,
Benjamin V.\ Rackham$^{3,5}$, % Spectro OK!!
Adam Burgasser$^{11}$, % Spectro   OK!!
Emma Softich$^{11}$, % Spectro   OK!!
Enric Pallé$^{4,6}$, % Muscat 1 OK!!
\newauthor
Akihiko Fukui$^{4,8}$, % Muscat 1 OK
Norio Narita$^{4,7,8}$, % Muscat 1 OK
Felipe Murgas$^{4,6}$,  % Muscat 1 OK
Steve~B.~Howell$^{17}$, %ok
Catherine~A.~Clark$^{24}$, %ok
\newauthor
Colin Littlefield$^{17,23}$, %OK
Allyson Bieryla$^{9,10}$, % KeplerCam
Andrew W. Boyle$^{18}$, % High-res
David Ciardi$^{12}$, % High-res
Karen Collins$^{9}$, %TESS + LCO   OK!!
\newauthor
Kevin I.\ Collins$^{20}$, %LCO OK!!
Jerome de Leon$^{21}$,
Courtney D. Dressing$^{13}$,  % High-res
Jason Eastman$^{9}$, 
Emma Esparza-Borges$^{4,6}$, %OK!! 
\newauthor
Steven Giacalone$^{13}$,  % High-res
Holden Gill$^{13}$,  % High-res OK!!
Micha\"el~Gillon$^{2}$, %OK
Kai Ikuta$^{21}$, % Muscat 1 OK!!
J. M. Jenkins$^{17}$, % TESS Architects OK!!
Taiki Kagetani$^{21}$,
\newauthor
David W. Latham$^{9}$,  % TESS Architects OK
Mayuko Mori$^{8,22}$, % Muscat 1
Hannu Parviainen$^{6,4}$, %OK!!
Emily Pass$^{5, 9}$, % TrES RVs OK
G. Ricker$^{5}$,  % TESS Architects OK
Boris S. Safonov$^{15}$, % High-res OK!!
\newauthor
Arjun B. Savel$^{14}$,  % High-res
Richard P. Schwarz$^{9}$, % LCO OK!!
Sara Seager$^{11,6,14}$, % TESS Architects OK!!
Ivan A. Strakhov$^{15}$,% High-res
Gregor Srdoc$^{19}$, %LCO OK!!
R. Vanderspek$^{5}$,  % TESS Architects OK
\newauthor
Noriharu Watanabe$^{21}$, % Muscat 1 OK!!
Cristilyn N.\ Watkins$^{9}$, %LCO OK!!
J. Winn$^{16}$  % TESS Architects OK!!
\newauthor
\\
{\it A list of affiliations is given at the end of the paper}
}

\date{Accepted XXX. Received YYY; in original form ZZZ}

\pubyear{2024}

\begin{document}

\label{firstpage}
%\tableofcontents
\pagerange{\pageref{firstpage}--\pageref{lastpage}}
\maketitle

\begin{abstract}
We present the discovery and validation of a super-Earth planet orbiting the M dwarf star TOI-1846 (TIC 198385543). The host star ($K_{\rm mag} = 9.6$) is located 47 pc away and has a radius of $R_\star = 0.41 \pm 0.01$\rsun, a mass of $M_\star = 0.40 \pm 0.02$\msun\, and an effective temperature of $T_{\rm eff} = 3568 \pm 44$K. Our analyses are based on joint modelling of \tess\ photometry and ground-based multi-color photometric data. We also use high-resolution imaging and archival images, as well as statistical validation techniques to support the planetary system nature. We find that TOI-1846\,b  is  a super-Earth sized planet with radius of $R_{p}=1.79 \pm 0.07$\re\  and a predicted mass of $M_p = 4.4^{+1.6}_{-1.0}$\me\ (from the Chen \& Kipping relation) on a 3.9~d orbit, with an equilibrium temperature of $ T_{\rm eq} = 589 \pm 20 K$ (assuming a null Bond Albedo) and an incident flux of $S_p = 17.6 \pm 2.0~S_\oplus$. Based on the two RV measurements obtained with the TRES spectrograph and  high-resolution imaging, a non-planetary transiting companion is excluded.
With a radius of $\approx $1.8\re,\, TOI-1846\,b is within the sparsely populated radius range around 2\re\ known as the radius gap (or radius valley). 
This discovery can contribute to refining the precise location of the radius valley for small planets orbiting bright M dwarfs, thereby enhancing our understanding of planetary formation and evolution processes.
\end{abstract}


%%%%%%%%%%%%%%%%%%%%%%%%%%%%%%%%%%%%%%%%%%%%%%%%%%

%%%%%%%%%%%%%%%%% BODY OF PAPER %%%%%%%%%%%%%%%%%%

\section{Introduction}
The past few decades have witnessed a remarkable surge in the discovery of exoplanets, enriching our understanding of the universe's diversity. Over 5700 exoplanets\footnote{\href{https://exoplanetarchive.ipac.caltech.edu/}{NASA Exoplanet Archive, }{\tt retrieved on 3 October 2024}} have been confirmed, many of them orbiting M dwarf stars. These discoveries, largely driven by missions such as NASA's Kepler \citep{howard2012planet, fressin2013false} and NASA's Transiting Exoplanet Survey Satellite (\tess; \citet{ricker2015transiting}), have revealed a bimodal distribution in the sizes of these exoplanets, with a noticeable gap known as the radius valley \citep{Fulton2017, van2018asteroseismic, cloutier2020toi, petigura2020two}. This gap, evident in planet radius versus orbital period (equally versus planet equilibrium temperature or stellar irradiation), is characterized by a notable paucity of planets with radii between 1.5 and 2$\re$ \citep{fulton2018california, berger2020gaia, cloutier2020evolution}. It distinctly separates two primary planetary categories: the smaller, typically rocky super-Earths, and the more substantial, gas-rich mini-Neptunes, raising intriguing questions about the formation processes and evolution of these planetary bodies. 

Various mechanisms have been proposed to explain the origin of this radius gap. One proposed explanation is photoevaporation, where X-ray and extreme ultraviolet (XUV) radiation from the host star can strip away the gaseous envelopes of smaller, nearby planets within the first $\approx$ 100 million years of their existence \citep{owen2013kepler, chen2016evolutionary, owen2017evaporation, LopezRice2018}. Another prominent theory is core-powered mass loss, where the heat from the core of a planet, originating from its formation, drives atmospheric loss over billions of years \citep{ginzburg2018core, gupta2019sculpting, gupta2020signatures}. Additional mechanisms, such as impact erosion from planetesimals, have also been proposed, with the potential to either strip atmospheres or contribute to the growth of secondary atmospheres \citep{schlichting2015atmospheric, wyatt2020susceptibility}. An alternative to atmospheric processing is the hypothesis of distinct rocky and non-rocky planet populations forming under different conditions, with the former group emerging from a gas-poor formation environment, where gas accretion is hindered until the protoplanetary disk nearly dissipates \citep{lee2014make, 2016ApJ...817...90L, LopezRice2018}.

Furthermore, a study by \citet{Luque&palle2022} analyzed 34 well-characterized exoplanets and observed a density gap rather than a radius valley when separating rocky and water-rich exoplanets. This challenges the conventional understanding of a radius valley and suggests a different classification based on planetary density. In their study, they categorized planets into three density regimes: rocky planets, water worlds, and puffy mini-Neptunes, each with distinct characteristics. This finding aligns with pebble accretion models for planet formation around low-mass stars \citep{venturini2020nature, brugger2020pebbles}. Notably, the larger radius dispersion among puffy planets hints at varied H/He accretion histories rather than atmospheric loss, suggesting that water worlds and puffy mini-Neptunes may belong to the same planetary group. Although these results are promising, the limited size of the exoplanet sample used in these studies means that more extensive research with larger and accurately characterized exoplanet samples is crucial to solidifying these conclusions.

In this paper, we present the discovery and characterization of a new super-Earth planet orbiting the M dwarf star TOI-1846, located at 47 parsecs. This star presents a compelling target for exoplanet exploration, given its proximity, its small size, and its infrared brightness. TOI-1846\,b is a super-Earth-sized planet, with a radius of approximately 1.8$\re$, placing it within the intriguing radius valley area of exoplanet sizes. Such planets are relatively rare, and their study can provide vital clues about planet formation and evolution processes. Additionally, M dwarfs are promising candidates for the search of small, temperate exoplanets using transit methods. The resulting transit signal is significantly more pronounced than that of similar planets orbiting Sun-like stars, making such planets easier to detect and characterize. Thus, planetary systems orbiting M dwarfs are prime targets for atmospheric characterization. \\

The paper is structured as follows. Section \ref{sec:2} presents the data from \tess\ and all ground-based observations. Stellar characterization is described in Section \ref{sec:3}, vetting and validation of the transit signals are covered in Section \ref{sec:4}, and data analyses and results are presented in Section \ref{sec:5}. We discuss our findings in Section \ref{sec:6} and give our conclusions in Section \ref{sec:7}.

\section{Observations}\label{observations}\label{sec:2}
\subsection{\tess\ photometry}

% \tar\   (TIC 198385543) was observed by \TESS\  with a 2-min cadence in the nominal mission in sectors 17 (08 October to 02 November 2019), 20 (25 December 2019 to 20 January 2020), 23--26 (19 March to 04 July 2020). In the extended mission, it was re-observed in sectors 40 (25 June to 23 July 2021), 47 (31 December 2021 to 27 June 2022) and 50--54 (26 March to 04 August 2022), 56--60 (02 September 2022 to 18 June 2023), 73--74 (7 Dec 2023 to 30 Jan 2024), and 77--81 (26 Mar to 10 Aug 2024). 

\begin{figure}
\centering
\includegraphics[width=0.5\textwidth]{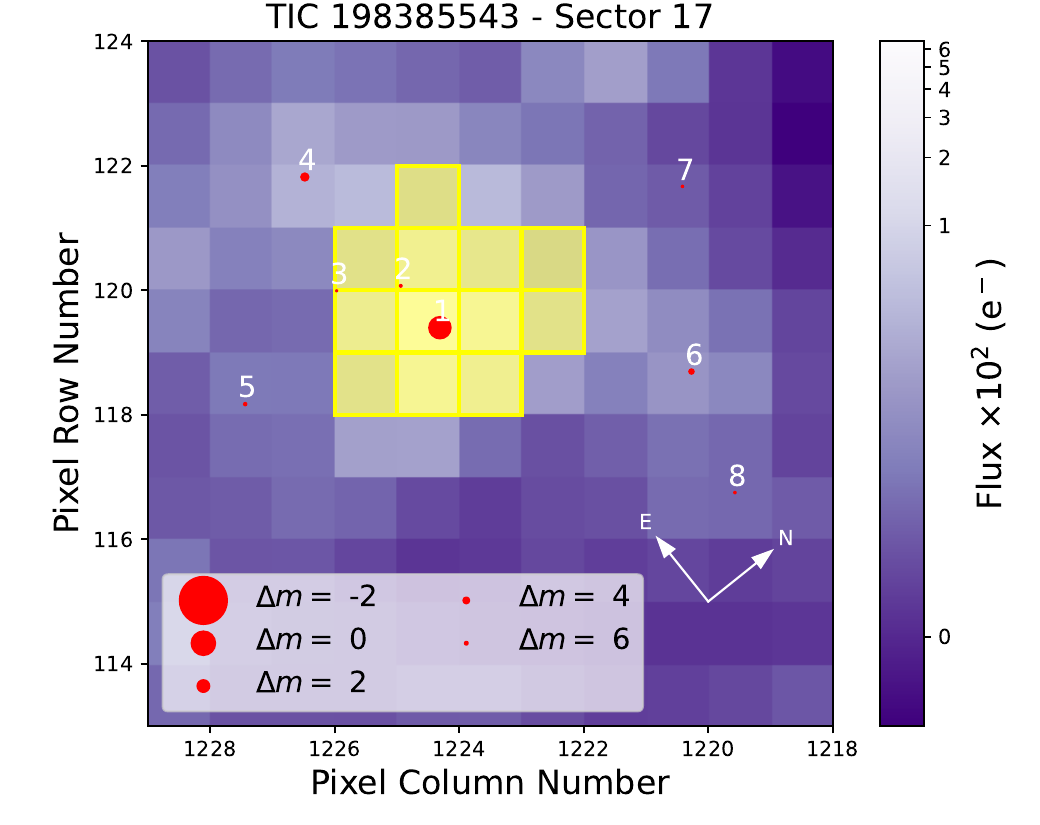}
\caption{
Target pixel files (TPF) of \tar\ in \tess\ Sector 17, made with \code{tpfplotter} \citep{Aller2020}. The yellow shaded region represents the aperture used to extract the photometry. The red circles are the \gaia\ DR2 sources. Different sizes represent different magnitudes in contrast with \tar. } 
\label{tess_fov}
\end{figure}

\tar\   (TIC 198385543) was observed by \TESS\  with a 2-min cadence in 6 sectors of the primary mission and 19 sectors in the extended missions (See Table~\ref{TESS_obs_table}).
Figure~\ref{tess_fov} shows the Target Pixel Files (TPFs) and the Simple Aperture Photometry (SAP) apertures used in each sector with the location of nearby Gaia DR2 \citep{GaiaCollaboration2018} sources overplotted. The primary contamination sources in the apertures are two faint stars, TIC 198385544 at 17.89\arcsec and TIC 1271044697 at 29.8\arcsec\ from the target. Both stars are fainter than the target by 4 mag in the \tess\ band.

\begin{figure}
%\centering
\includegraphics[width=0.49\textwidth]{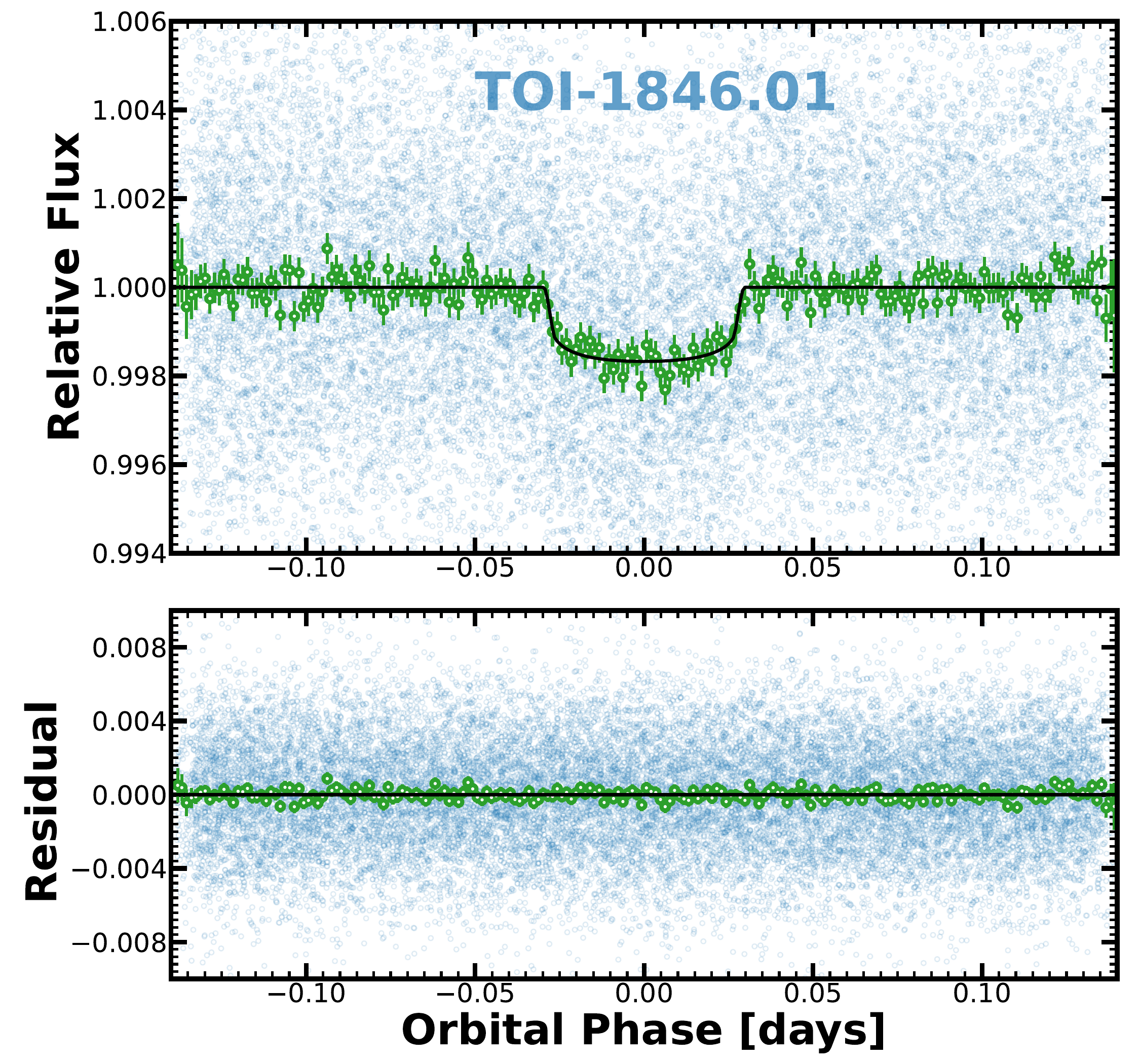}
\caption{TESS PDCSAP phase-folded light curves of TOI-1846. The blue and green  points are unbinned and binned (2-minutes) data. The solid line shows the best-fitting transit model.}
\label{TESS_phase_folded}
\end{figure}

\begin{figure}
\centering
\includegraphics[scale=0.4]{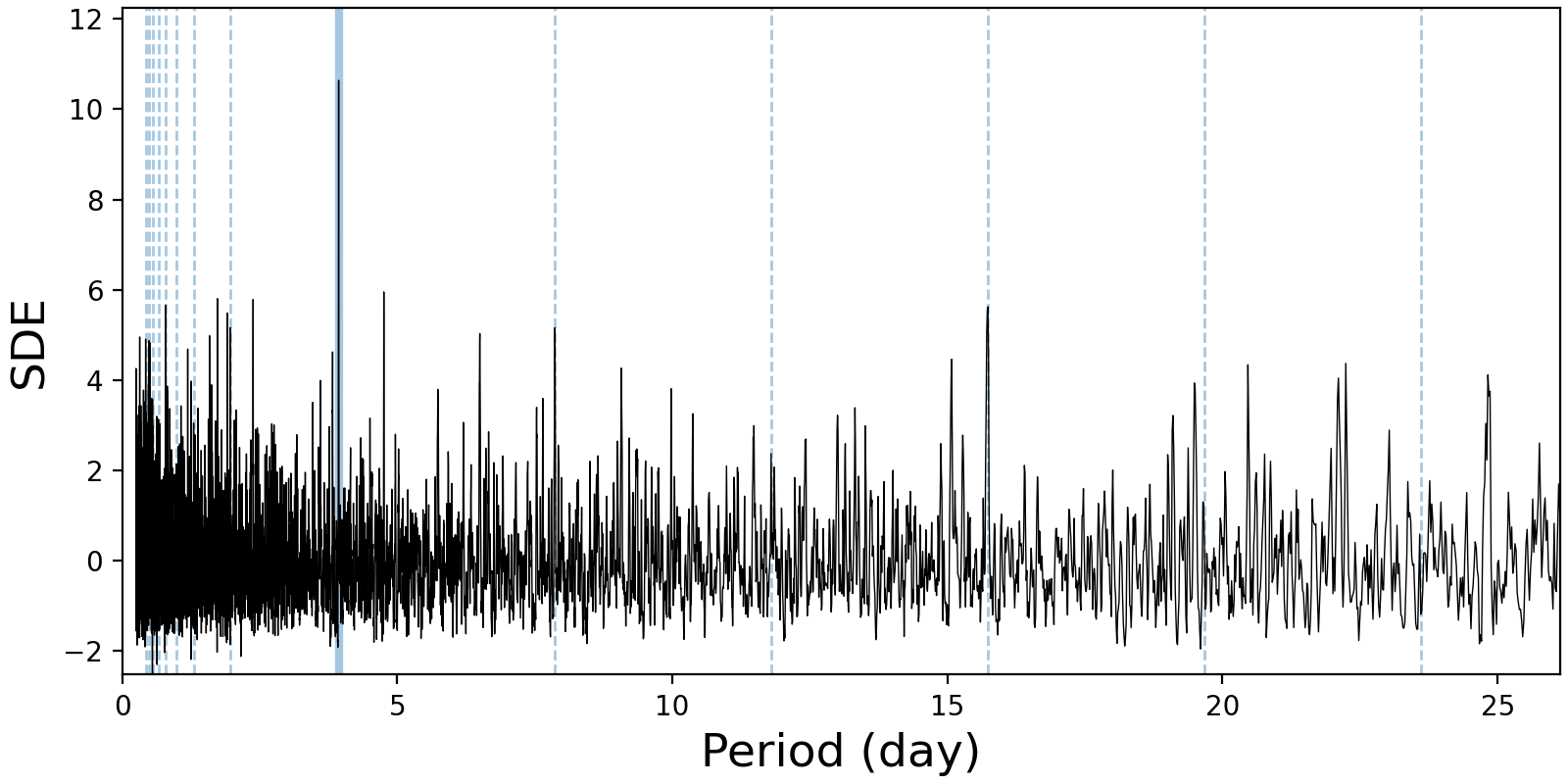}
\caption{TLS power spectra of the detrended \tess\  PDC light curves of TOI-1846.} 
\label{SDE_tess_photometry}
\end{figure}

For our analyses, we used \texttt{lightkurve} \citep{lightkurvecollaboration} to retrieve the 2-minute Presearch Data Conditioning light curves \citep[SPOC:][]{Stumpe2012,Stumpe2014,Smith2012} for all sectors from the Mikulski Archive for Space Telescopes. %Long-term trends are already removed in the PDC light curves but further detrending was highly needed. 
%We detrended our data using \todo{TODO by soubkiou}. We also removed all data points with non-zero quality flags. 
We removed all the data points flagged as "bad quality" and then detrended the light curves to remove stellar variability using a biweight time-windowed slider via wotan \citep{Hippke2019}. 
TESS photometric observations are listed in Table~\ref{TESS_obs_table}.
\tess\ transit light curves for TOI-1846 are  presented in Figure~\ref{TESS_phase_folded}, while Figure~\ref{SDE_tess_photometry} shows the periodogram of the TESS data.

%%%% TESS Observations Table
\begin{table}
 \begin{center}
 {\renewcommand{\arraystretch}{1.1}
 \resizebox{0.5\textwidth}{!}{% }
 \begin{tabular}{l c c c}
 % \toprule
 \hline
Sector  &  Camera & CCD &  Observation date\\ 
 \hline
 17  & 4 & 4 & 2019-Oct-07 -- 2019-Nov-02 \\
 20  & 4 & 3 & 2019-Dec-24 -- 2020-Jan-21 \\
 23  & 4 & 2 & 2020-Mar-18 -- 2020-Apr-16 \\
 24  & 2 & 1 & 2020-Apr-16 -- 2020-May-13 \\
 25  & 2 & 2 & 2020-May-13 -- 2020-Jun-08\\
 26  & 2 & 2 & 2020-Jun-08 -- 2020-Jul-04 \\
 40  & 2 & 2 & 2021-Jun-24 -- 2021-Jul-23 \\
 47  & 4 & 3 & 2021-Dec-30 -- 2022-Jan-28\\
 50  & 4 & 2 & 2022-Mar-26 -- 2022-Apr-22\\
 51  & 3 & 3 & 2022-Apr-22 -- 2022-May-18 \\
 52  & 2 & 2 & 2022-May-18 -- 2022-Jun-13 \\
 53  & 2 & 2 & 2022-Jun-13 -- 2022-Jul-09 \\
 54  & 4 & 1 & 2022-Jul-09 -- 2022-Aug-05 \\
 56  & 4 & 4 & 2022-Sep-01 -- 2022-Sep-30 \\
 57  & 4 & 4 & 2022-Sep-30 -- 2022-Oct-29 \\
 58  & 4 & 4 & 2022-Oct-29 -- 2022-Nov-26 \\
 59  & 4 & 3 & 2022-Nov-26 -- 2022-Dec-23 \\
 60  & 4 & 3 & 2022-Dec-23 -- 2023-Jan-18 \\
 73  & 4 & 3 & 2023-Dec-07 -- 2024-Jan-03  \\ 
 74  & 3 & 3 &  2024-Jan-03 -- 2024-Jan-30  \\
 77  & 2 & 1 &  2024-Mar-26 -- 2024-Apr-23 \\
 78  & 2 & 1 &  2024-Apr-23 -- 2024-May-21  \\
 79  & 2 & 2 &  2024-May-21 -- 2024-Jun-18   \\
 80  & 4 & 1 &  2024-Jun-18 -- 2024-Jul-15  \\
 81  & 4 & 1 &  2024-Jul-15 -- 2024-Aug-10   \\
\hline
 \end{tabular}}}
 \caption{TESS observations with 2-minute cadence for TOI-1846.}
 \label{TESS_obs_table}
 \end{center}
\end{table}%}

\subsection{Ground-Based photometry}\label{gbp}
\begin{figure}
%\centering
\includegraphics[width=0.5\textwidth]{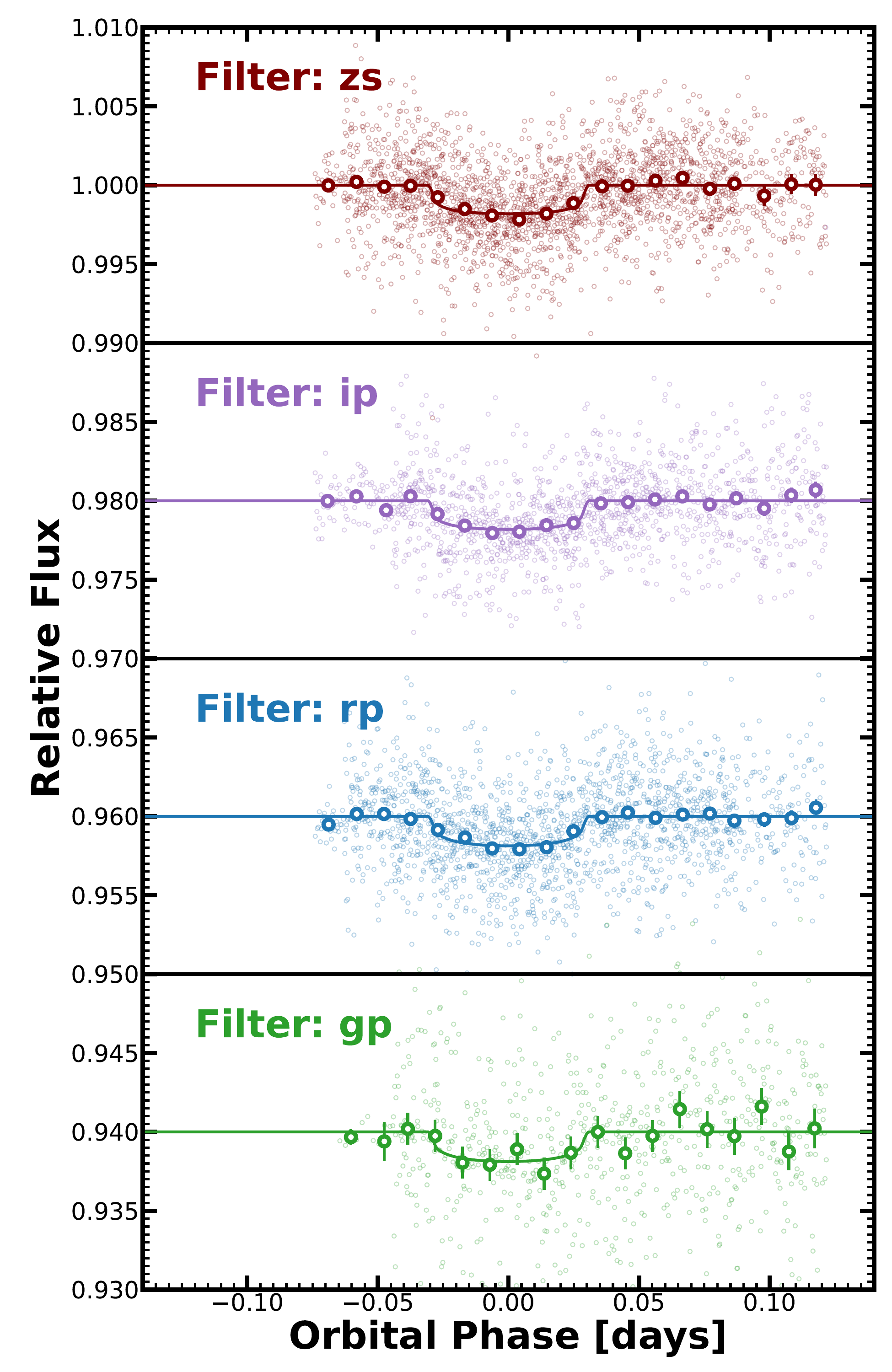}
\caption{Ground-based phase-folded transit light curves of TOI-1846\,b collected with MuSCAT, MuSCAT2, MuSCAT3 and KeplerCam.} The coloured lines are the best-fitting transit model. The light curves are shifted along the y-axis for visibility. 
\label{GB_phase_folded}
\end{figure}

\begin{figure*}
\centering
\includegraphics[width=\textwidth]{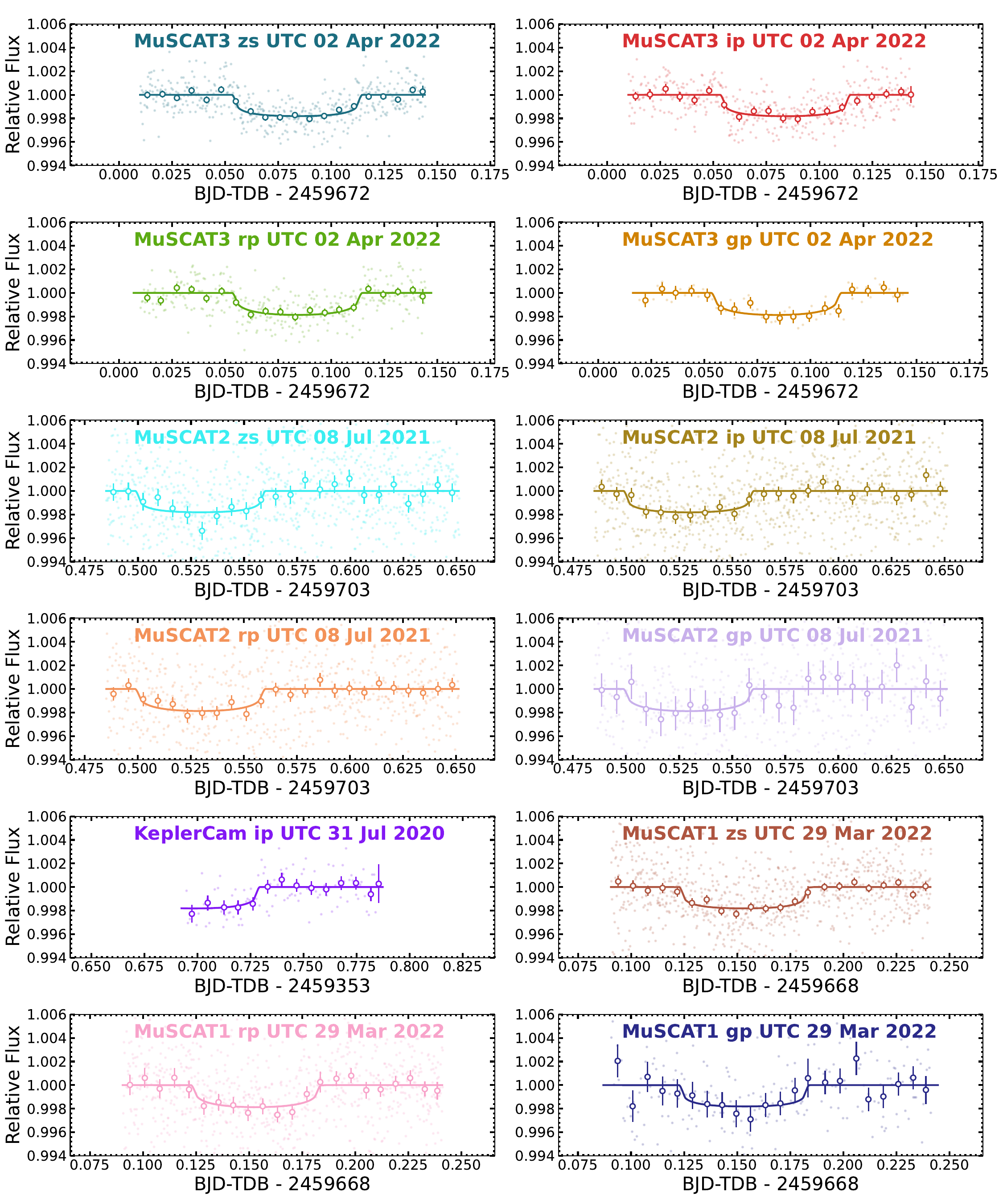}
\caption{Ground-based photometric follow-up for TOI-1846\,b obtained by MuSCAT, MuSCAT2, MuSCAT3 and KeplerCam. The solid lines show the superimposed best-fitting transit model.} 
\label{GB_photometry}
\end{figure*}

We collected a series of ground-based observations of \tar, as part of the \tess\ Follow-up Observing Program (TFOP\footnote{{\tt TFOP:} \url{https://tess.mit.edu/followup}}), to (1) confirm the transit signal is associated with the target star and rule out nearby eclipsing binaries as the source of the transit signal; (2) examine the chromaticity of the transit signal; and (3) refine the transit ephemeris and radius measurement.  We used the \tess\ Transit Finder, which is a customized version of the Tapir software package \citep{Jensen2013}, to schedule our transit observations. We summarize the details in Table \ref{po} and describe individual observations below. The observed transit light curves are shown in Figures~\ref{GB_phase_folded} and \ref{GB_photometry}.

\subsubsection{KeplerCam}
We obtained two lightcurves using the KeplerCam CCD on the 1.2m telescope at the Fred Lawrence Whipple Observatory (FLWO) in Arizona, USA. KeplerCam is a 4096x4096 pixel CCD camera with a Fairchild 486 detector. It is run in 2x2 bin mode with a resulting pixel scale of 0.672\arcsec/pixel. Data reduction and aperture photometry were performed using {\tt AstroImageJ} \citep[AIJ:][]{Collins2017} software.  No transit was detected and there was no sign of any nearby eclipsing binaries within $2.5'$ of the target star. Based on the newly updated ephemeris, these observations were all pre-ingress out-of-transit observations and consistent with a flat lightcurve. On UT2021-05-19 an egress was detected on the target star. Both observations used a 6-pixel ($4.03''$) uncontaminated target aperture.

\subsubsection{MuSCAT}

A full transit of \tar\ b was observed on UTC 29 March 2022 with the multicolor imager (MuSCAT) \citep{Norio:2015} on the 1.88m telescope of National Astronomical Observatory of Japan (NAOJ) located in Okayama, Japan. MuSCAT has three optical channels each equipped with a 1k $\times$ 1k CCD camera with a pixel scale of 0.361\arcsec. The observation was performed with exposure times of 60, 15, and 15\,s for the $g$, $r$, and $z_{\rm s}$ bands, respectively. The telescope was slightly defocused during the observation. The full width at half maximum (FWHM) of stellar point spread function was around 1\farcs5--2\farcs5 depending on the band and sky condition. The obtained images were calibrated for dark and flat-field patterns in a standard manner. We performed aperture photometry on the calibrated images using a custom pipeline \citep{Fukui:2011} with optimized uncontaminated aperture of radii of 10, 12, and 16 pixels for the $g$, $r$, and $z_{\rm s}$ bands, respectively.

\subsubsection{MuSCAT2}\label{MuSCAT2}

\tar\ b was observed on UTC 03 May 2022 with the multicolor imager MuSCAT2 \citep{Narita2019} mounted on the 1.5m Telescopio Carlos S\'{a}nchez (TCS) at Teide Observatory, Spain. MuSCAT2 has four CCDs with $1024 \times 1024$ pixels and each camera has a field of view of $7.4 \times 7.4$ arcmin$^2$ (pixel scale of 0.44\arcsec/pixel). The instrument is capable of taking images simultaneously in $g'$, $r'$, $i'$, and $z_s$ bands with little (1--4 seconds) read out time.
The observations were made with the telescope on focus, the exposure times were initially set to 20, 20, 15 and 15 seconds in $g'$, $r'$, $i'$ and $z_s$ respectively. The raw data were reduced by the MuSCAT2 pipeline \citep{Parviainen2019} which performs standard image calibration, photometric measurements. An aperture radius of 10.8\arcsec for the target star was selected by the pipeline as the best uncontaminated photometric aperture to extract the light curves.

\subsubsection{LCOGT-2m0/MuSCAT3}\label{MuSCAT3}

%%%% updated by Khalid on 2024.05.14

Two full transits of TOI-1846\,b were observed simultaneously in the $g'$, $r'$, $i'$ and $z_\mathrm{s}$ filters with the 2.0m Las Cumbres Observatory Global Telescope (LCOGT, \citet{Brown_2013})
on UTC 01 July 2021 and 06 April 2022. The telescope is equipped with the MuSCAT3 multi-band imager \citep{Narita_2020SPIE11447E}. 
We discarded the data from the first observation from our global analysis, 
because of a stellar flare during the pre-ingress. The exposure times were set to 23, 22, 23 and 22 seconds for the $g$, $r$, $i$ and $z_s$ bands, respectively. We extracted photometry with uncontaminated apertures of 2.65\arcsec--7.95\arcsec using the {\tt AstroImageJ} software. 
\begin{table}
    \centering
    \caption{Ground-based photometric follow-up observations log for \tar.}
    \resizebox{0.5\textwidth}{!}{% }
    \begin{tabular}{cccccl}
        \hline\hline
        Telescope &Filter &Pixel  & Aperture  &Coverage  &Date    \\
                  &       &  Scale (\arcsec)     &   Size (pixel) &      &  \\
        \hline
        KeplerCam & $i'$                  & 0.672 & 7  & Egress & 31 Jul  2020 \\
        MuSCAT    & $g'$,$r'$,$z_s$       & 0.36  & 16 & Full   & 29 Mar 2022 \\
        MuSCAT2   & $g'$,$r'$,$i'$,$z_s$  & 0.44  & 24 & Full   & 08 Jul  2021 \\ 
        MuSCAT3   & $g'$,$r'$,$i'$,$z_s$  & 0.265 & 18 & Full   & 02 Apr 2022 \\
        \hline
    \end{tabular} }
    \label{po}
\end{table}

\subsection{Spectroscopic Observations}
\subsubsection{TRES}

Two spectra of TOI-1846 were observed with the Tillinghast Reflector Echelle Spectrograph (TRES; \cite{gaborthesis_TrES}) mounted on the 1.5m Tillinghast Reflector telescope at the Fred Lawrence Whipple Observatory (FLWO) in Arizona, USA.
TRES is a fiber-fed optical (390--910 nm) spectrograph with a resolving power of R=44,000. This target was observed as part of the \tess\ reconnaissance follow-up. The spectra were observed at opposite quadratures assuming a circular orbit. The spectra were extracted using the standard TRES pipeline as outlined in \cite{buchhave2010}. The radial velocities were extracted using a pipeline optimized for mid- to late-M dwarfs. This pipeline also produces carefully calibrated radial-velocity uncertainties, taking into account the signal-to-noise ratio of the observed spectra, rotational broadening, template mismatch, the long-term stability of the spectrograph, and errors in the barycentric correction. Further details are provided in  \citet{Pass_2023}.
The two resulting RV measurements are $-20.720 \pm 0.056$~km/s on UTC 26 March 2021 and $-20.831 \pm 0.034$~km/s for UTC 6 May 2021, and the upper limit on the projected rotational velocity is found to be $v\sin(i) < 3.4$~km/s. The two RVs measurements allow us to rule out the non-planetary transiting companion.

\subsubsection{Shane/Kast}
\label{sec:kast}

TOI-1846 was observed with the Shane 3m Kast double spectrograph \citep{kastspectrograph} on UTC 6 2024 October 6 in clear conditions with 1$\farcs$3 seeing.
We used the 1$\farcs$5 (3.5~pixel) slit aligned with the parallactic angle and the 600/4310 grism on the blue channel to acquire 3750--5600~{\AA} spectra at a resolution of $\lambda/\Delta\lambda$ $\approx$ 1100, and the 
600/7500 grating on the red channel to acquire 5800--9000~{\AA} spectra at a resolution of $\lambda/\Delta\lambda$ $\approx$ 1500.
Total integrations of 300~s were acquired in both channels, split into two exposures of 150~s each in the red, at an average airmass of 1.15. The G2~V star HD 168009 ($V$ = 6.3) was observed immediately afterward at a similar airmass for telluric absorption correction, followed by the spectrophotometric flux standard HR~7596 ($V$ = 5.6; \citet{1992PASP..104..533H,1994PASP..106..566H}). Arclamps (HeHgCd in the blue, HeNeAr in the red), quartz flat field lamps, and bias frames were obtained at the start of the night for wavelength and pixel response calibration.
Data were reduced using the {\tt kastredux} package\footnote{\url{https://github.com/aburgasser/kastredux}} following standard procedures (cf.~\citet{Barkaoui_2024A&A}).
Our final data have signals-to-noise (S/N) of 60 at 5400~{\AA} and 150 at 7500~{\AA}.

\subsubsection{SpeX}

We obtained a medium-resolution near-infrared spectrum of TOI-1846 with the SpeX spectrograph \citep{Rayner2003} on the 3.2-m NASA Infrared Telescope Facility (IRTF) on UTC 19 Apr 2022. The conditions were clear with seeing of 0$\farcs$8--1$\farcs$0. We used the short-wavelength cross-dispersed (SXD) mode with the $0.3'' \times 15''$ slit aligned to the parallactic angle, which gives spectra covering 0.80--2.42\,$\mu$m with a resolving power of $R{\sim}2000$. We collected six exposures of 79.7\,s each, nodding in an ABBA pattern, totaling 8.0\,min on source. We collected a set of standard SXD flat-field and arc-lamp exposures immediately after the science frames, followed by a set of eight, 3.7-s exposures of the A0\,V star HD\,143187 ($V{=}6.3$). We reduced the data using Spextool v4.1 \citep{Cushing2004}, following the standard approach \citep[cf.][]{Delrez2022, Ghachoui2023, Barkaoui_2024A&A}. The final spectrum has a median SNR per pixel of 89 with peaks in the $J$, $H$, and $K$ bands of 112, 121, and 108, respectively, and an average of 2.5 pixels per resolution element.

\subsection{High Angular Resolution Imaging}\label{sec:high_res_obs}

%As part of our standard process for validating transiting exoplanets to assess the the possible contamination of bound or unbound companions on the derived planetary radii \citep{ciardi2015}. We observed  TOI~1846 with a combination of high-resolution resources including near-infrared adaptive optics (AO) imaging at Lick and Palomar Observatories, complemented by optical speckle observations at SAI-MSU and Gemini-North.  While the optical observations tend to provide higher resolution, the NIR AO tend to provide better sensitivity, especially to lower-mass stars. The combination of the observations in multiple filters enables better characterization for any companions that may be detected. The observations are described in detail in the following subsections and a summary is provided in Table~\ref{tab:high-res-imaging}. Gaia DR3 is also used to provide additional constraints on the presence of undetected stellar companions as well as wide companions.

As part of the validation process for TOI-1846, we employed high-resolution imaging to search for potential contamination from bound or unbound stellar companions, which could affect the derived planetary radii \citep{ciardi2015}. Observations were conducted using near-infrared (NIR) adaptive optics (AO) imaging at the Lick and Palomar Observatories, complemented by optical speckle observations at SAI-MSU and Gemini-North.

Each observational technique provides complementary advantages: optical observations offer higher spatial resolution, while NIR AO imaging achieves better sensitivity, especially for detecting lower-mass companions. Gaia DR3 astrometric data further constrains the presence of both wide and unresolved companions. A summary of the observational results is provided in Table \ref{tab:high-res-imaging}, and detailed descriptions of each observation are presented below.
\begin{table}[h!]
 \begin{center}
 {\renewcommand{\arraystretch}{1.1}
 \resizebox{0.5\textwidth}{!}{% }
 \begin{tabular}{l c c c l}
 % \toprule
 \hline
Facility & Instrument & Filter & Image Type & Date [UTC] \\
 \hline
    SAI-2.5& Speckle Plarimeter & $I_c$ & Speckle & 2020-02-03 \\
    Shane& ShARCS & $J$ & AO & 2021-03-29 \\
    Shane& ShARCS & $Ks$ & AO &2021-03-29 \\
	Palomar & PHARO & Br-$\gamma$ & AO & 2021-06-20 \\
    Gemini-North & 'Alopeke  &562 nm & Speckle & 2024-05-25 \\
    Gemini-North & 'Alopeke  & 832 nm & Speckle & 2024-05-25 \\
    \hline
\end{tabular}}}
\caption{A summary of the high-resolution imaging observations of TOI-1846.}
\label{tab:high-res-imaging}
\end{center}
\end{table}

\subsubsection{Palomar Observations}

The Palomar Observatory observations of TOI~1846 were made with the PHARO instrument \citep{hayward2001} behind the natural guide star AO system P3K \citep{dekany2013} on UTC 20 June 2021 in a standard 5-point quincunx dither pattern with steps of 5\arcsec\ in the narrow-band $Br-\gamma$ filter $(\lambda_o = 2.1686; \Delta\lambda = 0.0326~\mu$m).  Each dither position was observed three times, offset in position from each other by 0.5\arcsec\ for a total of 15 frames; with an integration time of 31.1 seconds per frame, respectively for total on-source times of 466 seconds. PHARO has a pixel scale of $0.025\arcsec$ per pixel for a total field of view of $\sim25\arcsec$.
    
The AO data were processed and analyzed with a custom set of IDL tools.  The science frames were flat-fielded and sky-subtracted.  The flat fields were generated from a median average of dark subtracted flats taken on-sky.  The flats were normalized such that the median value of the flats is unity.  The sky frames were generated from the median average of the 15 dithered science frames; each science image was then sky-subtracted and flat-fielded.  The reduced science frames were combined into a single combined image using a intra-pixel interpolation that conserves flux, shifts the individual dithered frames by the appropriate fractional pixels, and median-coadds the frames.  The final resolutions of the combined dithers were determined from the full-width half-maximum of the point spread functions: 0.102\arcsec.  
	
The sensitivities of the final combined AO image were determined by injecting simulated sources azimuthally around the primary target every $20^\circ $ at separations of integer multiples of the central source's FWHM \citep{furlan2017}. The brightness of each injected source was scaled until standard aperture photometry detected it with $5\sigma $ significance. The resulting brightness of the injected sources relative to TOI~1846 set the contrast limits at that injection location. The final $5\sigma $ limit at each separation was determined from the average of all of the determined limits at that separation and the uncertainty on the limit was set by the rms dispersion of the azimuthal slices at a given radial distance.  The final sensitivity curve for the Palomar data is shown in Figure~\ref{fig:high_res_ima}; no additional stellar companions were detected in agreement with observations from Shane, SAI, and Gemini (see below).

\subsubsection{Shane Observations}

We observed TOI-1846 on UTC 29 March 2021 using the ShARCS camera on the Shane 3m telescope at Lick Observatory \citep{2012SPIE.8447E..3GK, 2014SPIE.9148E..05G, 2014SPIE.9148E..3AM}. Observations were taken with the Shane adaptive optics system in natural guide star mode in order to search for nearby, unresolved stellar companions. We collected two sequences of observations, one with a $Ks$ filter ($\lambda_0 = 2.150$ $\mu$m, $\Delta \lambda = 0.320$ $\mu$m) and one with a $J$ filter ($\lambda_0 = 1.238$ $\mu$m, $\Delta \lambda = 0.271$ $\mu$m) and we reduced the data using the publicly available SImMER pipeline \citep{Savel:2020}. Our reduced images and corresponding contrast curves are shown in Figure~\ref{fig:high_res_ima}. We find no nearby stellar companions within our detection limits.

\subsubsection{SAI Observations}

We observed TOI-1846 on UTC 3 February 2021  with the Speckle Polarimeter \citep{Safonov2017} on the 2.5m telescope at the Caucasian Observatory of Sternberg Astronomical Institute (SAI) of Lomonosov Moscow State University. SPP uses Electron Multiplying CCD Andor iXon 897 as a detector. The atmospheric dispersion compensator allowed observation of this relatively faint target through the wide-band $I_c$ filter. The power spectrum was estimated from 4000 frames with 30 ms exposure. The detector has a pixel scale of $20.6$ mas pixel$^{-1}$, and the angular resolution was 89 mas. We did not detect any stellar companions brighter than $\Delta I_C=4$ and $6$ at $\rho=0\farcs25$ and $1\farcs0$, respectively, where $\rho$ is the separation between the source and the potential companion (see Figure~\ref{fig:high_res_ima}).

\subsubsection{Gemini Observations}

\begin{figure*}
\centering
\includegraphics[width=0.43\textwidth]{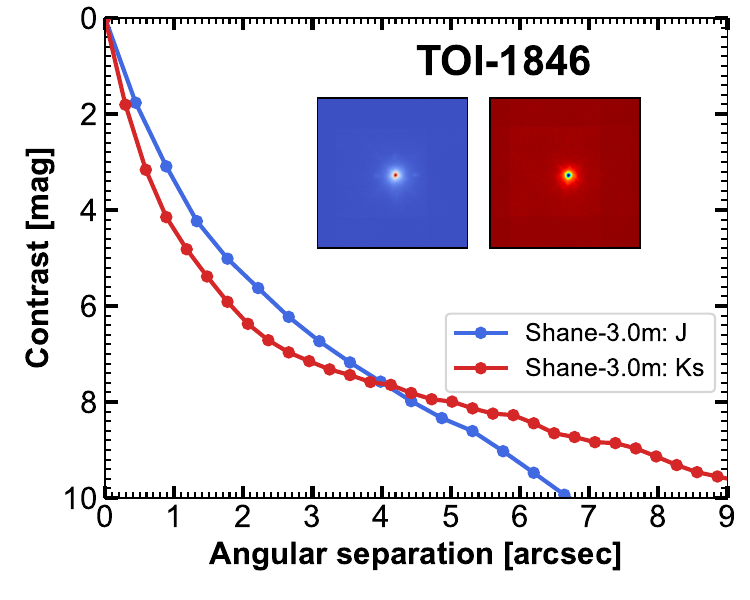}
\includegraphics[width=0.43\textwidth]{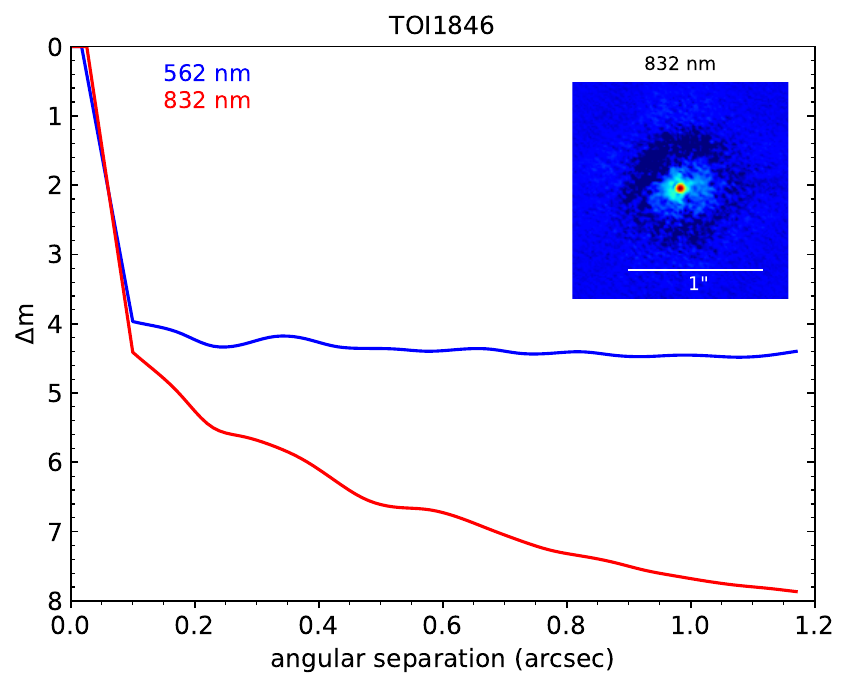}\\
\includegraphics[width=0.45\textwidth]{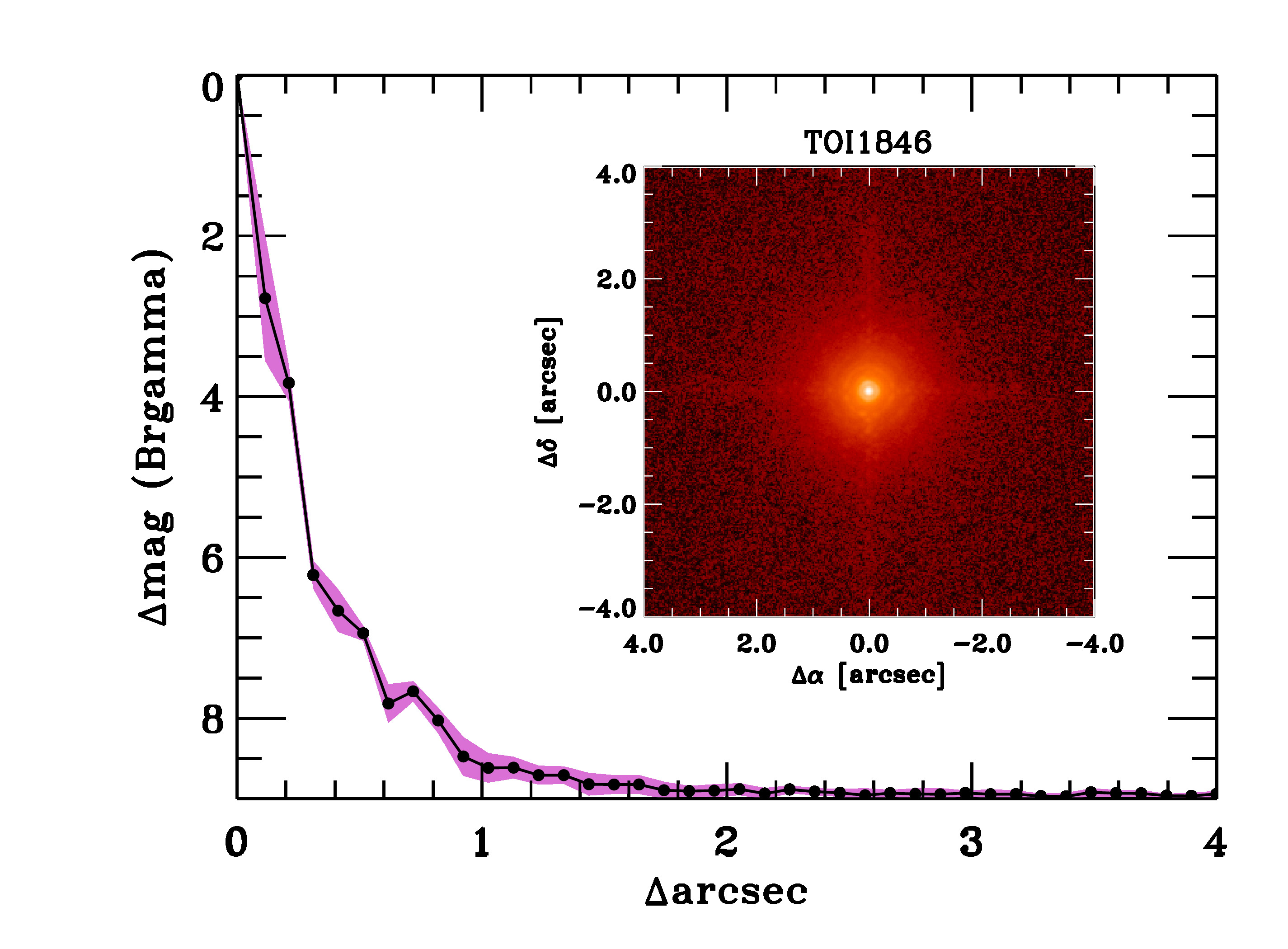} 
\includegraphics[width=0.45\textwidth]{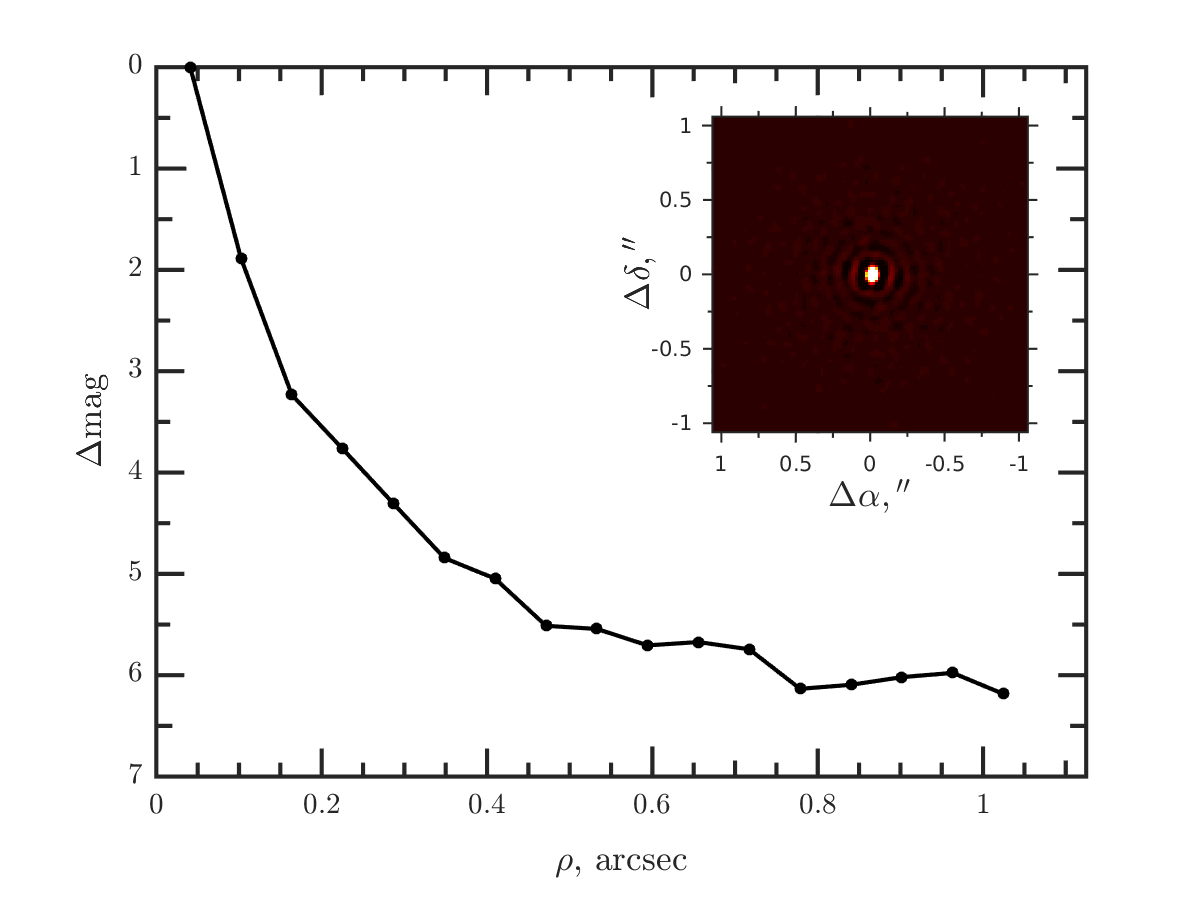}
\caption{High-resolution imaging of TOI-1846. {\bf Top left panel} shows the 3.0m-Shane high-resolution imaging in the $Ks$ and $J$ filters on UTC 29 March 2021. 
{\bf Top right panel} shows the  Gemini-North-8m/‘Alopeke high-resolution imaging taken simultaneously in two bands (562/54 nm and 832/40 nm) on UTC 25 2024 May.
{\bf Bottom left panel} shows the Palomar NIR AO high-resolution imaging taken in the Br$\gamma$ on UTC 20 June 2021. {\bf Bottom right panel} shows the SAI  high-resolution imaging obtained on UTC 3 February 2021 in the wide-band $I_c$ filter.} 
\label{fig:high_res_ima}
\end{figure*}

We took a speckle imaging observation of TOI-1846 on UTC 25 May 2024 using the ‘Alopeke speckle instrument on the Gemini North 8-m telescope \citep{scott2021}. ‘Alopeke provides simultaneous speckle imaging in two bands (562/54 nm and 832/40 nm) with output data products including a reconstructed image with robust contrast limits on companion detections. Eleven sets of $1000 \times 0.06$ second images were obtained and processed using our standard reduction pipeline (see \citet{Howell2011}). Figure~\ref{fig:high_res_ima} shows our final contrast curves and the 832 nm reconstructed speckle image. We find that TOI-1846 is a single star with no companion brighter than 4.5-8 magnitudes below that of the target star from the 8-m telescope diffraction limit (20 mas) out to 1.2$\arcsec$. At the distance of TOI-1846 (d=47 pc) these angular limits correspond to spatial limits of 0.94  to 56 AU.

\subsection{Gaia Assessment} \label{RUWE}

In addition to the high resolution imaging, we have utilized Gaia to identify any wide stellar companions that may be bound members of the system.  Typically, these stars are already in the \tess\ Input Catalog and their flux dilution to the transit has already been accounted for in the transit fits and associated derived parameters.  Based upon similar parallaxes and proper motions \citep{mugrauer2020,mugrauer2021}, there are no additional widely separated companions identified by Gaia. Moreover, the RV from Gaia DR3 archive is 20.64km/s, which is well consistent with the one derived by TRES observations, further supporting the presence of a planetary companion
    
Additionally, the Gaia DR3 astrometry provides additional information on the possibility of inner companions that may have gone undetected by either Gaia or the high resolution imaging. The Gaia Renormalised Unit Weight Error (RUWE) is a metric, similar to a reduced chi-square, where values that are $\lesssim 1.4$  indicate that the Gaia astrometric solution is consistent with the star being single whereas RUWE values $\gtrsim 1.4$  indicate excessive astrometric excess, possibly caused the presence of an unseen companion \citep[e.g., ][]{ziegler2020}.  TOI-1846 has a Gaia EDR3 RUWE value of 1.3 indicating that the astrometric fits are consistent with the single star model.

% \begin{figure}
% \centering
% \includegraphics[width=0.5\textwidth]{figures/TOI1846_20240525_562_832_final_Gemini}
% \caption{The 5$\sigma$ speckle imaging contrast curves in both filters as a function of the angular separation out to 1.2$\arcsec$. The inset shows the reconstructed 832 nm image of TOI-1846 with a 1$\arcsec$ scale bar. TOI-1846 was found to have no close companions from the diffraction limit out to 1.2$\arcsec$ to within the contrast levels achieved.} 
% \label{fig:high_res_ima_gemini}
% \end{figure}

\section{Stellar Characterization}\label{stellar_properties}\label{sec:3}
\subsection{Stellar Characterization}\label{stellar_characterization}

%empirical relation

To characterize the host star \tar, we used empirical relationships convenient for M dwarfs. We first calculated the $M_{K}$ absolute magnitude from 2MASS $m_k$ visual magnitude and parallax from \Gaia\ DR2. We find {$M_k$ = 0.408 $\pm$ 0.01 }. We then used $R-M_k$ empirical relation of \cite{Mann2015}, where we find R$_{\star}$ = 0.4115 $\pm$ 0.0119 \rsun, assuming a typical uncertainty of 3\%.

As for the stellar mass, we used the mass–luminosity relation of \cite{Mann2019}. We get $M_{\star} = 0.3997 \pm 0.008 \msun$. We also used the analogous relation of \cite{Benedict2016} where we found $ M_{\star} = 0.4123 \pm 0.001$ \msun. The uncertainty is dominated by the scatter in the mass--Ks empirical relation.

For the stellar effective temperature, we used four of the different empirical color magnitude relations (Equations 1--2 and 11--12 of Table 2) of  \citet{Mann2015}. Taking the weighted average of the four temperatures, we get T$_{\rm eff}$ = 3554 $\pm$ 100 K for TOI-1846. For the set of calculations, the standard deviation of the four temperatures was $\sim$56~K. We also calculated stellar luminosity using V-band bolometric correction based on the V--J empirical relation in \citet{Mann2015}. This gives a luminosity of $ L_{\star} = 0.02456 \pm 0.00049$~L$_{\sun}$.

\begin{figure}
	\centering
	\includegraphics[width=0.5\textwidth]{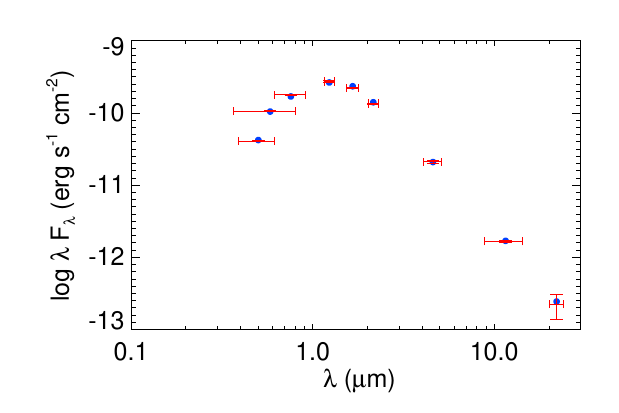}
	\caption{Spectral energy distribution (SED) fit of flux as a function of wavelength for TOI 1846. Blue points are the best-fit values, and red points are the corresponding model values and errors.}
	\label{fig:sed}
\end{figure}

We pulled the  magnitudes from {\it 2MASS} \citep{Cutri2003,Skrutskie_2006AJ_2MASS}, the W1--W3 magnitudes from {\it WISE} \citep{wright2010}, and three \gaia\ magnitudes $G, G_{\rm BP}, G_{\rm RP}$ \citep{GaiaEDR3}. Together with the available photometry spans the full stellar SED over the wavelength range 0.4\,--\,10~$\mu$m (see Figure~\ref{fig:sed}). We placed a gaussian prior on the \Gaia\ ERD3 parallax that was correted for systematics by subtracting -0.019984479 from the nominal value and 0.0147000 from the uncertainty as recommended by \citep[see, e.g.,][]{Stassun2021}. We set an upper limit on the extinction ($A_V$) from the dust maps of \citet{schlegel1998}. 
Table~\ref{stellarinfor} shows the basic stellar information of TOI-1846. 

We also determined the stellar radius and effective temperature through broad-band Spectral Energy Distribution (SED) fitting using the \texttt{EXOFASTv2} package \citep{Eastman2019}. For the SED fitting, we employed the MIST method, the favored approach reported in \cite{Eastman2019} (-MISTSEDFILE), which interpolates the 4D grid of $\log g$, $T_{\rm eff}$, [Fe/H], and an extinction grid from Conroy et al., (in preparation) to determine the bolometric corrections in each observed band. We utilized the VB Photoelectric Catalog \cite{Zacharias_2012yCat.1322}, $JHK_S$ magnitudes from {\it 2MASS} \cite{Cutri2003, Skrutskie_2006AJ_2MASS}, the W1--W3 magnitudes from {\it WISE} \cite{wright2010}, and the three {\it Gaia} magnitudes $G, G_{\rm BP}, G_{\rm RP}$ \cite{GaiaEDR3}. Together, the available photometry covers the full stellar SED over the wavelength range 0.4--10~$\mu$m (see Figure~\ref{fig:sed}). We placed a Gaussian prior on the {\it Gaia} EDR3 parallax, corrected for systematics by subtracting -0.019984479 from the nominal value and 0.0147000 from the uncertainty as recommended by \cite{Stassun2021}. We set an upper limit on the extinction ($A_V$) based on the dust maps by \cite{schlegel1998}. The results of our EXOFASTv2 stellar characterization, presented in Table~\ref{exofaststarparam}, are in excellent agreement with our previous determinations.
%We using MIST stellar models \citep{Dotter_2016,Choi2016} with stellar mass, radius and effective temperature being free parameters to be determined \todo{TODO}

\begin{table}
    %\centering
    \caption{Basic stellar information of \tar.}
    {\renewcommand{\arraystretch}{1.34}
    \begin{tabular}{lll}
        \hline\hline
        Parameter       &Value       \\\hline
        \it{Main identifiers}                    \\
         TOI                     &$1846$         \\
         TIC                     &$198385543$   \\
         \gaia\ ID            &$1420011162670761600$ \\
         2MASS               & J17114930+5431369 \\
         \it{Equatorial Coordinates} \\
         $\rm R.A.\ (J2015.5)$                    & 17:11:49.19 \\
         $\rm DEC.\ (J2015.5)$                    &+54:31:40.25    \\
         \it{Photometric properties}\\
         \tess\ (mag)                 &$11.847\pm 0.008$   &$\rm TIC\ V8^{[1]}$     \\
         \gaia\ (mag)                 &$12.981\pm0.011$   &\gaia\ EDR3$^{[2]}$   \\
         \gaia\ BP\ (mag)               &$14.237\pm0.012$   &\gaia\ EDR3$^{[2]}$   \\
         \gaia\ RP\ (mag)               &$11.875\pm0.012$   &\gaia\ EDR3$^{[2]}$   \\
         $B$\ (mag)                     &$15.556\pm0.050$         &UCAC4$^{[3]}$  \\
         $V$\ (mag)                     &$13.971\pm0.011$          &UCAC4$^{[3]}$ \\
         $J$\ (mag)                     &$10.385 \pm0.022$   &2MASS$^{[4]}$ \\
         $H$\ (mag)                     &$9.804\pm0.023$   &2MASS$^{[4]}$ \\
         $K$\ (mag)                     &$9.596\pm0.016$    &2MASS$^{[4]}$ \\
         \wise1 (mag)                   &$9.439\pm0.024$   &\wise$^{[5]}$ \\
         \wise2 (mag)                   &$9.309\pm0.020$   &\wise$^{[5]}$ \\
         \wise3 (mag)                   &$9.251\pm0.022$   &\wise$^{[5]}$ \\
         \wise4 (mag)                   &$9.290\pm0.359$   &\wise$^{[5]}$ \\
         \it{Astrometric properties}\\
         $\varpi$ (mas)              &$21.1667\pm0.01468$  &\gaia\ EDR3$^{[2]}$  \\
         $\mu_{\rm \alpha}\ ({\rm mas~yr^{-1}})$     &$-61.533\pm0.022$   &\gaia\ EDR3$^{[2]}$   \\
         $\mu_{\rm \delta}\ ({\rm mas~yr^{-1}})$     &$+211.257\pm0.022$   &\gaia\ EDR3$^{[2]}$  \\
         RV\ (km~s$^{-1}$)                          &$-25.93\pm2.00$ &This work  \\
         \it{Derived parameters} \\
         Distance (pc)                & $47.244\pm0.033$  &This work     \\
         $U_{\rm LSR}$ (km~s$^{-1}$)       &$48.99\pm4.59$     &This work\\
         $V_{\rm LSR}$ (km~s$^{-1}$)       &$-20.10\pm1.92$     &This work\\
         $W_{\rm LSR}$ (km~s$^{-1}$)       &$-6.73\pm0.56$     &This work\\
         %$v\sin i\ ({\rm km~s^{-1}})$ &$xx\pm xx$ &This work \\
         \hline\hline 
    \end{tabular}}
    \begin{tablenotes}
    \item[]  [1]\ \cite{Stassun2017tic,Stassun2019tic} 
    \item[]  [2]\ \cite{GaiaEDR3}
    \item[]  [3]\ \cite{Zacharias_2012yCat.1322}
    \item[]  [4]\ \cite{Skrutskie_2006AJ_2MASS}
    \item[]  [5]\ \cite{Cutri_2014yCat.2328}
    \end{tablenotes}
    \label{stellarinfor}
\end{table}

\begin{table}
	%\centering
{\renewcommand{\arraystretch}{1.34}
	\begin{tabular}{lll}
		\hline\hline
		Parameter & Units      & Value       \\ \hline
		$M_*$  & Mass ($M_{\odot}$)  &$0.418\pm0.025$\\
		$R_*$  & Radius ($R_{\odot}$)  &$0.397^{+0.011}_{-0.012}$\\
		$R_{*,SED}$  & Radius$^{1}$ ($R_{\odot}$)  &$0.4083^{+0.0086}_{-0.0089}$\\
		$L_*$  & Luminosity ($L_{\odot}$)  &$0.02306^{+0.00072}_{-0.00070}$\\
		$F_{Bol}$  & Bolometric Flux (cgs 10$^{-10}$)  &$3.31\pm0.10$\\
		$\rho_*$  & Density (cgs)  &$9.43^{+0.82}_{-0.71}$\\
		$\log{g}$  & Surface gravity (cgs)  &$4.862\pm0.027$\\
		$T_{\rm eff}$  & Effective Temperature (K)  &$3568^{+47}_{-41}$\\
		$T_{\rm eff,SED}$  & Effective Temperature$^{1}$ (K)  &$3521\pm27$\\
		$[{\rm Fe/H}]$  & Metallicity (dex)  &$0.169^{+0.051}_{-0.055}$\\
		$[{\rm Fe/H}]_{0}$  & Initial Metallicity$^{2}$  &$0.155^{+0.051}_{-0.053}$\\
		$Age$  & Age (Gyr)  &$7.2^{+4.6}_{-5.1}$\\
		$EEP$  & Equal Evolutionary Phase$^{3}$  &$291^{+15}_{-32}$\\
		$A_V$  & V-band extinction (mag)  &$0.034^{+0.020}_{-0.022}$\\
		$\sigma_{SED}$  & SED photometry error scaling  &$2.75^{+0.92}_{-0.59}$\\
		$\varpi$  & Parallax (mas)  &$21.167\pm0.015$\\
		$d$  & Distance (pc)  &$47.244\pm0.033$\\
		\hline\hline 
	\end{tabular}}
 \caption{Median values and 68\% confidence interval for TOI-1846 from the
SED fit alone.}
	\label{exofaststarparam}
\end{table}

\subsection{Spectroscopic analysis}

\subsubsection{Kast observations}

\begin{figure}
    \centering
    \includegraphics[width=0.5\textwidth]{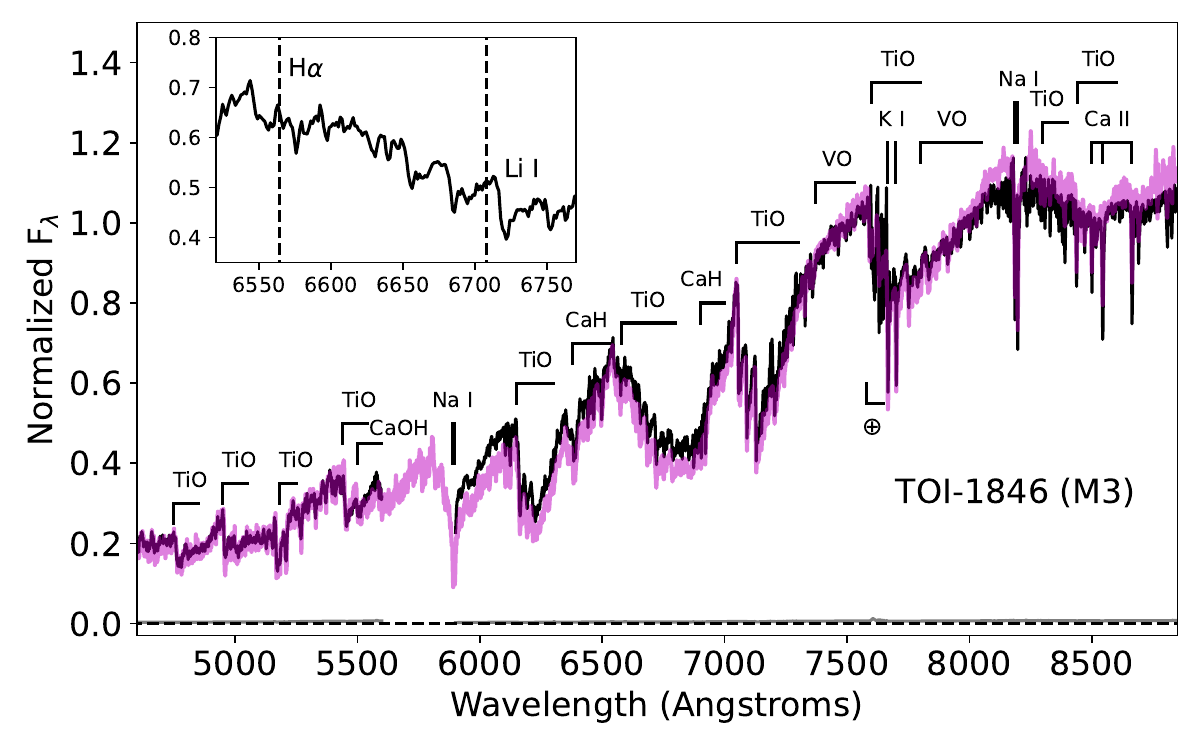}
    \caption{Kast spectrum (black lines) of TOI-1846 (black line) compared to the best-fit M3 spectral templates from \citet[magenta line]{2007AJ....133..531B}. Both spectra are normalized at 7500~{\AA}, with the blue and red orders of Kast are relatively scaled to match the spectral standard. Key atomic and molecular spectral features are labeled.
    The inset box highlights the spectral region around H$\alpha$ emission (6563~{\AA}) and Li~I absorption (6708~{\AA}).}
    \label{fig:kast}
\end{figure}

The reduced Kast optical spectrum of TOI-1846 is shown in Figure~\ref{fig:kast}.
Comparison to SDSS spectral templates from \citet{2007AJ....133..531B} indicate a best match to an M3 dwarf, while index-based classifications range over M2--M3  \citet{1997AJ....113..806G,2003AJ....125.1598L,2007MNRAS.381.1067R}. These are consistent with previously reported classifications from, e.g., LAMOST (M2--M3; \citet{2019ApJS..243...28L}), and we adopt an optical classification of M2.5$\pm$1.
Evaluation of the metallicity-sensitive TiO and CaH bands in the 6800--7100~{\AA} region yield $\zeta$ = 1.065$\pm$0.003  \citet{2013AJ....145..102L}, consistent with a 
%slightly metal-rich 
solar metallicity system, with an estimated [Fe/H] = +0.09$\pm$0.20 based on \citet{Mann2013}, consistent with the SED analysis.
We note that \citet{2022ApJS..260...45D} report a significantly subsolar metallicity for this source, [Fe/H] = $-$0.61$\pm$0.22, based on LAMOST observations of this source, although they note a systematic underestimate of 0.26~dex for dwarf stars in the sample compared to near-infrared APOGEE measurements \citep{2020AJ....160..120J}. 
Finally, we see weak H$\alpha$ in emission with an EW = $-$0.68$\pm$0.22~{\AA}, consistent with $\log{L_{H\alpha}/L_{bol}}$ = $-$4.61$\pm$013 using the $\chi$ relation of \citet{2014ApJ...795..161D}, and consistent with an activity age $\lesssim$2~Gyr \citep{2008AJ....135..785W}.

\subsubsection{SpeX observations}

\begin{figure}
\centering
\includegraphics[width=\linewidth]{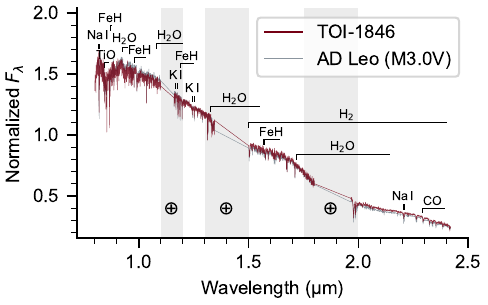}
\caption{SpeX SXD spectrum of TOI-1846 (red) compared to that of the M3V standard AD Leo \citep[grey;][]{Rayner2009}. Strong spectral features of M dwarfs are highlighted, and regions of strong telluric absorption are shaded. } 
\label{fig:spex}
\end{figure}

%% by Ben
The SpeX SXD spectrum of TOI-1846 is shown in Figure\,\ref{fig:spex}. We used the SpeX Prism Library Analysis Toolkit \citep[SPLAT, ][]{splat} to compare the spectrum to that of single-star spectral standards in the IRTF Spectral Library \citep{Cushing2005, Rayner2009}. We find the M2.5 standard Gl\,581 to be the best spectral match using the full wavelength range, while the M3.5V standard Gl\,273 is the best match using only the 0.9--1.4\,$\micron$ region following \citet{Kirkpatrick2010}. Accordingly, we adopt an infrared spectral type of M3.0 $\pm$ 1.0 for TOI-1846, consistent with its optical type. 

We also used SPLAT to estimate the stellar metallicity with the SpeX spectrum. We measured the equivalent widths of the $K$-band Na\,\textsc{i} and Ca\,\textsc{i} doublets and the H2O--K2 index \citep{Rojas-Ayala2012}. We then used the \citet{Mann2013} relation between these observables and [Fe/H] to estimate the stellar metallicity, propagating uncertainties using a Monte Carlo approach (see \citet{Delrez2022}). This yields $\mathrm{[Fe/H]} = -0.23 \pm 0.13$, distinct from the optical spectral metallicity, although both measurements are consistent with near solar abundances.

\subsection{Galactic Component}
%\todo{Abderahmane will fill in}
To identify the Galactic population membership of \tar, we first calculate the three-dimensional space motion with respect to the LSR based on \cite{Johnson1987}. We adopt the astrometric values ($\varpi$, $\mu_{\alpha}$, $\mu_{\delta}$) from \gaia\ EDR3 and the spectroscopically determined systemic RV from the SpeX spectrum, and we find $U_{\rm LSR}=48.99\pm4.59$ km s$^{-1}$, $V_{\rm LSR}=-20.10\pm1.92$ km s$^{-1}$, $W_{\rm LSR}=-6.73\pm0.56$ km s$^{-1}$. Following the procedure described in \cite{Bensby2003}, we compute the relative probability $P_{\rm thick}/P_{\rm thin}=0.02$ of \tar\ to be in the thick and thin disks by taking use of the recent kinematic values from \cite{Bensby2014}, indicating that \tar\ belongs to the thin-disk population. We further integrate the stellar orbit with the ``MWPotential2014'' Galactic potential using \code{galpy} \citep{Bovy2015} following \cite{Gan2020}, and we estimate that the maximal height $Z_{\rm max}$ of \tar\ above the Galactic plane is about $109$ pc, which agrees with our thin-disk conclusion. 
%rotation:
\subsection{Stellar rotation}
%\todo{update}

%We used the SAP and PDC-SAP light curves to search for signals of the rotation period of the star. We began by visually examining the PDC-SAP light curves of the star. We found some clear rotational modulation in sectors 23, 24, 25, and 26. We then we applied the systematics-insensitive periodogram (SIP) method \citep{Angus_2016,Hedges_2020} for SAP and the Gaussian process (GP) model on PDC-SAP light curves. The SIP method detrends the SAP light curve from \tess\ instrument systematics and also calculates the Lomb-Scargle periodogram and without requiring a predetrending of the light curves as other methods (e.g., the autocorrelation function, ACF, by \cite{McQuillan2013MNRAS}). This method was initially used for the Kepler mission and has recently been successfully applied for \tess\ data such as TOI-1259A \citep{Martin_2021}  and TOI-700 \citep{Hedges_2020}. Searching for rotational periods between 2 and 40 days, we found a significant signal of $\approx28.75$ days. The SIP powers spectrum is presented in Fig. \ref{fig:TESS-SIP}. This rotational signal was also found in the ZTF lightcurves. %We also searched for rotational periods longer than 40 days as the target is observed in 14 TESS sectors. 

\begin{figure} %%%
	\includegraphics[width=\columnwidth]{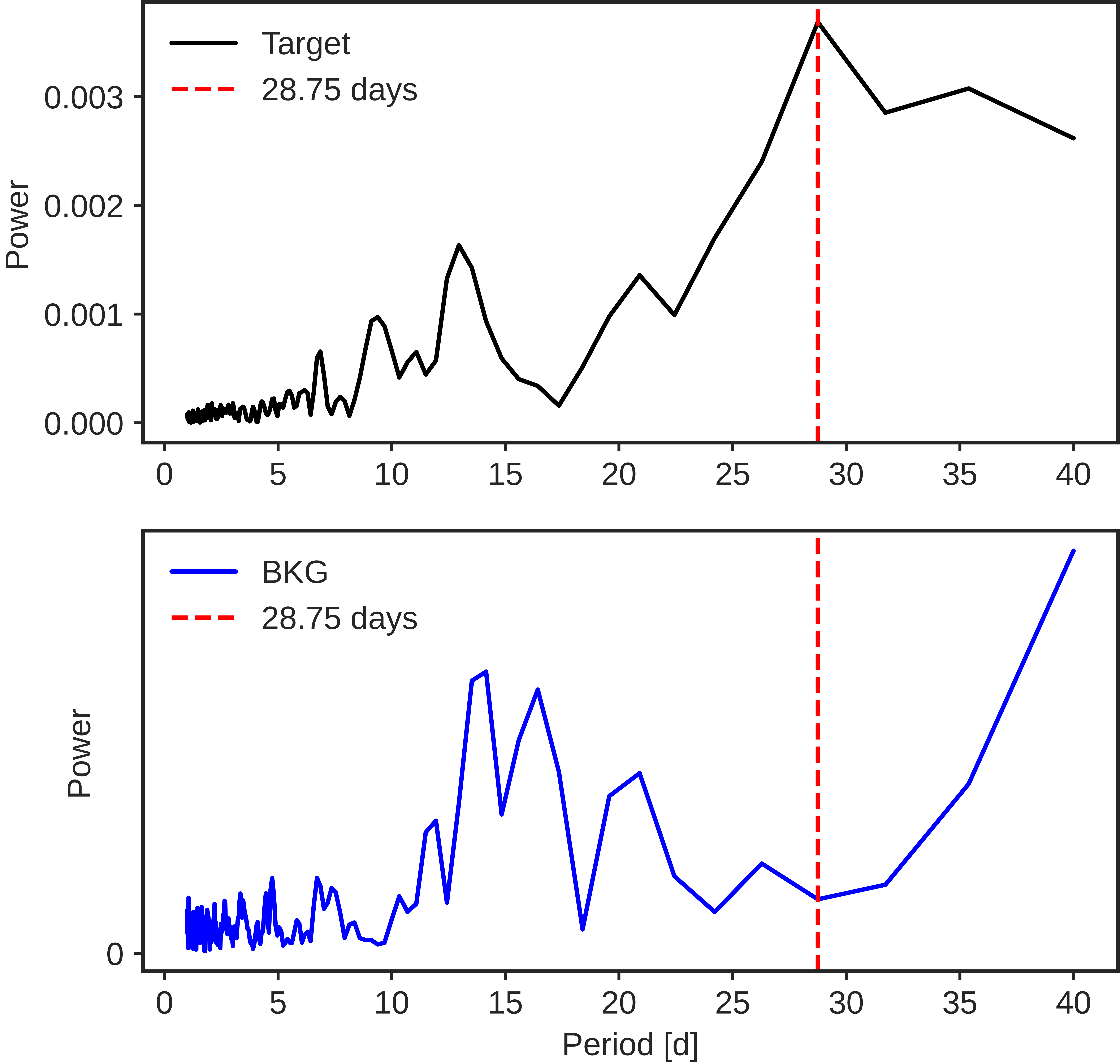}
	\caption{\texttt{TESS-SIP} power spectrum of TOI-1846. We calculate the periodogram for both the corrected light curve (top panel) and the background (BKG) pixels (bottom panel). The star's periodogram shows a rotational signal of $\approx28.75$ days not existing in the lightcurve of the background pixels.}
	\label{fig:TESS-SIP}
\end{figure}

We used both the SAP and PDC-SAP light curves to search for signatures of the stellar rotation period. We began by visually inspecting the PDC-SAP light curves and identified clear rotational modulation in Sectors 23, 24, 25, and 26. We then applied the systematics-insensitive periodogram (SIP) method \citep{Angus_2016,Hedges_2020} to the full SAP light curve, combining all available TESS sectors. The SIP method removes known TESS instrumental systematics without requiring prior detrending and computes a Lomb-Scargle periodogram. Unlike methods such as the autocorrelation function (ACF; \citealt{McQuillan2013MNRAS}), SIP does not require manually pre-processed light curves. Originally developed for Kepler data, SIP has been successfully used for TESS targets such as TOI-1259A \citep{Martin_2021} and TOI-700 \citep{Hedges_2020}. 
We searched for rotation periods between 2 and 40 days and identified a strong peak at $\sim$28.75 days in the SIP power spectrum (see Figure~\ref{fig:TESS-SIP}). To validate this result, we analyzed ground-based photometry from the Zwicky Transient Facility (ZTF; \citet{Masci2019}) in the r-band. The Generalized Lomb-Scargle (GLS) periodogram revealed a prominent peak at 27.75 days, further supporting the TESS-based detection.
%Searching for rotational periods between 2 and 40 days, we identified a strong peak at ≈28.75 days in the SIP power spectrum (see Fig.~\ref{fig:TESS-SIP}). This periodic signal is consistent with the modulation observed in the PDC-SAP light curves and is further supported by an independent analysis of the ZTF r-band photometry, which yielded a best-fit rotation period of 27.75 days.}}

\section{Vetting and Validation}\label{sec:4}
\subsection{TESS data validation report}
%% by khalid
TOI-1846 (TIC~198385543) was observed by \tess\ in Sectors 17, 20, 23--26, 40, 47, 50--54, 56--60, 73--74, and 77--81 with a cadence of 2-minutes. The SPOC (Science Processing Operations Center, \cite{SPOC_Jenkins_2016SPIE}) extracted the photometric measurements and performed a transit search \citep{jenkins2002,jenkins2010,Jenkins_2020TPSkdph}, yielding a  threshold crossing event with an orbital period of 3.9-days and a signal-to-noise ratio of  $S/N = 26.1$. 
The SPOC DV report was reviewed by the TOI vetting team on  Sept 11 2022 \citep{guerrero2021}. A second report was issued on Feb 05 2023. 
The measured transit depth was found to be $dF = 1.603 \pm 0.76$~ppt, corresponding to a planet radius of $R_p = 1.7 \pm 0.6~R_\oplus$, and with an orbital period of $P = 3.93069\pm 0.00001$~days. 
Given the large pixel scale of 21~arcsecs, three neighboring stars were fully or partially included in the \emph{TESS} aperture (see \autoref{fov}), with TOI-1846 was identified as the likely source of the events. 
All additional validation tests, including centroid offset, difference-imaging centroid, as well as bootstrap and ghost were successfully passed \citep{Twicken2018}.

\subsection{Archival imaging}
\begin{figure*}
\centering
\includegraphics[width=0.99\textwidth]{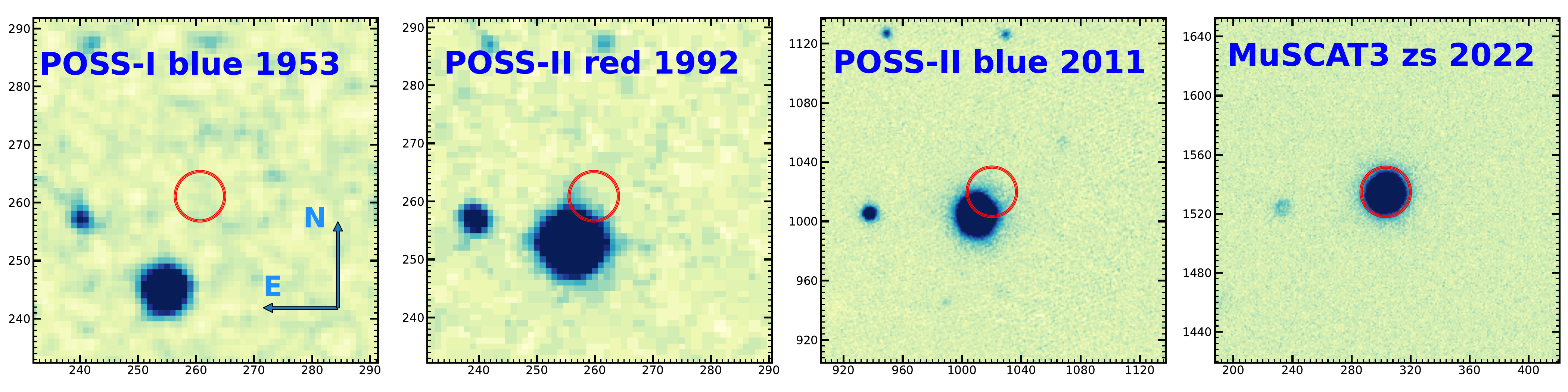}
\caption{POSS I 1953 and POSS II (1922 and 2011) archival images cropped with a field of view of 1'$\times$1’ around TOI-1846. The central red circle marks the current position of TOI-1846.} 
\label{fig:archivalimage}
\end{figure*}

%% by Mourad and Khalid
We used archival images of TOI-1846 in order to check for background stellar objects that could be blended with the target at its current position. This kind of object might introduce the same transit event that we observed in our data, and skew the physical parameters of the system that we obtained  from the global analysis. We pulled archival images, in blue and red filters, from POSS I/DSS \citep{1963POSS-I} and POSS II/DSS \citep{1996DSS_POSS-II}. In addition to our own image from LCO-HAL-2m0/MuSCAT3 taken in 2022, these images spans 69 years as depicted on Figure~\ref{fig:archivalimage}. TOI-1846 has a mean proper motion of 220~$mas/yr$ and has moved by an angular distance of $\sim$15.2\arcsec\ from 1953 to 2022. No other stellar object is observed at the current position of the target TOI-1846. 

\subsection{Discarding possible false-positives}
\begin{figure}
\includegraphics[width=0.5\textwidth]{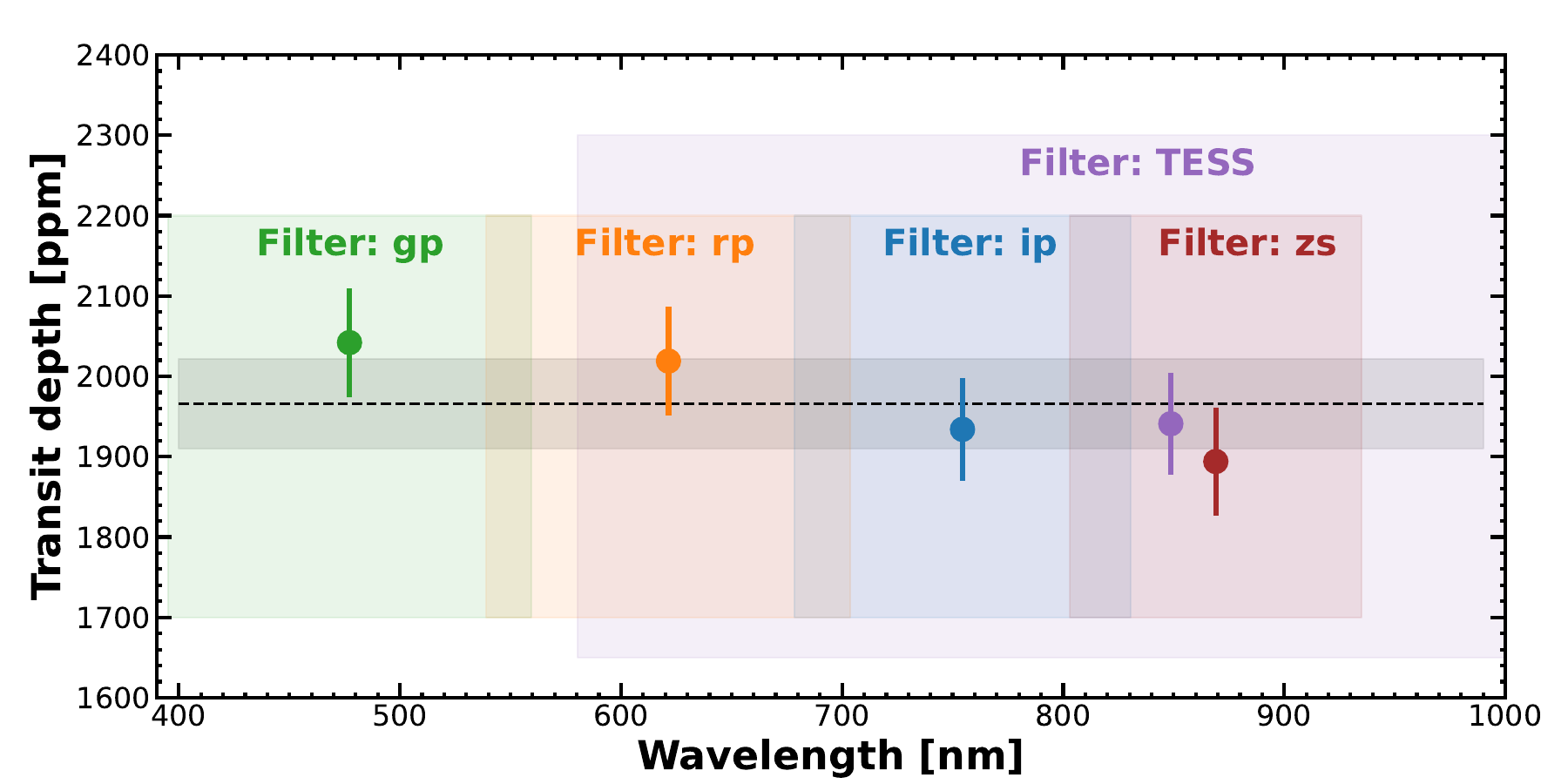}
\caption{Transit depths measured for multi-band photometric follow-up of TOI-184\,b (colored points). 
Horizontal dashed line correspond to the transit depth measured by \tess. Colored regions show the bands coverage.
}  
\label{fig:transit_depths}
\end{figure}

Three cases involving EBs can result in a false positive scenario in \tess\ data \citep{Lillo-Box_2024AA}.
The first is that of nearby eclipsing binaries (NEBs) contaminating the \tess\ apertures. As explained in details in Sec.~\ref{gbp}, the apertures used in the ground-based photometry excluded all the nearby stars that contaminated the \tess\ photometry. This rules out the the possibility of NEBs and confirms the transit event on TOI-1846. The second case is that of an unresolved background Eclipsing Binary. Many factors ruled out this possibility: (1) The archival imaging reveals that there was no bright star in the current position of the target over 69 years; (2) the high resolution adaptive optic imaging (see Sec.~\ref{sec:high_res_obs}) detect no source of light close to ($\lesssim 1\arcsec$) the target; the RUWE value of 1.3 from Gaia DR3 indicates that TOI-1846 is a single star (see Sect.~\ref{RUWE}). The third case is that of hierarchical eclipsing binary. This case is discarded by many factors: (1) the spectral energy distribution of the star is well-fitted by single-model star (see Figure~\ref{fig:sed}); (2) the spectroscopic observations detect no secondary spectrum that can be attributed to additional star in the system; (3) the ground-based observations cover a wavelengths range from 477 to 1000\,nm, where the transit depths in all bands were consistent within 2$\sigma$ (see Figure~\ref{fig:transit_depths}) with that in the \tess\ band, confirming the achromatic nature of the transit.

\subsection{False-positive likelihood}

To calculate the false positive probability FPP for TOI-1846\,b, we used of the open-source software package TRICERATOPS \citep{Giacalone2021}, which uses a Bayesian tool that incorporates prior knowledge of the target star, planet occurrence rates, and stellar multiplicity to calculate the probability that the transit signal is due to a transiting planet or another astrophysical source. The criteria for statistical validation of a planetary candidate is stated as FPP$<$0.015 and NFPP$<$ 10$^{-3}$, which is the sum of probabilities for all false positive scenarios.
Using ground-based photometric observations, the transit event is well detected on the target star, and thus we excluded other nearby sources, means that NFPP = 0 for TOI-1846.01.
We applied TRICERATOPS to the \tess\ 2min cadence phase-folded light curves supplied with the contrast curve obtained by the PHARO and Gemini speckle imaging in section~\ref{sec:high_res_obs}. The resulting   value is $FPP = (1.17 \pm 0.58)\times 10^{-4}$. Consequently, the candidate is a statistically validated planet.

\section{Data analysis and results}\label{analysis}\label{sec:5}
\subsection{Transit modeling} \label{Exofast_fit} %\todo{Abderahmane will write up the modeling part}
To refine the physical parameters, we used the IDL software package EXOFASTv2 \citep{Eastman2019} to perform a global modeling of the TESS and ground-based transit light curves. EXOFASTv2 provides an integrated framework to jointly analyze multiple exoplanet data sets, drawing from the IDL astronomy library \citep{Landsman1993}. It simultaneously fits a wide range of stellar, planetary, orbital, and instrumental parameters in a self-consistent manner, leveraging the rich complementarity of modern data sets.

EXOFASTv2's transit model is generated using \citet{Mandel2002} and \citet{Agol_2020}, with limb-darkening parameters constrained by \citet{Claret2011} and \citet{Claret2018}. The exoplanet mass-radius relation from \citet{Chen2017} can be referenced within EXOFASTv2 to estimate the mass or radius of the exoplanet (and all relevant derived parameters) in the absence of an RV data set or transit. In this case, the transit data constrain the radius, and the mass is estimated using this relation.

We imposed Gaussian priors on the stellar parameters:the effective temperature T$_{\rm eff}$, the metallicity [Fe/H], the stellar radius and mass, the quadratic limb-darkening coefficients (LDCs), and the transit period.  Additionally, we used uniform priors on the period and transit epoch based on the values reported in ExoFOP.

The fits fully converged according to two different statistics: the Gelman–Rubin statistic R$_z$ and the number of independent samples T$_z$. We set very stringent thresholds of R$_z$<1.01 and  T$_z$>1000.For precise details on the internal operations of EXOFASTv2, consult the primary paper by \citet{Eastman2019}.
The median and 1$\sigma$ uncertainties of the derived physical parameters are listed in Table \ref{tab:198385543}.

\subsection{Independent transit modeling}
%% by khalid
An independent analysis of \tess\ and ground-based photometric data was performed using  the Metropolis-Hastings \citep{Metropolis_1953,Hastings_1970} algorithm implemented in {\tt TRAFIT}, a revised version of Markov-chain Monte Carlo (MCMC) code described in \cite{Gillon2010AA,Gillon2012,Gillon2014AA}.  We followed the same strategy as described in \citet{Barkaoui2023A&A,Barkaoui_2024A&A}.
The transit light curves  were modeled using the quadratic limb-darkening model of \cite{Mandel2002}, multiplied by a baseline model to correct for several external systematic effects related to the time, FWHM, background, x \& y position and airmass.\\
For each transit light curve, the baseline model was selected based on minimizing the  Bayesian information criterion (BIC; \cite{schwarz1978}), and photometric error bars were re-scaled using the correction factor $CF = \beta_{w} \times \beta_{r}$, where, $\beta_{w}$ and $\beta_{r}$ are white and red noises, respectively \citep{Gillon2012}. \\
The global free parameters for the photometric observations are  the transit depth, the impact parameter, the total transit duration, the orbital period of the planet, as well as the transit timing, and the quadratic limb-darkening coefficients $u_1$ and  $u_2$, which are computed from \cite{Claret_2012AA,Claret2018}. During the analysis, we used the combination of the the quadratic limb-darkening coefficients, $q_1 = (u_1 + u_2)^2$ and $q_2 = 0.5u_1(u_1 + u_2)^{-1}$  proposed by \citep{Kipping_2013MNRAS.435.2152K}. The results are consistent with the ones obtained using ExoFASTv2 (see Section~\ref{Exofast_fit}).

\subsection{Planet search and detection limits from the TESS photometry} \label{sec:tkmatrix}
%\todo{Sherlock and matrix}

We performed our own search for candidates using the \texttt{SHERLOCK}\footnote{\url{https://github.com/franpoz/SHERLOCK}}(Searching for Hints of Exoplanets fRom Lightcurves Of spaCe-based seeKers) pipeline presented in \citep{Pozuelos2020, Demory2020, Devora2024}. The \texttt{SHERLOCK} pipeline is an open-source package that provides six different modules to find and validate transit signals in photometric time-series observations: (1) acquiring automatically the data from an online database, such as MAST in the case of TESS data; (2) searching for planetary transit signals; (3) performing a vetting of the detected signals; (4) conducting a statistical validation of the vetted signals; (5) modeling them to retrieve their ephemerides; and (6) finding the upcoming transit windows observable from ground-based observatories. See \cite{Delrez2022} and \cite{Pozuelos2023} for recent applications and further details.

We conducted two separate transit search runs using short-cadence TESS data. The first run covered Sectors 17, 20, 23, 24, 25, 26, 40, and 47, while the second run focused on Sectors 50, 51, 52, 53, 54, 58, 59, 60, 73, 74, 77, 78, 79, 80, 81, and 83. For both runs, we analyzed periods between 0.5 and 40 days, using a signal-to-noise ratio (S/N) threshold of 5 or higher and allowing a maximum of three runs.

In particular, we applied a multidetrend approach that uses a biweight filter several times with different window sizes, combined with the Transit Least Squares (TLS) algorithm \citep{Hippke2019}, to optimize the transit search by focusing on signals with the highest S/N and signal detection efficiency (SDE). This method has been shown to be effective in transit searches (e.g., \citealt{Delrez2022, Pozuelos2023}). During our analysis, we successfully recovered the signal of TOI-1846b, with an S/N of 31.98 in the first run and 49.203 in the second run. No additional signals were detected that could be attributed to other planets.

In addition, to assess the detectability of other planets in the data from Sectors 73 and 74, we performed an injection-recovery test using \texttt{MATRIX}\footnote{\url{https://github.com/PlanetHunters/tkmatrix}} \citep{Pozuelos2020,tkmatrix2022}. Synthetic planetary signals were injected into the PDC-SAP fluxes, representing planets with varying radii and orbital periods. We then detrended the light curve using the previously identified best method, which is the bi-weight approach with a window size of 0.5 days. Before initiating the search for planets, the known candidate planet with a period of 3.93 days was masked.

The $R_{\rm p}$–$P$ parameter space was explored in the ranges of 0.5–3.5 R$_{\oplus}$ (with steps of 0.20 R${\oplus}$) and 1–20 days (with steps of 1 day). For each combination of $R_{\rm p}$ and $P$, \texttt{MATRIX ToolKit} examined four different orbital phases, resulting in a total of 1200 scenarios. A recovery was considered successful if the detected period was within 5\% of the injected period and the transit duration was within 1 hour of the set value.

\begin{figure}
\centering
\includegraphics[width=0.49\textwidth]{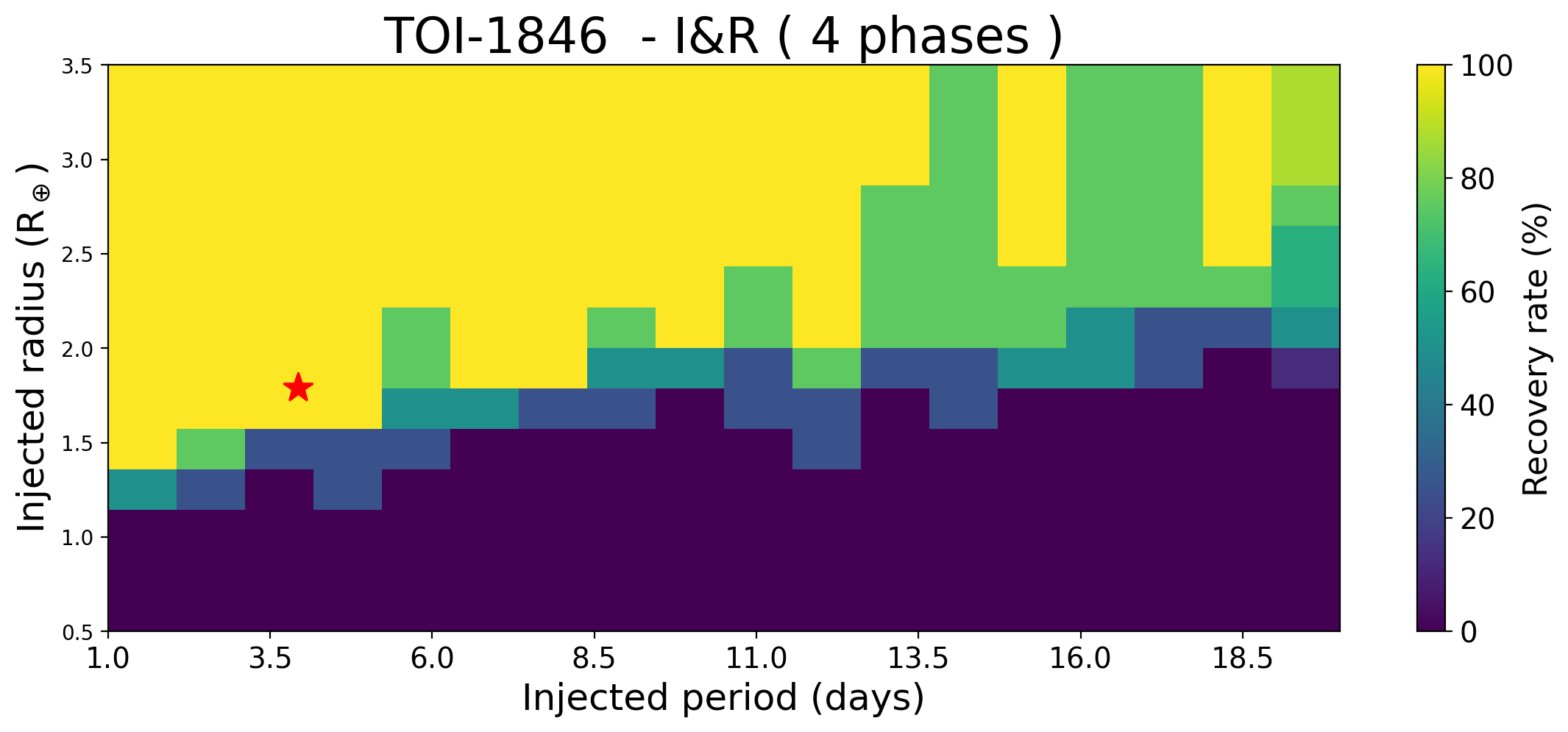}
\caption{Injection-and-recovery experiment performed to test the detectability of extra planets in the system using the tow TESS sectors described in Sect \ref{sec:tkmatrix}. The red star marks the position of \hbox{TOI-1846\,b}}. 
\label{inj_recov}
\end{figure}

The detectability map derived from these injection-recovery experiments is shown in Figure \ref{inj_recov}. We found that an Earth-sized planet ($R_{\rm p} \sim 1~R_\oplus$) would remain undetected across the entire range of periods explored. However, planets with radii R$_{\mathrm{p}} \gtrsim 1.5\ R_{\oplus}$ and periods of up to 15 days could be ruled out with a recovery rate ranging from $\geq 80\%$ to 100\%. Notably, TOI-1846\,b lies in a region with a 100\% recovery rate.

\subsection{ Transit timing variation}\label{Transit_timing_variation}
%TO BE UPDATES, fig removed, should run the code with all new sectors.

We searched for transit timing variations (TTVs) using the full TESS photometric dataset and \texttt{EXOFASTv2}. \texttt{EXOFASTv2} employs a Differential Evolution Markov Chain Monte Carlo (DE-MCMC) method to estimate the stellar and planetary parameters along with their uncertainties. It fits a linear ephemeris to the transit times and applies a penalty for deviations from the best-fit linear ephemeris.

For the TTV analysis of \hbox{TOI-1846\,b}, we fixed the stellar parameters to the values listed in Table~\ref{exofaststarparam}, and the orbital parameters to those derived from the joint fit. We found no evidence of significant TTVs in the current photometric data, likely due to the low signal-to-noise ratio of individual transits.

\section{Discussion}\label{sec:6}
\subsection{TOI-1846 b in the radius/density valley} %By Ghachoui Mourad
The ``radius valley'' \citep{Fulton2017, Cloutier2018} and the ``density valley'' \citep{Luque&palle2022} are some of most important features identified recently in the radius and density distributions of small exoplanets ($R_p<4\re$) in orbits shorter than 100 days around M dwarf and FGK stars. These features can shed light on the formation mechanisms of exoplanets and their evolutionary pathways.

\begin{figure}
    \centering
    \includegraphics[width=\columnwidth]{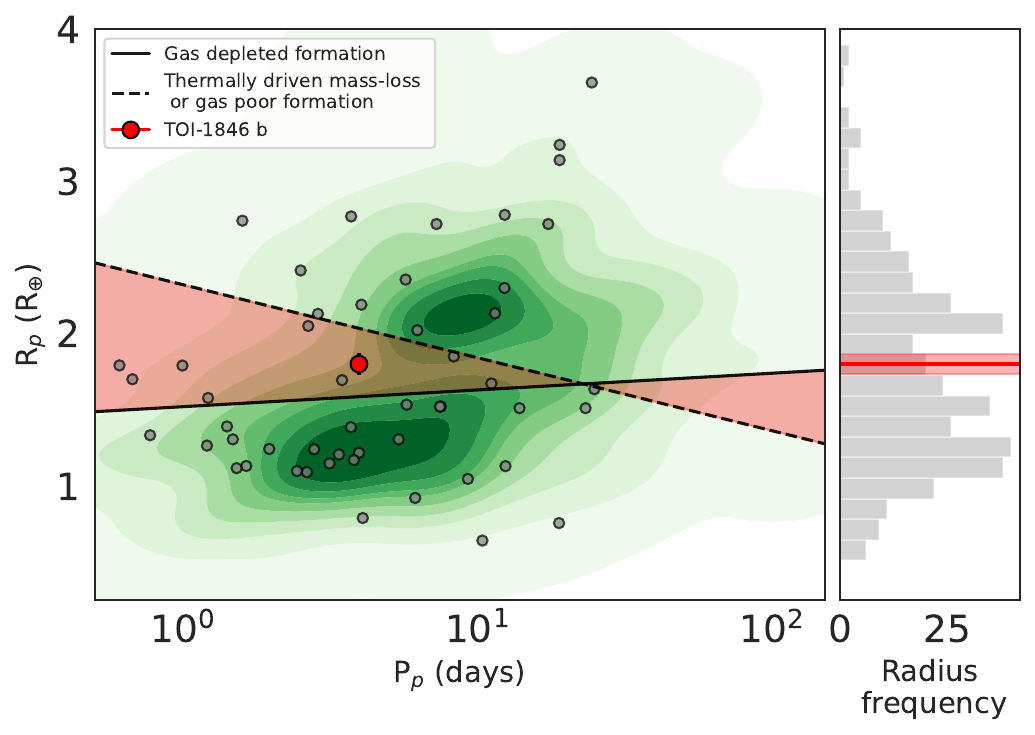}
    \caption{ The planet radius and orbital period diagram of all confirmed small
planets hosted by low mass stars. The green contours are
the density distribution of planets. The solid and dashed lines depict the locations of radius
valley for low mass stars predicted by the gas poor, gas depleted and photoevaporation models, taken from \protect \cite{cloutier2020evolution}. TOI-1846 b is marked as a red dot.}
    \label{fig:radius_valley}
\end{figure}

The ''radius valley'' refers to the dip in occurrence in the radius distribution separating rocky super-Earths and gaseous sub-Neptunes. This dearth of planets is observed around $\sim$1.8 \re\ \citep[see, e.g.,][]{Fulton2017,cloutier2020evolution}. As shown in Figure \ref{fig:radius_valley}, the location of the radius valley depends on the orbital periods and the spectral types of the host stars, raising important questions on the mechanisms controlling the formation and evolution of these planets. For FGK stars, the radius valley location shifts towards small planets for long periods \citep[see, e.g.,][]{Fulton2017,VanEylen_2018,LopezRice2018,Lee_2021}. Two explanations for this slope have been proposed. The first is thermally-driven mass-loss, which proposes that planets are formed with their primordial gaseous envelopes, which can be evaporated for some of the planets because of the high-energetic XUV radiation from their host stars, or released by their cooling pletary cores. In this theory, the effectiveness of atmospheric outflow decreases for large orbits and high planetary masses, which explains the decreasing slope of the radius valley. The second is gas-poor formation, where some rocky super-Earths planets are formed later in a gas-poor environment after dissipation of most of the primordial gas of a protoplanetary disc. For M dwarf stars, the radius valley location has been found tending towards large planets for increasing orbital periods \citep[see, e.g.,][]{LopezRice2018,Cloutier_Menour_2020}, supporting the gas-depleted formation explanation. In this theory, planets in short orbits have less matter to accrete than planets in large orbits do, which explains the increasing slope of the radius valley. However, a new study by \cite{Gaidos2024}, performed for M dwarfs with refined ages and radii of their transiting planets, found that the radius valley is similar to that of FGK stars, but with significantly weaker slope. In this study, it has been found a decline in sub-Neptunes and rise in super-Earths for M dwarfs with ages older than $\sim 3 Gyr$, hinting that thermally-driven mass-loss can still be effective for these type of stars.

The density valley features three peaks in the density distribution of exoplanets in orbits around M dwarf stars. These peaks were observed at $\rho=0.94\pm0.13~\rho_{\oplus}$, $\rho=0.47\pm0.05~\rho_{\oplus}$ and $\rho=0.24\pm0.04~\rho_{\oplus}$, which were attributed to purely rocky planets, water-rich planets and gas-rich planets, respectively. No planets with intermediate composition have been found between purely rocky and water-rich planets, favoring the pebble accretion model for forming these types of planets around M dwarfs. This study suggests that mini-Neptunes can either be of water- or gas-rich compositions. This is in line with coupled formation models which predict that the mini-Neptunes are rich in water formed outside the ice-line and migrate toward the host stars \citep{Venturini2020,Burn2024,Venturini2024}. Contrarily, other studies argue against the existence of distinct population of water-rich worlds around M dwarf stars \citep[see e.g.][]{Parviainen2024}. 

These debates on both radius and density valleys are pushing the scientific community to enlarge the sample of small planets with precise determination of their masses and radii to have strong statistical inferences on the relative dominance of the different formation and evolution models. As shown in Figure~\ref{fig:radius_valley}, the difference in the slopes of in radius valley resulted in a region of contradicting predictions, known as the region of ``keystone planets'' \citep{Cloutier_Menour_2020,Cherubim2023}. These planets are expected to be of purely rocky composition in the thermally-driven mass-loss and gas-poor formation mechanisms, and rocky with likely small gaseous envelope in the gas-depleted mechanism. However, the keystone planets with RV mass measurements have been found to have different bulk compositions, making it challenging to firmly conclude which mechanisms dominate over the others. Having more planets, with accurate radii and masses, in this region of interest is thus necessary for tighter statistical inferences in the future.

With a radius of $R_{p}=1.79 \pm 0.07 $ \re\ and a period of $P = 3.93067$~days, TOI-1846\,b falls squarely in the region of keystone planets. This makes it a promising target for both mass measurements from RV observations and atmospheric studies. It most probably has a water-rich bulk composition based on its radius, according to the results of \cite{Luque&palle2022}. RV observations, as elaborated in Sect~\ref{RV:OBS}, will reveal its composition, making it interesting additional object in the sample of keystone planets. 

\subsection{Prospects for RV follow-up} \label{RV:OBS}
\begin{figure}
    \centering
    \includegraphics[width=\columnwidth]{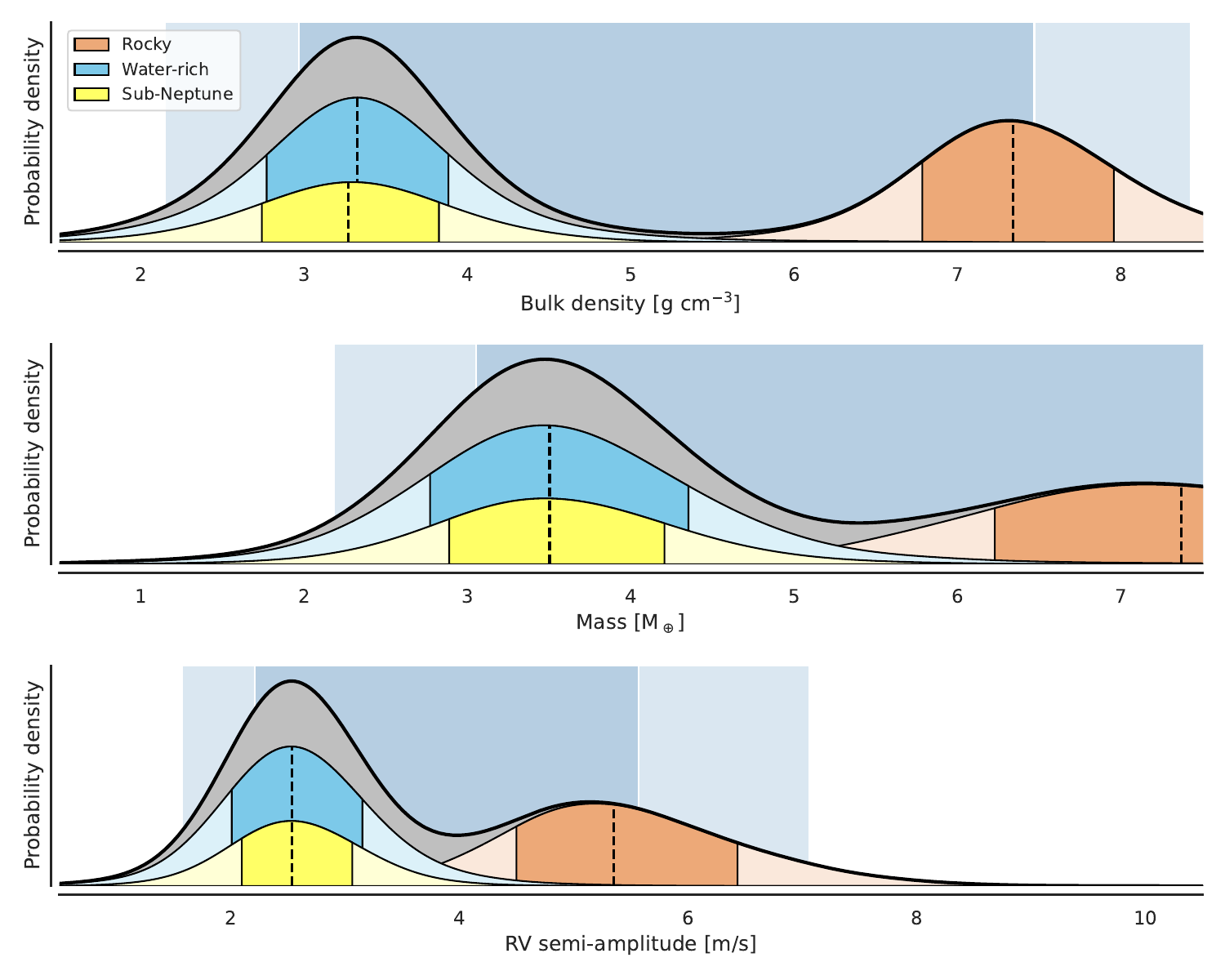}
    \caption{The probability distributions for TOI-1846b's bulk density, mass, and radial velocity semi-amplitude, as predicted by the spright package based on a posterior radius estimate of 1.79 $\pm$ 0.07 R$_{\oplus}$. The complete probability distribution is represented by the thick black line and gray shading. The individual contributions of the three spright model components—rocky planets (light brown), water-rich planets (light blue), and sub-Neptunes (yellow)—are also shown. The blue shading in the background highlights the 68\% and 95\% central posterior intervals for these distributions.}
    \label{fig:rv_semi_amplitude}
\end{figure}

\begin{figure}
    \centering
    \includegraphics[width=\columnwidth]{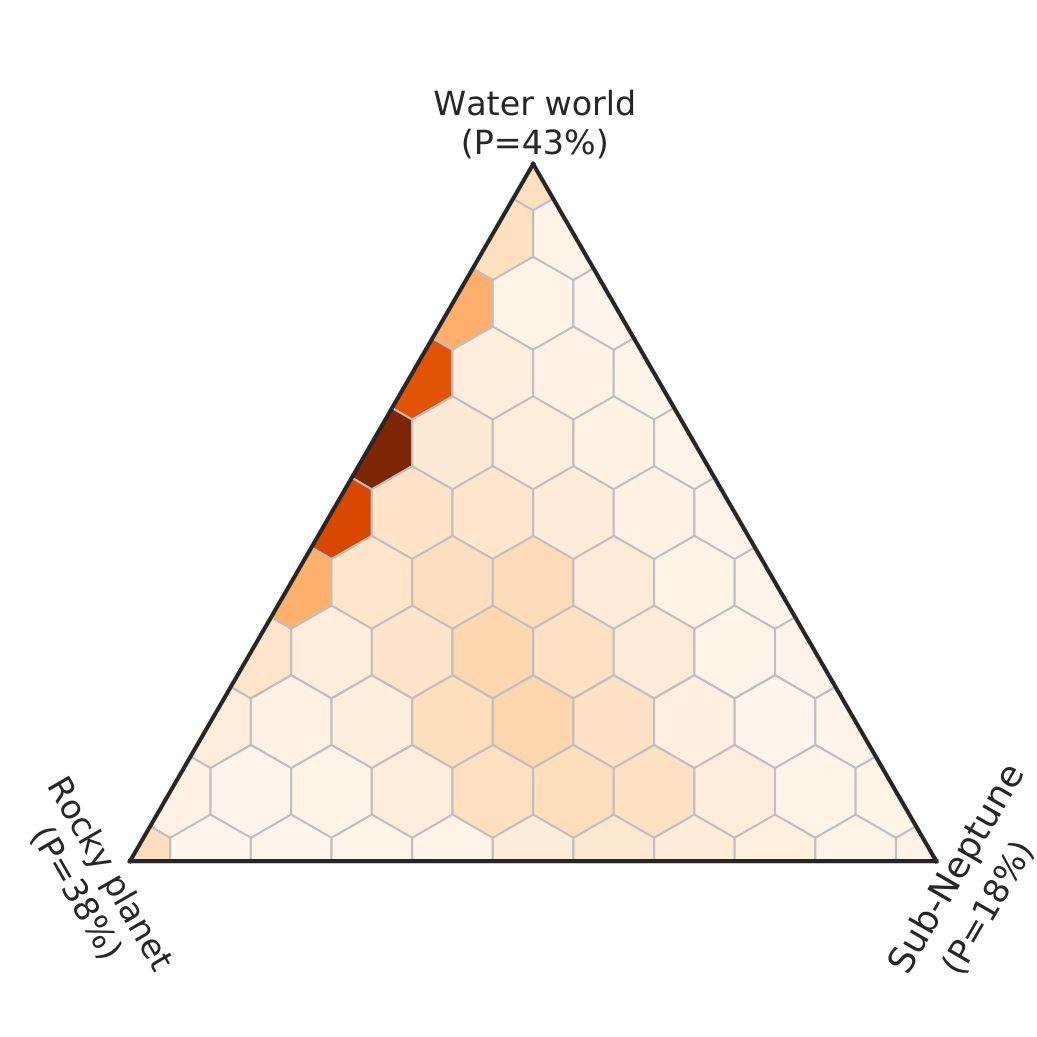}
    \caption{TOI-1846 b's composition class based on its radius predicted by the spright package.}
    \label{fig:spright_class}
\end{figure}

To estimate the mass and potential composition of TOI-1846\,b, we employed the numerical radius-mass relationship provided by the spright package \cite{Parviainen2023}. This package implements a probabilistic model that represents the joint distribution of radius and bulk density for small planets. Spright utilizes a three-component mixture model, accounting for rocky planets, water-rich planets, and sub-Neptunes. The final radius--mass relation is derived by marginalizing over all model solutions consistent with the observational data. Importantly, this approach allows for scenarios where a distinct water-rich planet population might not exist, making the spright model agnostic to their presence. Applying this model, we generated mass and radial velocity semi-amplitude distributions for TOI-1846b (Figure~\ref{fig:rv_semi_amplitude}), along with probabilities for different compositional classes (see Figure~\ref{fig:spright_class}).

Applying the spright model to TOI-1846b, we obtain a predicted radial velocity semi-amplitude (K) distribution ranging from 1.57 to 7.06 m~s$^{-1}$ (95\% central posterior limits, Figure~\ref{fig:rv_semi_amplitude}). Based on the planet's radius, the model suggests a K value of 5.4 $\pm$ 1.0 m~s$^{-1}$ if TOI-1846\,b is primarily rocky in composition. In contrast, if the planet is water-rich or a sub-Neptune, we expect K values of 2.5 $\pm$ 0.6 and 2.5 $\pm$ 0.5 m~$s^{-1}$, respectively.
By combining the faintness of the star ($V_{\rm mag} = 13.97$) with predicted RV semi-amplitude, a MAROON-X \citep{Seifahrt_MaroonX} like spectrograph is required. Assuming an exposure times of 900--1800s and good observing conditions, the predict RV precision of $\sim1$~m/s per spectrum, which will allow for constraining the mass, bulk density and the eccentricity of the orbit of the planet (e.g. \citet{Barkaoui_2025AA_TOI2015}; \citet{Bonfanti_2024AA_TOI732b}). 
% \autoref{fig:rv_semi_simulation} shows RV observations simulation of TOI-1846\,b for MAROON-X. 

% \begin{figure}
%     \centering
%     \includegraphics[width=\columnwidth]{figures/RVs_TOI-1846.01.pdf}
%     \caption{Simulated RV observations for a predicted mass of $M_p = 4.4 ^{+1.6}_{-1.0}~M_\oplus$ and a circular orbit. Each data point has precision calculated using an empiric relation for MAROON-X.}
%     \label{fig:rv_semi_simulation}
% \end{figure}

%\todo{a plot of TSM vs. Eq temperature (Planets with $1.5\ R_{\oplus}<R_{p}<4\ R_{\oplus}$ around M dwarfs with $3\sigma$ significance mass measurements).}

%%%%%%%%%%%%%%%%%%%%%%%%%%%%%%%%%%%%%%%%%%%%%%%%%%%%%%%%%%%%%%%%%%%%%%%%%%
%

%%%%%%%%%%%%%%%%%%%%%%%%%%%%%%%%%%%%%%%%%%%%%%%%%%%%%%%%%%%%%%%%%%%%
%%%%%%%%%%%%%%%%%%%%%%%%%%%%%%%%%%%%%%%%%%%%%%%%%%%%%%%%%%%%%%%%%%%%
%%%%%%%%%%%%%%%%%%%%%%%%%%%%%%%%%%%%%%%%%%%%%%%%%%%%%%%%%%%%%%%%%%%%
\subsection{Atmospheric characterization}

\begin{figure}
	\centering
	\includegraphics[width=0.5\textwidth]{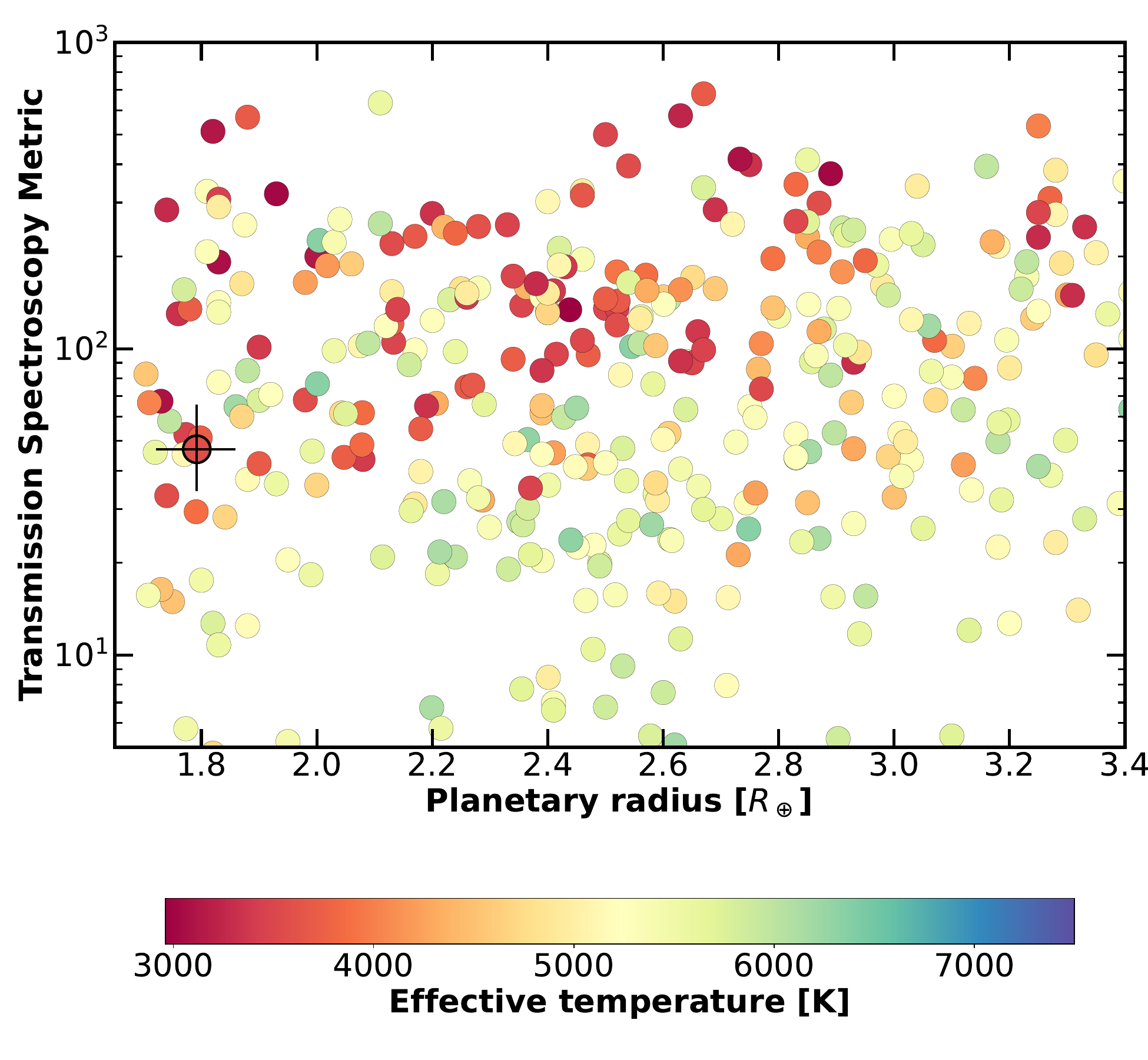}
	\caption{The transmission spectroscopy metric as defined by \citep{kempton2018} for all known transiting exoplanets with both a measured mass and radius as a function of equilibrium temperature. We only considered planets with mass uncertainties less than 20\% and a radius of less than 3 R$_{\oplus}$ but larger than 1.5 R$_{\oplus}$. TOI-1846\,b is highlighted with the black circle and error bar.}
	\label{fig:TSM}
\end{figure}

To assess the potential for atmospheric characterization of TOI-1846\,b, we first evaluated its Transmission Spectroscopy Metric (TSM) \citep{kempton2018}. We find a TSM of $47^{+18}_{-12}$ for TOI‑1846 b  (Figure~\ref{fig:TSM}), which is below the $\approx$ 90 benchmark for small sub‑Neptunes (1.5–2.6$ $R$_{\oplus}$). This calculation is based on an estimated planetary mass derived from empirical mass–radius relationships \citet{Chen2017}. While this metric suggests that TOI‑1846 b may be less favorable for atmospheric study, it’s important to note that TSM values are guidelines based on expected signal strength and not strict rules. Scientific interest—such as exploring a planet in the sparsely populated radius valley, orbiting a bright, quiet M‑dwarf (J = 10.38) can justify observations even when the TSM is modest.
A precise mass measurement would significantly refine the TSM and the planet’s prioritization for future atmospheric studies.
%

%To assess the potential for atmospheric characterization of TOI-1846\,b using the JWST,
Then we explored various atmospheric scenarios and simulated JWST observations for each using the VULCAN photochemical model \citep{vulcan} and petitRADTRANS \citep{mol}. We selected the temperature profile based on a modern Earth’s atmosphere and we then increased the surface temperature for it to be consistent with the equilibrium temperature of TOI-1846\,b \citep{luque2019planetary}. We followed a similar approach as in \citet{chouqar}, we scaled the TOI-1846's UV flux from AD Leo’s observed UV spectrum according to
Equation (1) derived by \citet{rugheimer2015effect}. We then combined the synthetic PHOENIX spectrum with the scaled observations for AD Leo. To set up the models, we used the N-C-H-O photochemical network with about 700 reactions and over 50 molecular species. The network is available as part of the VULCAN package. VULCAN is implemented with the FastChem equilibrium chemistry code \citep{fastchem} to initialize a state in chemical equilibrium.\\

\begin{figure*}
	\centering
	\includegraphics[width=\columnwidth]{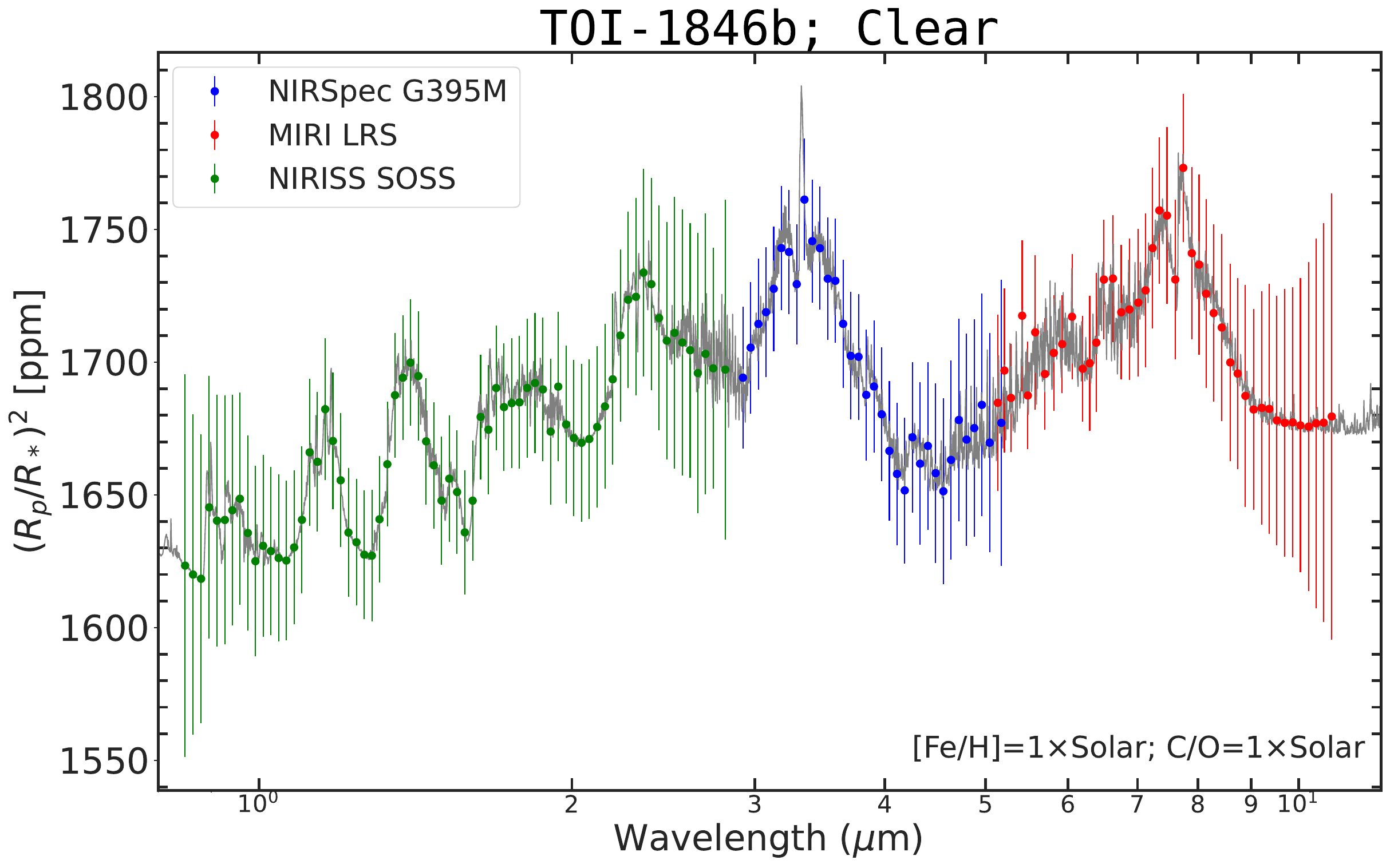} \hfill
        \includegraphics[width=\columnwidth]{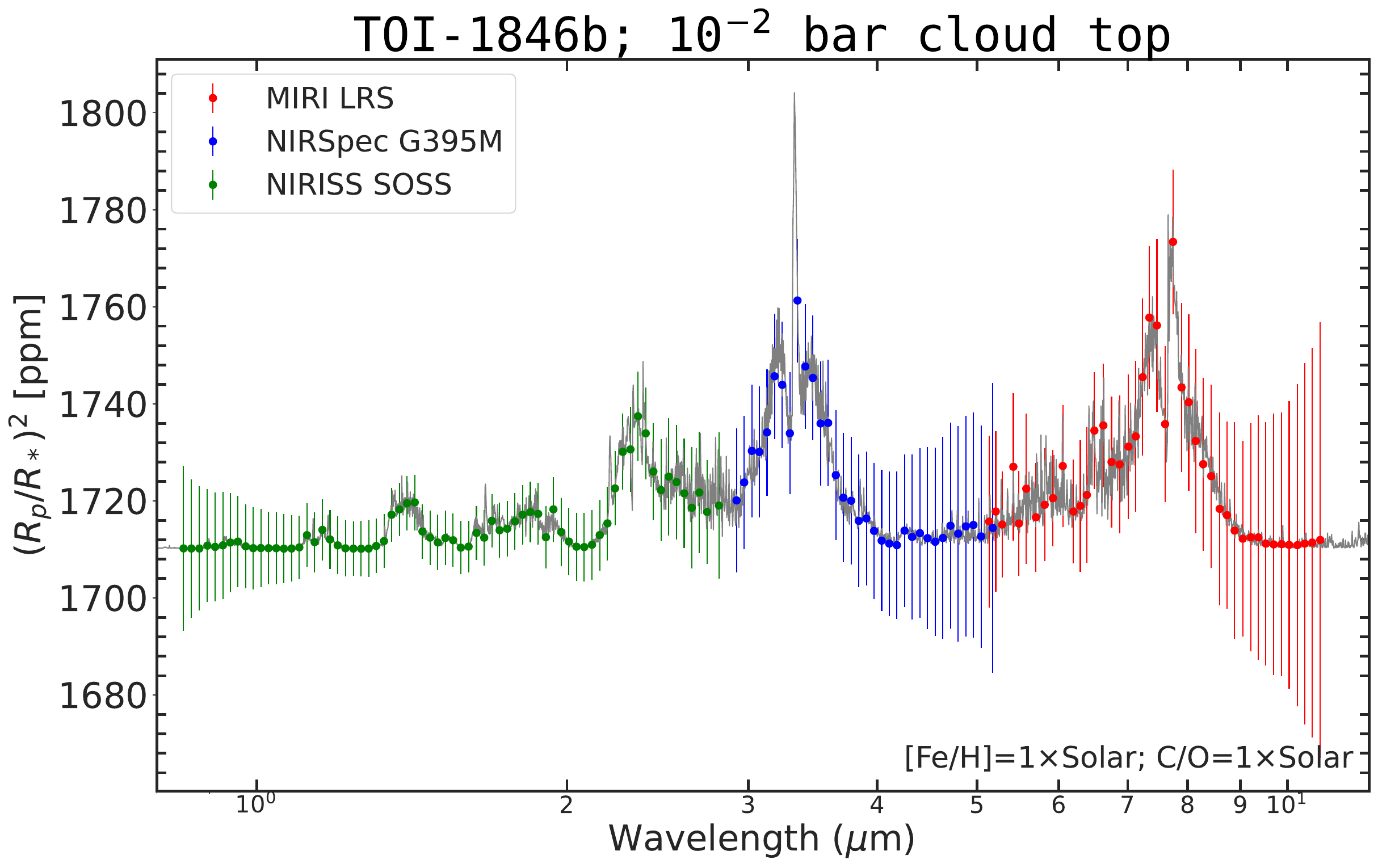} \hfill
        \includegraphics[width=\columnwidth]{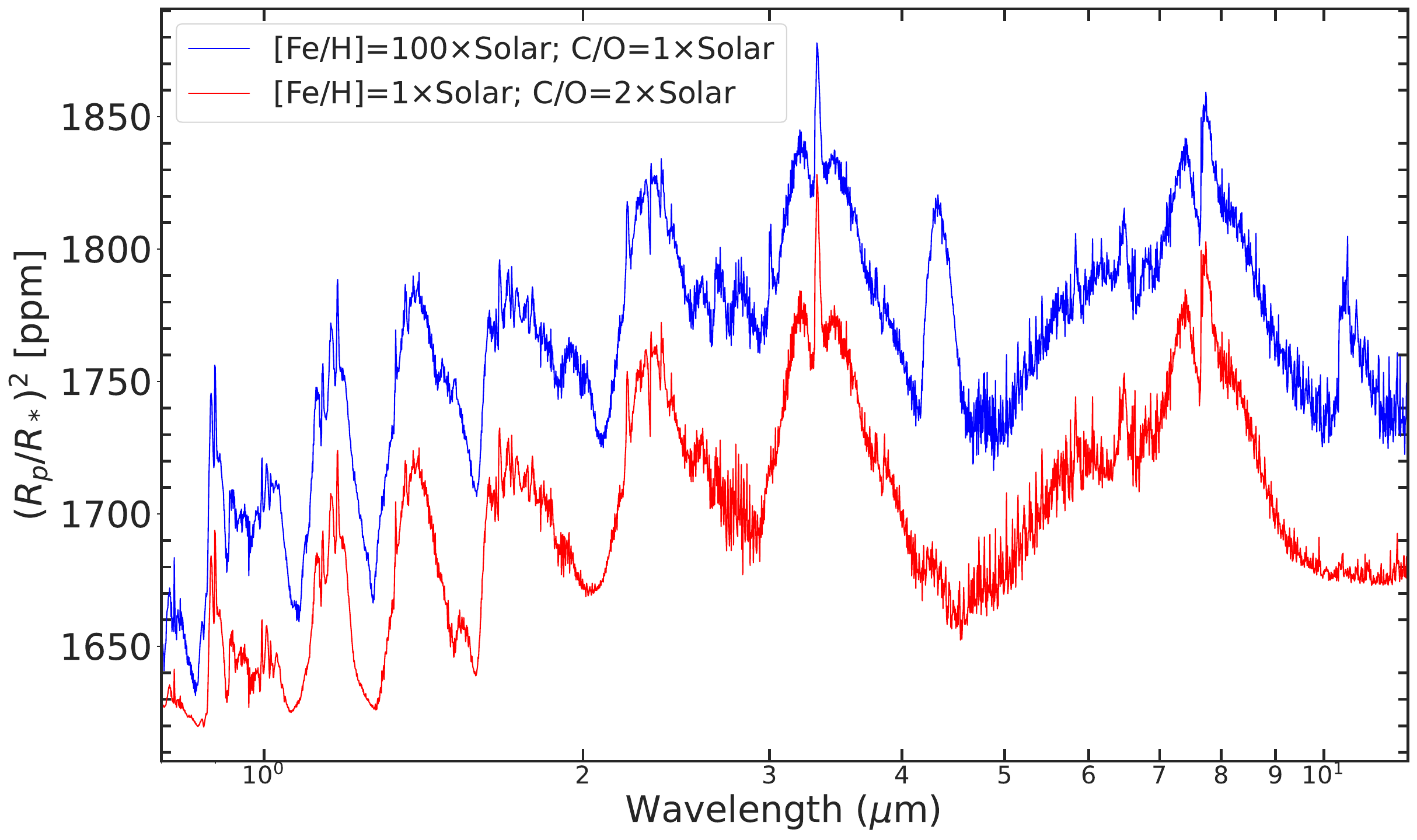} \hfill
        \includegraphics[width=\columnwidth]{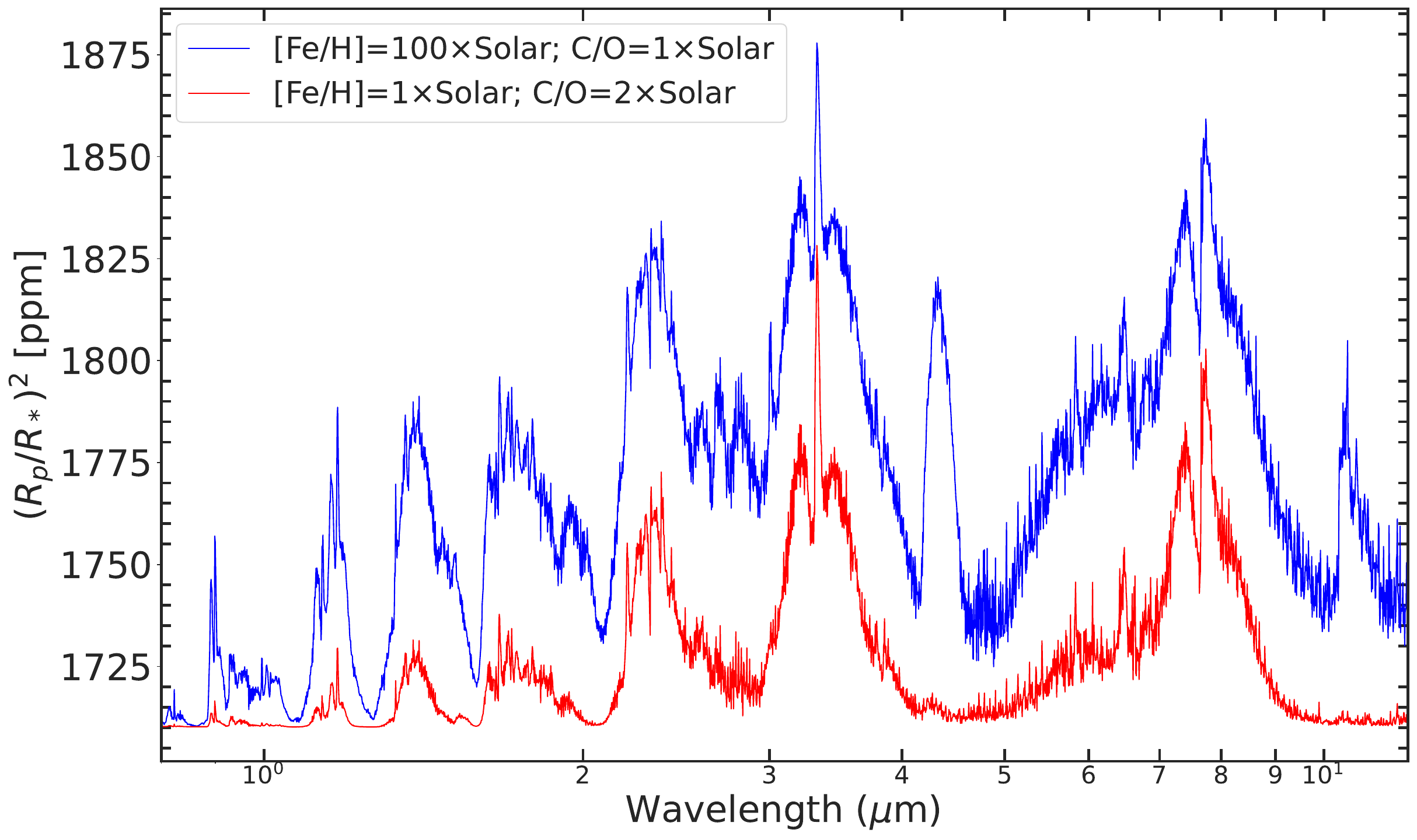}
\caption{Top panel: Models with solar abundance are shown as solid lines. PandExo simulated observations for JWST NIRISS-SOSS, NIRSpec-G395M, and MIRI-LRS modes are also depicted, with wavelength coverage denoted by colored error bars. Bottom panel: Models include one with metallicity enhanced by a factor of 100 and another with the carbon-to-oxygen ratio enhanced by a factor of two. Cloud-free spectra are represented in the left column, while the right column features cloudy spectra.}
	\label{tr1}		
\end{figure*}

Figure~\ref{tr1} displays our synthetic JWST transmission spectra of TOI-1846\,b assuming different metallicities, carbon-to-oxygen ratios, and cloud effects. For our models assuming atmospheric compositions similar to solar abundance (top panel), the major sources of opacity are H$_2$O and CH$_4$, with notable contributions from HCN and CO$_2$. The spectral features attributed to water and methane are well above the JWST’s noise floor; 20 ppm and 50 ppm, for NIRISS SOSS and MIRI LRS, respectively \citep{greene2016characterizing}. To simulate observations of our planet, we employed the JWST instrument noise simulator PandExo \citep{bat}. In this scenario, the cloudy case exhibits smaller absorption features due to the suppression of contributions from deeper atmospheric layers. The features are significantly muted with 10$^{-2}$ bar cloud top.\\

Additionally, we explored two variations from our baseline model ([Fe/H]=1xSolar; C/O=1xSolar) : (1) a carbon-to-oxygen ratio (C/O) that is two times the solar value, and (2) a metallicity that is a hundred times greater than that of the sun. Enhancing the C/O by itself does not significantly alter the composition or the spectral characteristics, as illustrated in the bottom panel of Figure \ref{tr1}. In contrast, enhanced metallicity leads to more pronounced spectral features, including notable signatures from CO$_2$ at 4.2 $\mu$m. Consistent with expectations, CH$_4$ and CO$_2$ play a more significant role in smaller planets with higher metallicities \citep{molaverdikhani2019cold}.

\section{Conclusion}\label{sec:7}

We have validated TOI-1846 b using \tess\ and multicolor ground-based photometric data, high-resolution imaging, and spectroscopic observations. Our analyses found that TOI-1846 b is a super-Earth with a radius of $1.792_{-0.068}^{+0.065}$\re\ and an orbit of $P = 3.93067$~days around an M dwarf star. With these parameters, TOI-1846 b is located in the wedge of keystone planets in the radius valley. According to \cite{Luque&palle2022}, TOI-1846 b is first guessed to be a water-rich world. These findings make TOI-1846 b well-suited for mass determination via RV observations. This could be possible with the MAROON-X instrument. When it is done, this planet will join the sample of keystone planets, necessary to test the various mechanisms thought to be at work
in the formation and evolution of super-Earths and sub-Neptunes around M dwarf stars. 

%$$$$$$$$$$$$$$$$$$$$$$$$$$$$$$$$$$$$$$
%$$$$$$$$$$$$$$$$$$$$$$$$$$$$$$$$$$$$$$
%$$$$$$$$$$$$$$$$$$$$$$$$$$$$$$$$$$$$$$
%$$$$$$$$$$$$$$$$$$$$$$$$$$$$$$$$$$$$$$
\include{table_results}
%$$$$$$$$$$$$$$$$$$$$$$$$$$$$$$$$$$$$$$
%$$$$$$$$$$$$$$$$$$$$$$$$$$$$$$$$$$$$$$
%$$$$$$$$$$$$$$$$$$$$$$$$$$$$$$$$$$$$$$

%is well-suited for ground-based RV mass determination with instruments such as MAROON-X. A precise mass measurement, combined with our radius, will make TOI-1846\,b a valuable addition to the study of small-planet populations and planet formation scenarios.

%Given its radius, TOI-1846 b joins a small group of planets located within a transition zone where \cite{luque2022density} have identified both rocky planets and water worlds. Measuring the planet's density could provide deeper insights into the different formation histories of these two populations. Additionally, in the context of rocky planet formation scenarios, TOI-1846 b occupies a sparsely populated region in the period-radius plane, known as the 'keystone' wedge, as defined by \cite{Cloutier_Menour_2020}. If TOI-1846\,b is confirmed to be a water world or a sub-Neptune, it would lend further support to the gas-depleted formation scenario.

%%%%%%%%%%%%%%%%%%%%%%%%%%%%%%%%%%%%%%%%%%%%%%%%%
%\begin{comment}
\section*{Acknowledgements}
% \tess\
Funding for the \tess\ mission is provided by NASA's Science Mission directorate. 
% \tess\ Alerts
We acknowledge the use of \tess\ public data from pipelines at the \tess\ Science Office and at the \tess\ Science Processing Operations Center.
%SPOC
Resources supporting this work were provided by the NASA High-End Codmputing (HEC) Program through the NASA Advanced Supercomputing (NAS) Division at Ames Research Center for the production of the SPOC data products.
% ExoFOP
This research has made use of the Exoplanet Follow-up Observation Program website, which is operated by the California Institute of Technology, under contract with the National Aeronautics and Space Administration under the Exoplanet Exploration Program. 
% MAST
This paper includes data collected by the \tess\ mission, which are publicly available from the Mikulski Archive for Space Telescopes\ (MAST). 
%Gaia
This work has made use of data from the European Space Agency (ESA) mission
{\it Gaia} (\url{https://www.cosmos.esa.int/gaia}), processed by the {\it Gaia} Data Processing and Analysis Consortium (DPAC,
\url{https://www.cosmos.esa.int/web/gaia/dpac/consortium}). Funding for the DPAC has been provided by national institutions, in particular the institutions participating in the {\it Gaia} Multilateral Agreement.
%SAI-2m
%TAP
This research uses data obtained through the Telescope Access Program (TAP), which has been funded by the TAP member institutes.
%MUSCAT
This work is partly supported by JSPS KAKENHI Grant Numbers JP17H04574 , JP18H05439, 20K14521, JST PRESTO Grant Number JPMJPR1775, and the Astrobiology Center of National Institutes of Natural Sciences (NINS) (Grant Number AB031010).
%MUSCAT2
This article is based on observations made with the MuSCAT2 instrument, developed by ABC, at Telescopio Carlos Sánchez operated on the island of Tenerife by the IAC in the Spanish Observatorio del Teide.
%%
%Kast
The paper is based on observations made with the Kast spectrograph on the Shane 3m telescope at Lick Observatory.
A major upgrade of the Kast spectrograph was made possible through
generous gifts from the Heising-Simons Foundation and William and Marina Kast. 
We acknowledge that Lick Observatory sits on the unceded ancestral homelands of the Chochenyo and Tamyen Ohlone peoples, including the Alson and Socostac tribes, who were the original inhabitants of the area that includes Mt. Hamilton.
% Gemini RES
Some of the observations in this paper made use of the High-Resolution Imaging instrument ‘Alopeke and were obtained under Gemini LLP Proposal Number: GN-2024-DD-101. ‘Alopeke was funded by the NASA Exoplanet Exploration Program and built at the NASA Ames Research Center by Steve B. Howell, Nic Scott, Elliott P. Horch, and Emmett Quigley. Alopeke was mounted on the Gemini North telescope of the international Gemini Observatory, a program of NSF’s OIR Lab, which is managed by the Association of Universities for Research in Astronomy (AURA) under a cooperative agreement with the National Science Foundation. on behalf of the Gemini partnership: the National Science Foundation (United States), National Research Council (Canada), Agencia Nacional de Investigación y Desarrollo (Chile), Ministerio de Ciencia, Tecnología e Innovación (Argentina), Ministério da Ciência, Tecnologia, Inovações e Comunicações (Brazil), and Korea Astronomy and Space Science Institute (Republic of Korea).
% IRTF
Visiting Astronomer at the Infrared Telescope Facility, which is operated by the University of Hawaii under contract 80HQTR24DA010 with the National Aeronautics and Space Administration.
B.S.S. and I.A.S. acknowledge the support of M.V. Lomonosov Moscow State University Program of Development.
%%% K Barkaoui and M. Gillon
The postdoctoral fellowship of KB is funded by F.R.S.-FNRS grant T.0109.20 and by the Francqui Foundation. MG is F.R.S.-FNRS Research Director.
% B Rackham
This material is based upon work supported by the National Aeronautics and Space Administration under Agreement No.\ 80NSSC21K0593 for the program ``Alien Earths''.
The results reported herein benefited from collaborations and/or information exchange within NASA’s Nexus for Exoplanet System Science (NExSS) research coordination network sponsored by NASA’s Science Mission Directorate.
% Norio
This work is partly supported by JSPS KAKENHI Grant Number JP24H00017 and JSPS Bilateral Program Number JPJSBP120249910.
%tpfplotter
This work made use of \texttt{tpfplotter} by J. Lillo-Box (publicly available in \url{https://www.github.com/jlillo/tpfplotter})
and \texttt{kastredux} by A.\ Burgasser 
(publicly available in \url{https://github.com/aburgasser/kastredux}),
both of which also make use of the python packages \texttt{astropy}, \texttt{lightkurve}, \texttt{matplotlib} and \texttt{numpy}.
%\end{comment}

%%%%%%%%%%%%%%%%%%%%%%%%%%%%%%%%%%%%%%%%%%%%%%%%%%
\section*{Data Availability}
This paper includes photometric data collected by the \tess\ mission and ground instruments, which are publicly available in ExoFOP, at \url{https://exofop.ipac.caltech.edu/tess/target.php?id=336128819}. All spectroscopy data underlying this article are listed in the text. All of the high-resolution speckle imaging data is available at the NASA exoplanet Archive with no proprietary period.

%%%%%% List of affiliations

\section*{List of affiliations}
% List of institutions \\
$^{1}$Oukaimeden Observatory, High Energy Physics and Astrophysics
Laboratory, Cadi Ayyad University, Marrakech, Morocco\\
$^{2}$Astrobiology Research Unit, Universit\'e de Li\`ege, All\'ee du 6 Ao\^ut 19C, B-4000 Li\`ege, Belgium \\
$^{3}$Department of Earth, Atmospheric and Planetary Science, Massachusetts Institute of Technology, 77 Massachusetts Avenue, Cambridge, MA 02139, USA\\
$^{4}$Instituto de Astrof\'isica de Canarias (IAC), Calle V\'ia L\'actea s/n, 38200, La Laguna, Tenerife, Spain\\
$^{5}$Department of Physics and Kavli Institute for Astrophysics and Space Research, Massachusetts Institute of Technology, Cambridge, MA 02139, USA\\
$^{6}$Departamento de Astrof\'isica, Universidad de La Laguna (ULL), E-38206 La Laguna, Tenerife, Spain\\
$^{7}$Komaba Institute for Science, The University of Tokyo, 3-8-1 Komaba, Meguro, Tokyo 153-8902, Japan\\
$^{8}$Astrobiology Center, 2-21-1 Osawa, Mitaka, Tokyo 181-8588, Japan\\
$^{9}$Center for Astrophysics \textbar  Harvard \& Smithsonian, 60 Garden St, Cambridge, MA 02138, USA\\
$^{10}$ University of Southern Queensland, Centre for Astrophysics, West Street, Toowoomba, QLD 4350 Australia\\
$^{11}$Department of Astrophysics \& Astrophysics, University of California San Diego, La Jolla, CA 92093, USA\\
$^{12}$NASA Exoplanet Science Institute-Caltech/IPAC, Pasadena, CA 91125, USA\\
$^{13}$Department of Astronomy, California Institute of Technology, Pasadena, CA 91125, USA\\
$^{14}$Department of Astronomy, University of Maryland, College Park, College Park, MD 20742 USA\\
$^{15}$Sternberg Astronomical Institute Lomonosov Moscow State University
119992, Moscow, Russia, Universitetskii prospekt, 13\\
$^{16}$Department of Astrophysical Sciences, Princeton University, Princeton, NJ 08544, USA\\
$^{17}$NASA Ames Research Center, Moffett Field, CA 94035, USA\\
$^{18}$Department of Astronomy, California Institute of Technology, 1200 E. California Blvd, Pasadena, CA 91125, USA\\
$^{19}$Kotizarovci Observatory, Sarsoni 90, 51216 Viskovo, Croatia\\
$^{20}$George Mason University, 4400 University Drive, Fairfax, VA, 22030 USA\\
$^{21}$Department of Multi-Disciplinary Sciences, Graduate School of Arts and Sciences, The University of Tokyo, 3-8-1 Komaba, Meguro, Tokyo 153-8902, Japan\\
$^{22}$National Astronomical Observatory of Japan, 2-21-1 Osawa, Mitaka,
Tokyo 181-8588, Japan\\
$^23$Bay Area Environmental Research Institute, Moffett Field, CA 94035, USA\\
$^24$NASA Exoplanet Science Institute, IPAC, California Institute of Technology, Pasadena, CA 91125 USA\\
%##############################
%\tableofcontents



\bibliographystyle{mnras}
\bibliography{planet} % if your bibtex file is called example.bib

%\begin{thebibliography}{99}
%\end{thebibliography}

%%%%%%%%%%%%%%%%%%%%%%%%%%%%%%%%%%%%%%%%%%%%%%%%%%

%%%%%%%%%%%%%%%%% APPENDICES %%%%%%%%%%%%%%%%%%%%%

%xxx


%%%%%%%%%%%%%%%%%%%%%%%%%%%%%%%%%%%%%%%%%%%%%%%%%%

% Don't change these lines
\bsp	% typesetting comment
\label{lastpage}
\appendix
\begin{figure*}
\centering
\includegraphics[width=0.33\textwidth]{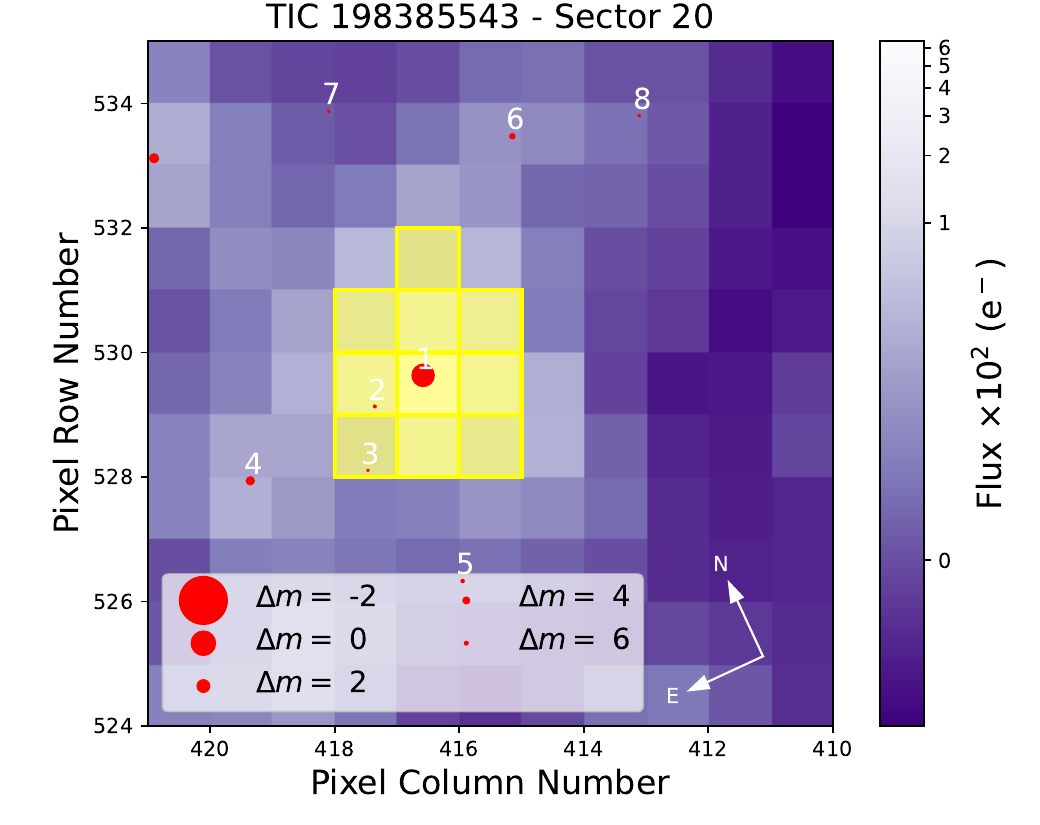}
\includegraphics[width=0.33\textwidth]{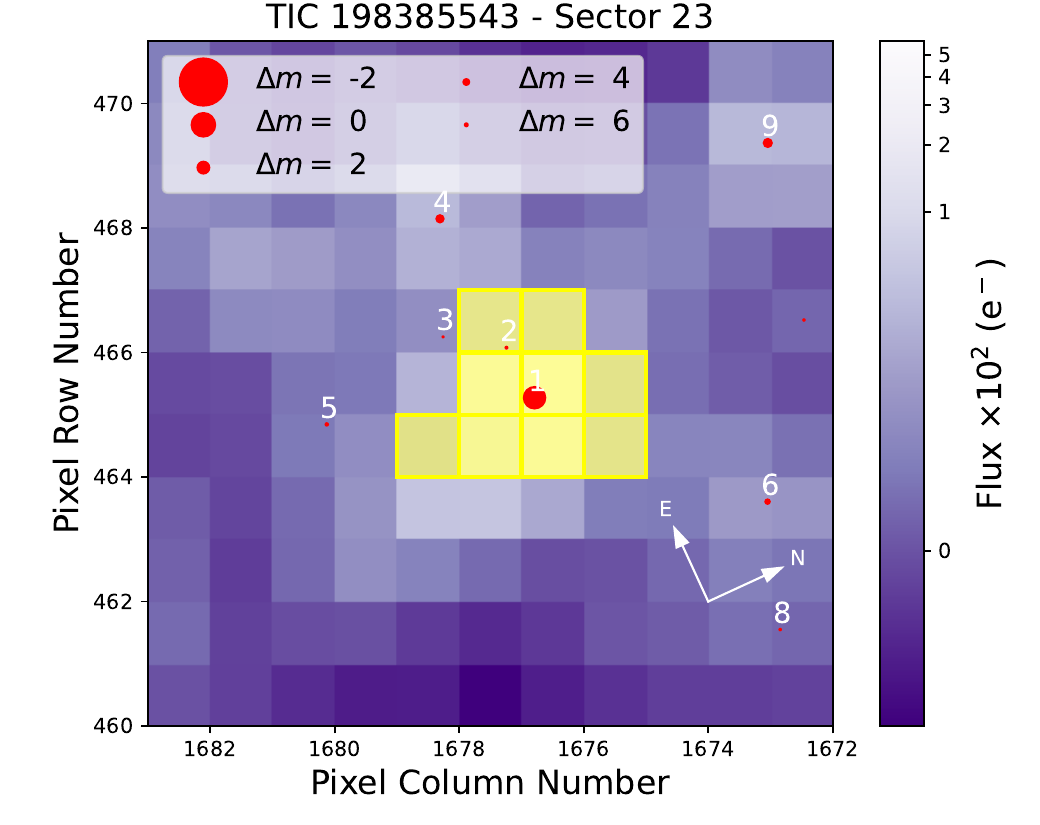}
\includegraphics[width=0.33\textwidth]{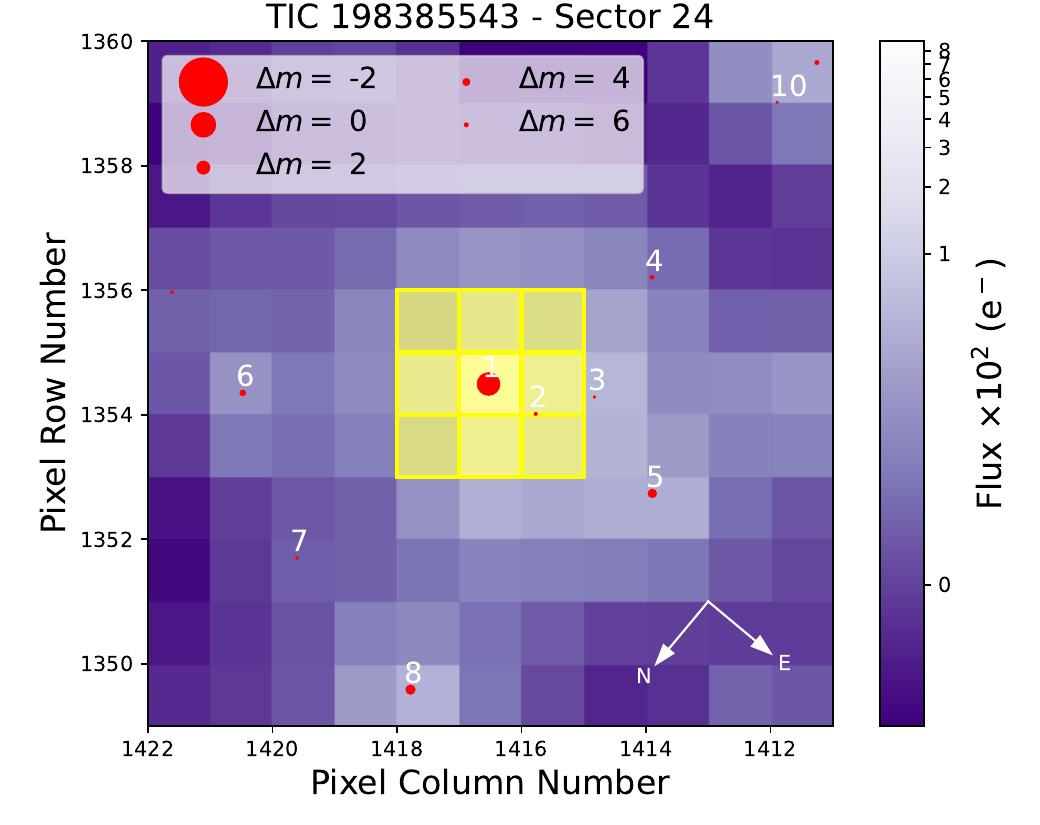}

\includegraphics[width=0.33\textwidth]{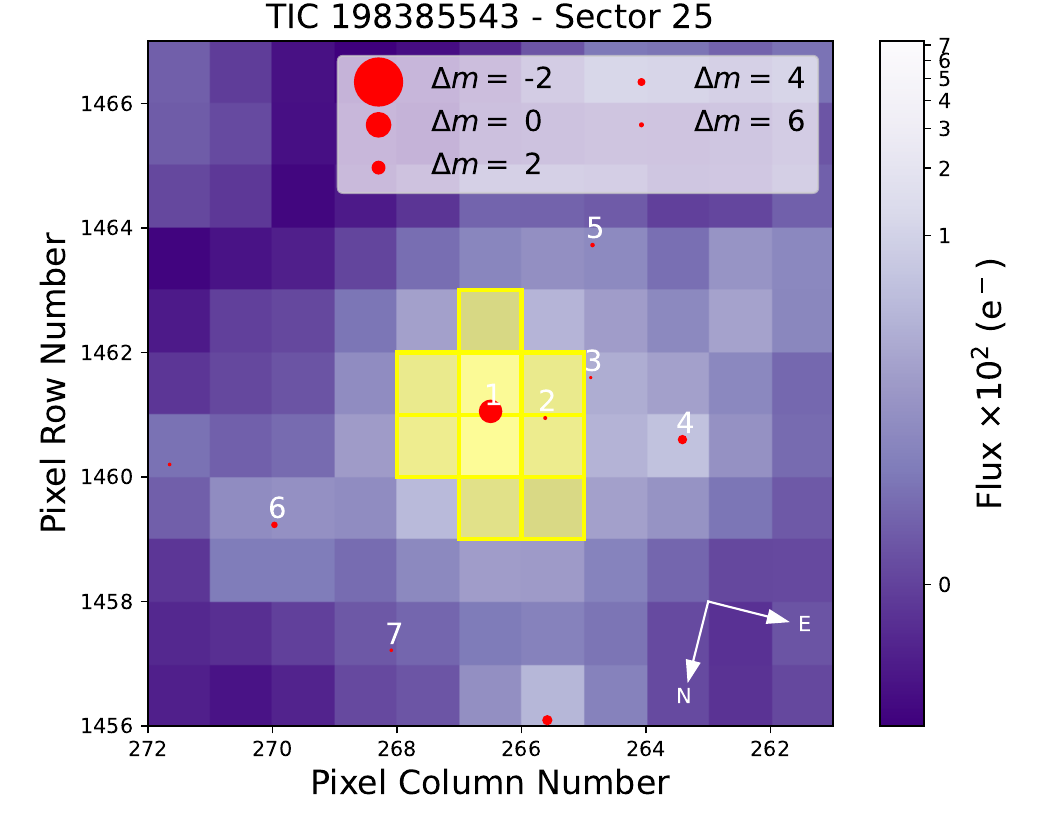}
\includegraphics[width=0.33\textwidth]{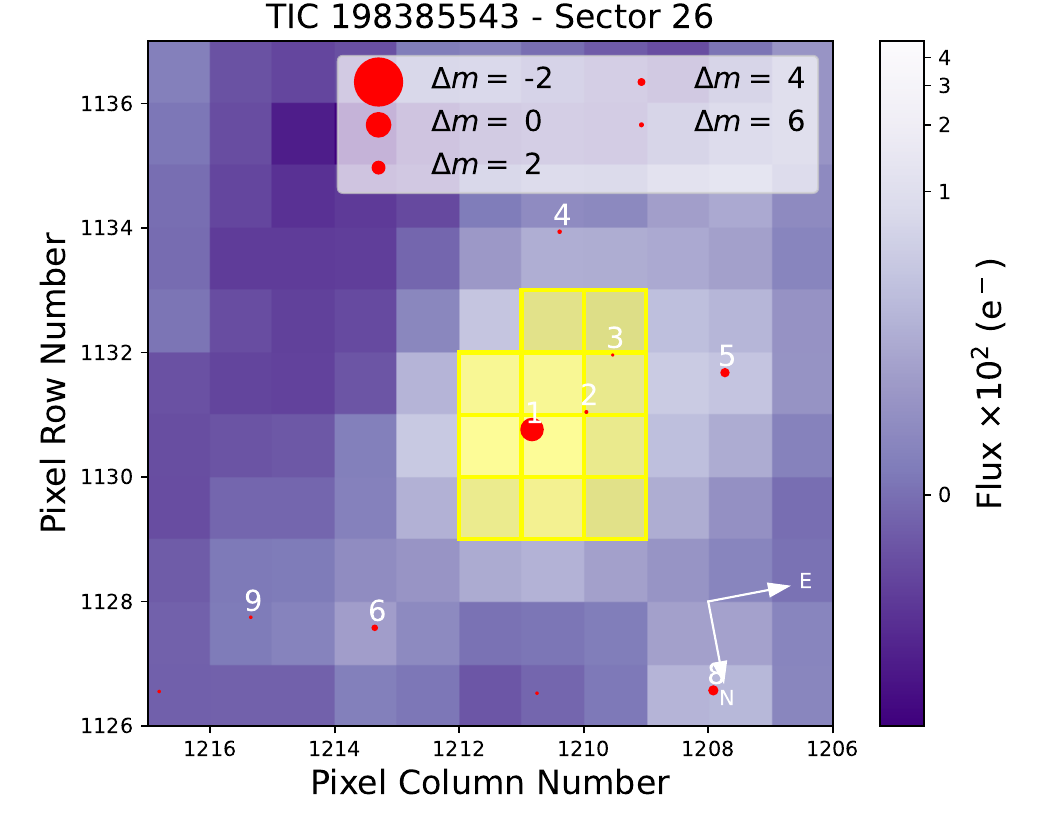}
\includegraphics[width=0.33\textwidth]{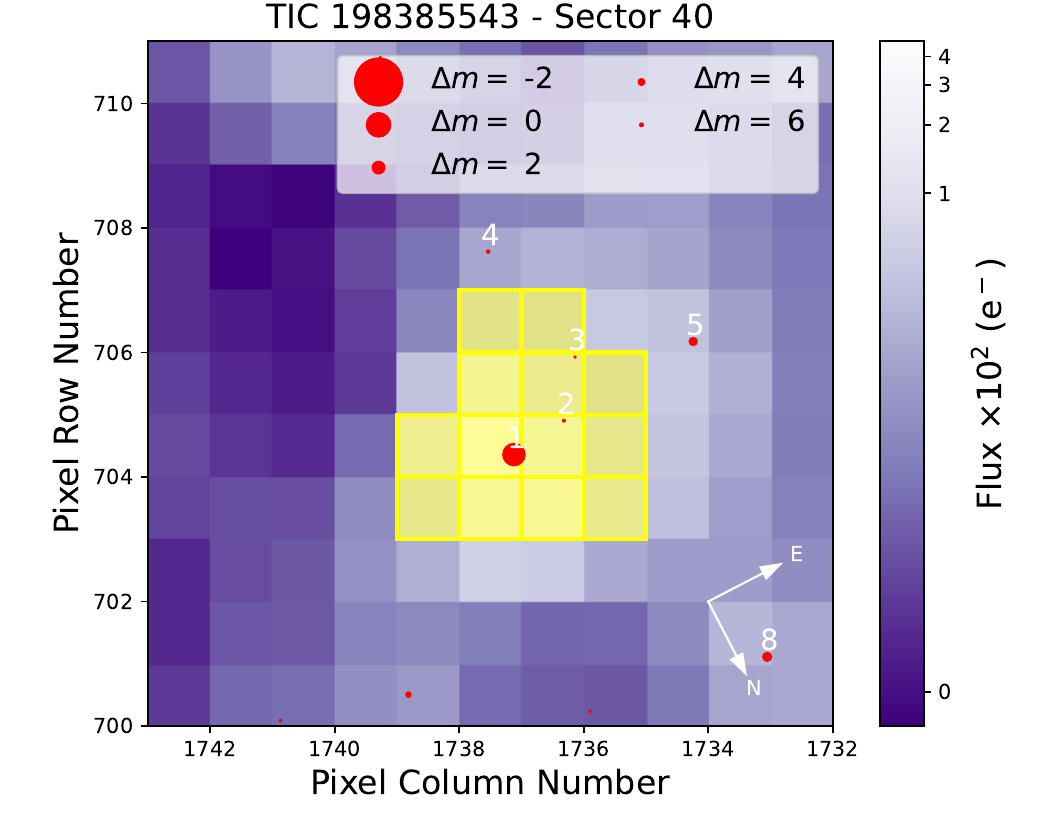}

\includegraphics[width=0.33\textwidth]{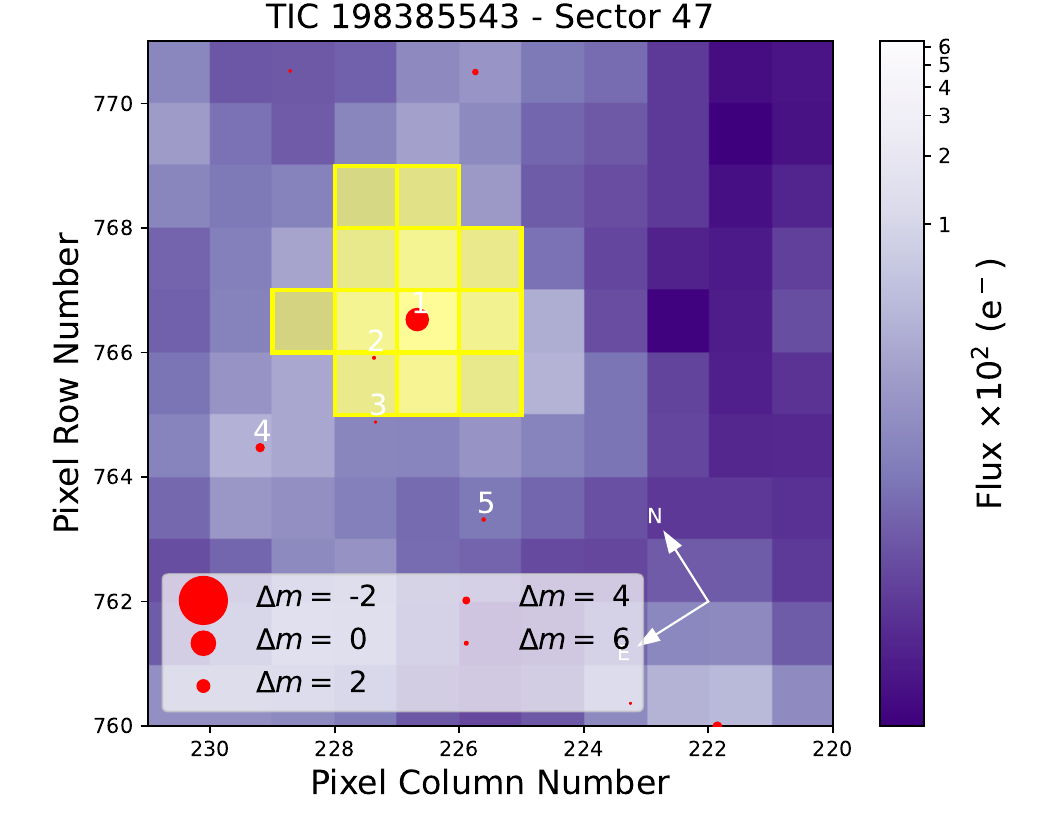}
\includegraphics[width=0.33\textwidth]{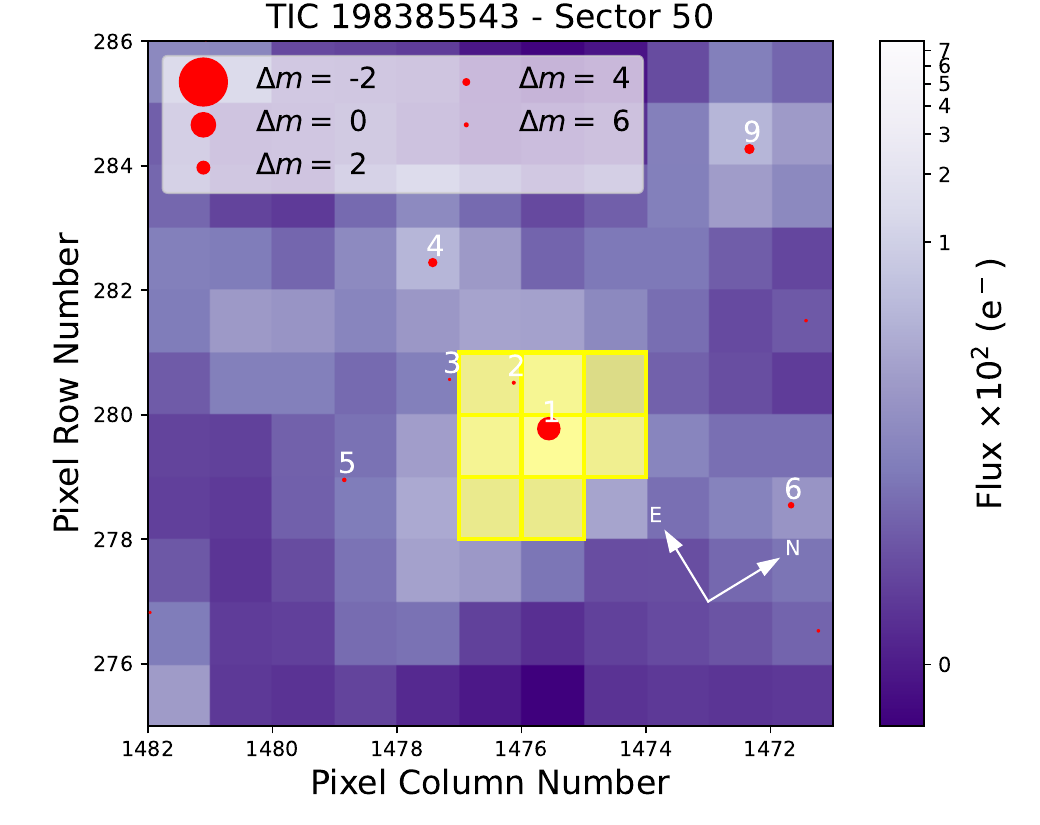}
\includegraphics[width=0.33\textwidth]{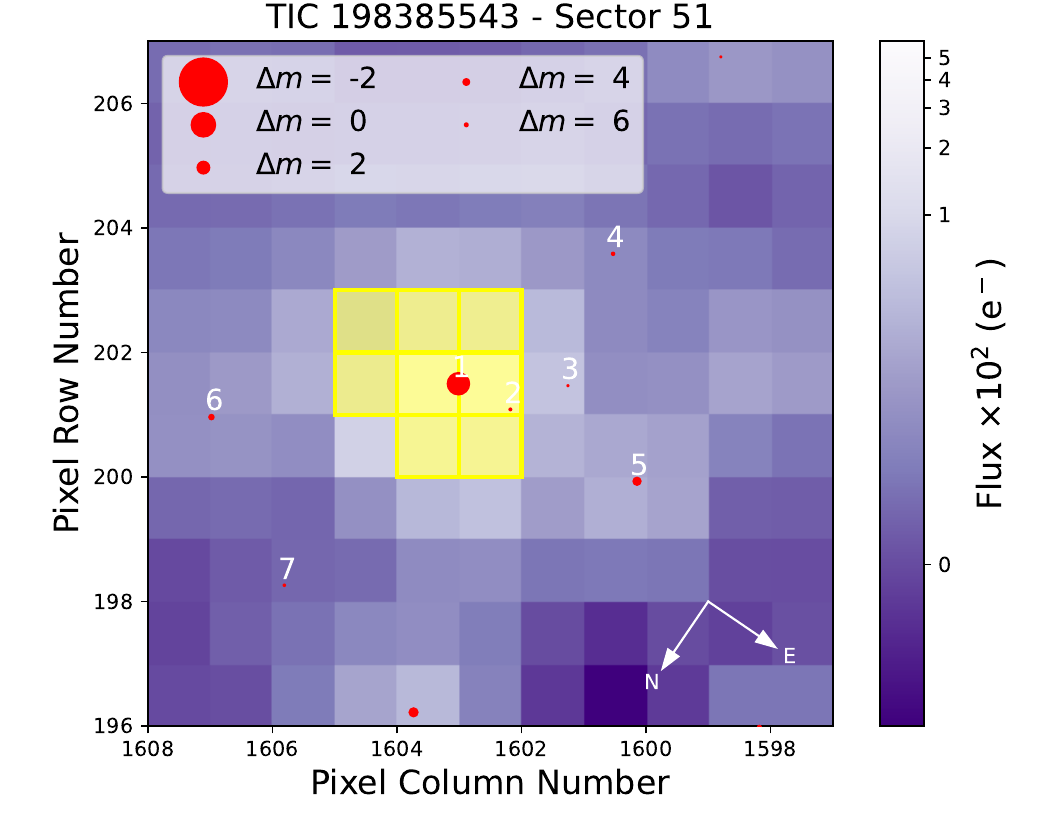}

\includegraphics[width=0.33\textwidth]{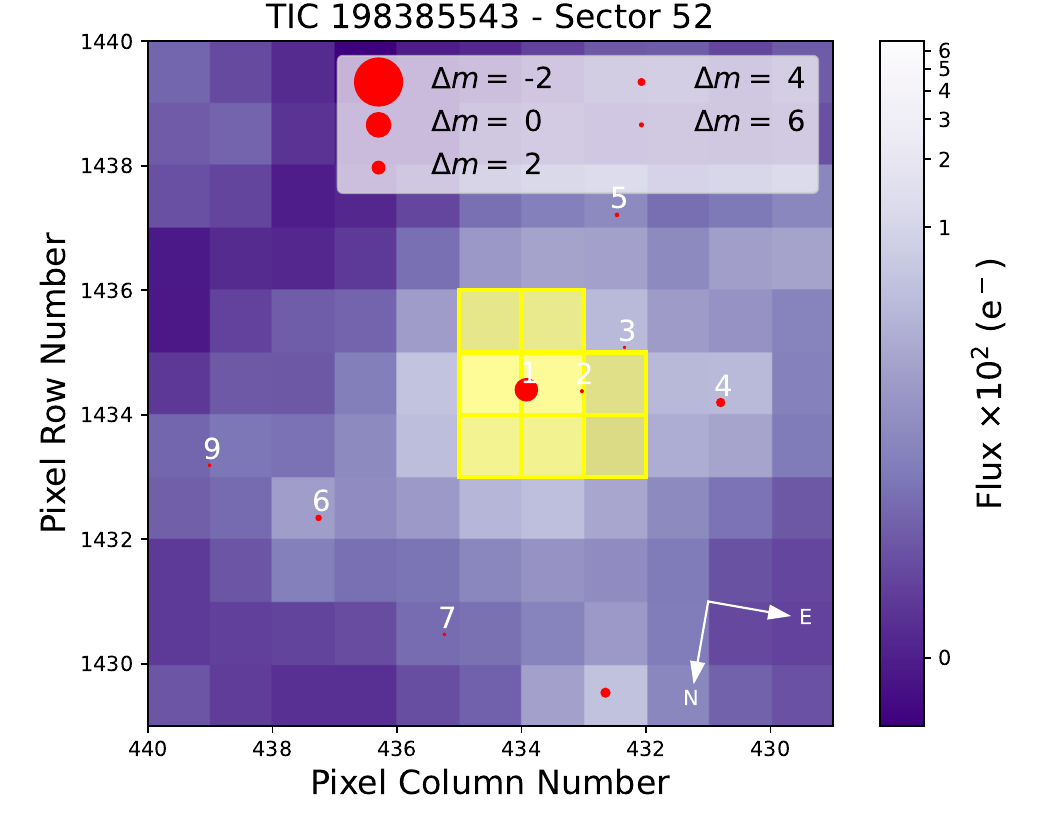}
\includegraphics[width=0.33\textwidth]{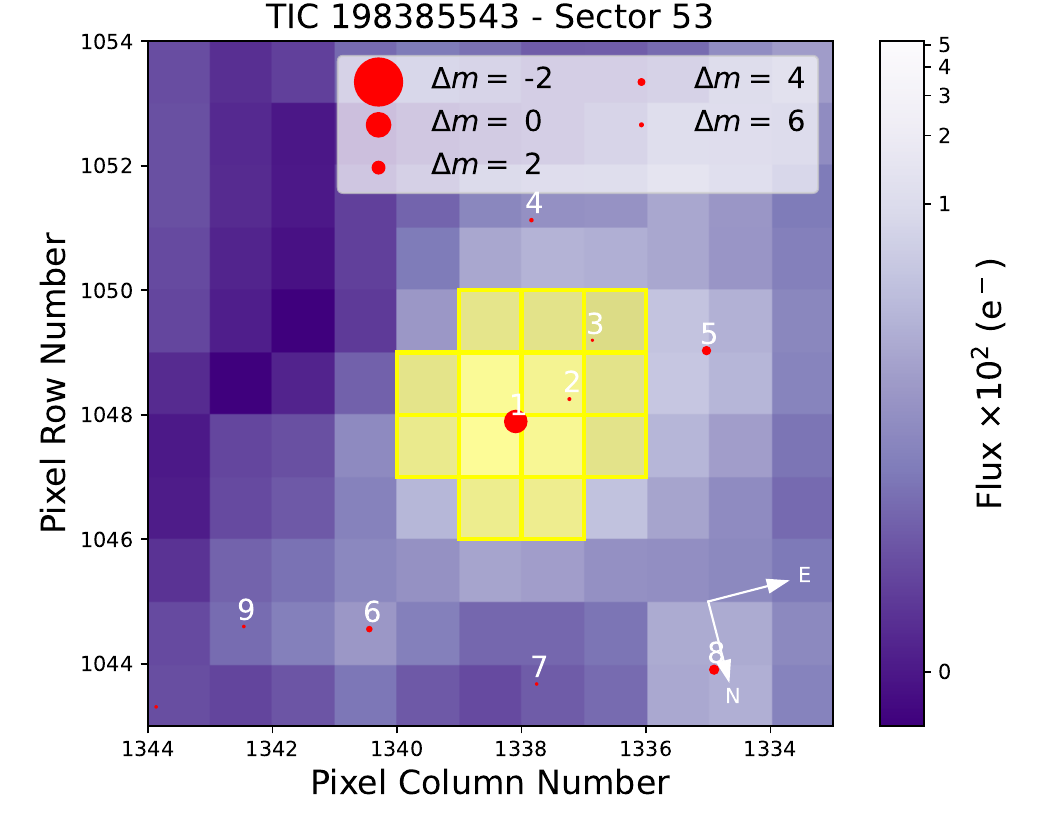}
\includegraphics[width=0.33\textwidth]{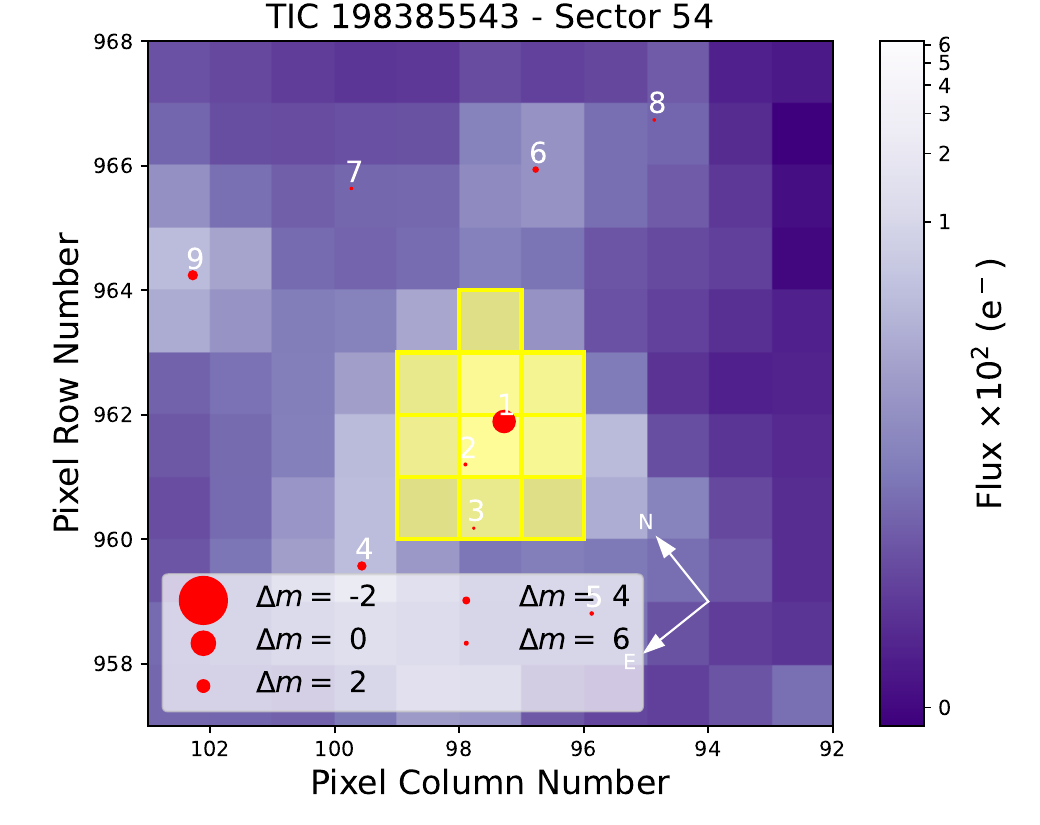}

\caption{{\it Left panel:} \tess\ target pixel file images of TOI-1846 observed in Sectors 20, 23, 24, 25, 26, 40, 47, 50, 51, 52, 53, and 54. The red circles show the sources in the field identified by the Gaia DR2 catalogue with scaled magnitudes. The position of the targets is indicated by white crosses and the mosaic of yellow squares show the mask used by the pipeline to extract photometry. These plots were made with \code{tpfplotter}.} 
\label{fov}
\end{figure*}

\begin{figure*}
 \centering
 \includegraphics[width=\textwidth]{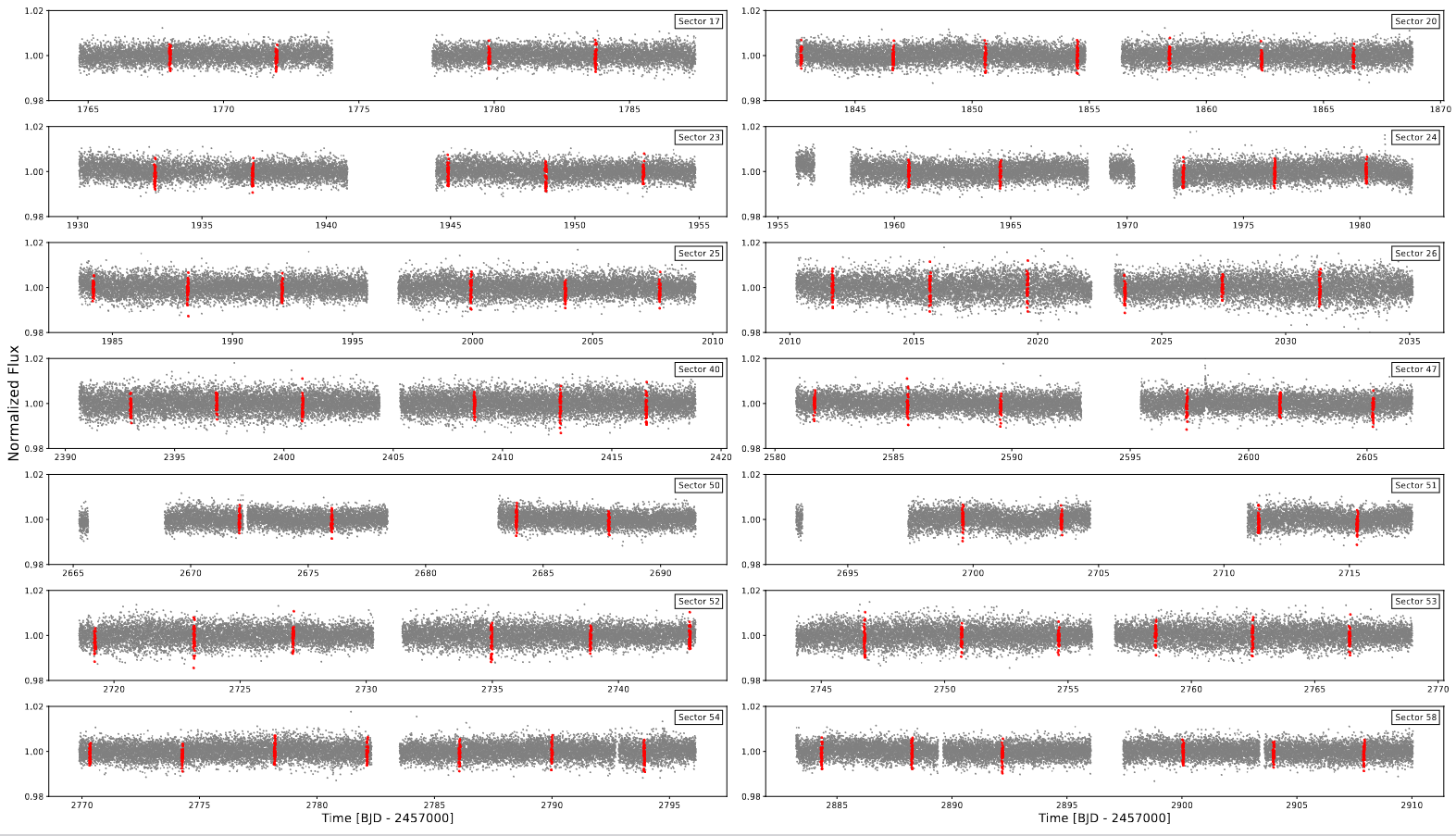}
 \caption{\tess\ photometric data of TOI-1846. The gray points show the PDSAP fluxes obtained from the SPOC pipeline. The red points correspond to the location of the transit for the candidate TOI-1846.} 
\label{TESS_photometry}
\end{figure*}

\end{document}

%% file: table_results.tex
\begin{table*}
\centering
\caption{Median values and 68\% confidence intervals for the parameters of TOI-1846\,b obtained using EXOFASTv2.}
{\renewcommand{\arraystretch}{1.2}
\resizebox{0.99\textwidth}{!}{% }
\begin{tabular}{lccccc}
\hline
\hline
~~~Parameter & \multicolumn{3}{c}{Units} & Values & \\
\hline
\multicolumn{2}{l}{Planetary Parameters:} & & & & \\
$P$ & \multicolumn{3}{c}{Period (days)} & $3.9306737^{+0.0000042}_{-0.0000044}$ &  \\
$R_p$ & \multicolumn{3}{c}{Radius (\re)} & $1.792^{+0.065}_{-0.068}$ &  \\
$M_p$ & \multicolumn{3}{c}{Mass (\me)} & $4.4^{+1.6}_{-1.0}$ &  \\
$T_c$ & \multicolumn{3}{c}{Time of conjunction (\bjdtdb)} & $2458768.02844^{+0.00100}_{-0.00089}$ &  \\
$T_T$ & \multicolumn{3}{c}{Time of minimum projected separation (\bjdtdb)}  & $2458768.02844^{+0.00100}_{-0.00089}$  & \\
$T_0$ & \multicolumn{3}{c}{Optimal conjunction time(\bjdtdb)} & $2459565.95523^{+0.00032}_{-0.00030}$ &  \\
$a$ & \multicolumn{3}{c}{Semi-major axis (AU)} & $0.03646^{+0.00072}_{-0.00074}$ &  \\
$i$ & \multicolumn{3}{c}{Inclination (Degrees)} & $88.65^{+0.20}_{-0.17}$ &  \\
$T_{eq}$ & \multicolumn{3}{c}{Equilibrium temperature (K)} & $568.1^{+6.1}_{-5.9}$ & \\
$\tau_{\rm circ}$ & \multicolumn{3}{c}{Tidal circularization timescale (Gyr)} & $85^{+32}_{-21}$ &  \\
$K$ & \multicolumn{3}{c}{RV semi-amplitude (m/s)} & $3.17^{+1.1}_{-0.73}$ &  \\
$R_P/R_*$ & \multicolumn{3}{c}{Radius of planet in stellar radii} & $0.04137\pm0.00072$ &  \\
$a/R_*$ & \multicolumn{3}{c}{Semi-major axis in stellar radii} & $19.75^{+0.56}_{-0.51}$ &  \\
$\delta$ & \multicolumn{3}{c}{Transit depth (fraction)} & $0.001712^{+0.000060}_{-0.000059}$ & \\
$Depth$ & \multicolumn{3}{c}{Flux decrement at mid transit} & $0.001712^{+0.000060}_{-0.000059}$ &  \\
$\tau$ & \multicolumn{3}{c}{Ingress/egress transit duration (days)} & $0.00296\pm0.00019$ & \\
$T_{14}$ & \multicolumn{3}{c}{Total transit duration (days)} & $0.05912^{+0.00065}_{-0.00066}$ & \\
$T_{FWHM}$ & \multicolumn{3}{c}{FWHM transit duration (days)} & $0.05616^{+0.00068}_{-0.00069}$ &  \\
$b$ & \multicolumn{3}{c}{Transit impact parameter} & $0.464^{+0.045}_{-0.057}$ &  \\
$\delta_{S,2.5\mu m}$ & \multicolumn{3}{c}{Blackbody eclipse depth at 2.5$\mu$m (ppm)} & $0.274^{+0.034}_{-0.030}$ &  \\
$\delta_{S,5.0\mu m}$ & \multicolumn{3}{c}{Blackbody eclipse depth at 5.0$\mu$m (ppm)} & $13.47^{+0.96}_{-0.91}$ & \\
$\delta_{S,7.5\mu m}$ & \multicolumn{3}{c}{Blackbody eclipse depth at 7.5$\mu$m (ppm)} & $43.0^{+2.5}_{-2.4}$ & \\
$\rho_P$ & \multicolumn{3}{c}{Density  (cgs)} & $4.20^{+1.5}_{-0.96}$ & \\
$logg_P$ & \multicolumn{3}{c}{Surface gravity} & $3.13^{+0.13}_{-0.11}$ & \\
$\Theta$ & \multicolumn{3}{c}{Safronov number} & $0.0150^{+0.0053}_{-0.0034}$ &  \\
$\fave$ & \multicolumn{3}{c}{Incident flux (\fluxcgs)} & $0.02362^{+0.0010}_{-0.00097}$ & \\
$T_P$ & \multicolumn{3}{c}{Time of periastron (\bjdtdb)} & $2458768.02844^{+0.00100}_{-0.00089}$ & \\
$T_S$ & \multicolumn{3}{c}{Time of eclipse (\bjdtdb)} & $2458766.06311^{+0.00100}_{-0.00089}$ & \\
$T_A$ & \multicolumn{3}{c}{Time of ascending node (\bjdtdb)} & $2458770.97645^{+0.00099}_{-0.00089}$ &  \\
$T_D$ & \multicolumn{3}{c}{Time of descending node (\bjdtdb)} & $2458769.01111^{+0.00099}_{-0.00089}$ &  \\
$V_c/V_e$ & \multicolumn{3}{c}{} & $1.00$ &  \\
$M_P\sin i$ & \multicolumn{3}{c}{Minimum mass (\me)} & $4.4^{+1.6}_{-1.0}$ &  \\
$M_P/M_*$ & \multicolumn{3}{c}{Mass ratio} & $0.0000314^{+0.000011}_{-0.0000073}$ & \\
$d/R_*$ & \multicolumn{3}{c}{Separation at mid transit} & $19.75^{+0.56}_{-0.51}$ &  \\
$P_T$ & \multicolumn{3}{c}{A priori non-grazing transit prob} & $0.0485\pm0.0013$ &  \\
$P_{T,G}$ & \multicolumn{3}{c}{A priori transit prob} & $0.0527^{+0.0014}_{-0.0015}$ & \\
\hline
\multicolumn{2}{l}{Wavelength Parameters:} & & & & \\
 & g' & i' & r' & z' & TESS \\
$u_{1}$  & $0.464\pm0.022$ & $0.298 \pm 0.017$ & $0.453\pm0.026$ & $0.194\pm0.010$ & $0.288\pm0.011$ \\
$u_{2}$  & $0.329\pm0.027$ & $0.296\pm0.030$ & $0.289 \pm 0.030$ & $0.335 \pm 0.021$ & $0.329\pm0.014$ \\
Dilution from neighboring stars: &    &  &   &  &  \\
$A_D$ &  -- & -- & -- & -- & $0.04\pm0.14$\\
\hline
%\multicolumn{2}{l}{Transit Parameters:} & \multicolumn{4}{c}{} \\
%$\sigma^{2}$ & Added variance & $-0.0000096^{+0.0000005}_{-0.0000004}$ & $-0.0000214^{+0.0000004}_{-0.0000003}$ & $-0.00000632^{+0.0000004}_{-0.0000004}$ & $-0.000003^{+0.0000005}_{-0.00000048}$ \\
%$\sigma^{2}$ & Added variance & $-0.00000419^{+0.00000047}_{-0.00000045}$ & $-0.00000388^{+0.00000068}_{-0.00000064}$ & $-0.00000155^{+0.00000062}_{-0.00000059}$ & $0.00000152^{+0.00000047}_{-0.00000044}$ \\
%$\sigma^{2}$ & Added variance & $0.00000053^{+0.00000027}_{-0.00000026}$ & $0.00000044^{+0.00000024}_{-0.00000019}$ & $0.00000051^{+0.00000014}_{-0.00000013}$ & \\
%$F_0$ & Baseline flux & $1.00039\pm0.00011$ & $1.000332^{+0.000086}_{-0.000087}$ & $1.00028\pm0.00010$ & $1.00042\pm0.00011$ \\
%$F_0$ & Baseline flux & $1.00024\pm0.00010$ & $1.00038^{+0.00013}_{-0.00012}$ & $1.00026\pm0.00012$ & $1.000262^{+0.000100}_{-0.000099}$ \\
%$F_0$ & Baseline flux & $1.00047\pm0.00011$ & $1.00024\pm0.00011$ & $1.00012\pm0.00011$ & $1.00028\pm0.00011$ \\
%$F_0$ & Baseline flux & $1.000300^{+0.000089}_{-0.000088}$ & $1.000255^{+0.000092}_{-0.000091}$ & $1.00113\pm0.00044$ & $1.00056\pm0.00011$ \\
%$F_0$ & Baseline flux & $1.000595^{+0.000085}_{-0.000084}$ & $1.00013\pm0.00014$ & $0.999980\pm0.000075$ & $1.000046\pm0.000066$ \\
%$F_0$ & Baseline flux & $0.99995\pm0.00018$ & $1.000023\pm0.000100$ & $1.00007\pm0.00012$ & $0.999876^{+0.000100}_{-0.000099}$ \\
\hline
\end{tabular} }}
\label{tab:198385543}
\begin{tablenotes}
\item See Table 3 in \citet{Eastman2019} for a detailed description of all parameters.
%\item[1] This value ignores the systematic error and is for reference only.  Uses measured radius and estimated mass from \citet{Chen2017}.
%\item[5] Time of conjunction is commonly reported as the "transit time".  Time of minimum projected separation is a more correct "transit time".
%\item[7] Optimal time of conjunction minimizes the covariance between $T_C$ and period.
\item  Assumes no albedo and perfect redistribution.
\item[4] The estimated mass from \protect \citet{Chen2017}
\end{tablenotes}
\end{table*}